\documentclass{article}
\usepackage{epubtk}
\usepackage{epsf}
\usepackage{longtable}
\usepackage{lscape}
\usepackage{booktabs}
\usepackage{amssymb}
\usepackage{ifpdf}

\newcommand{\vsp}{\rule{0.0 em}{1.2 em}}
\newcommand{\z}{\phantom{0}}
\newcommand{\zz}{\phantom{00}}
\newcommand{\zzz}{\phantom{000}}
\newcommand{\zzzz}{\phantom{0000}}
\newcommand{\zzzzz}{\phantom{00000}}

\newcommand{\zzzzzzz}{\phantom{0000000}}
\newcommand{\dz}{\phantom{.0}}
\newcommand{\dzz}{\phantom{.00}}

\begin{document}

\title{Binary and Millisecond Pulsars}

\author{\epubtkAuthorData{Duncan R.\ Lorimer}
        {Department of Physics\\
         West Virginia University\\
         P.O. Box 6315\\
         Morgantown\\
         WV 26506, U.S.A.}
        {Duncan.Lorimer@mail.wvu.edu}
        {http://astro.wvu.edu/people/dunc}
}

\date{}
\maketitle


\begin{abstract}
  We review the main properties, demographics and applications of
  binary and millisecond radio pulsars. Our knowledge of these
  exciting objects has greatly increased in recent years, mainly due
  to successful surveys which have brought the known pulsar population
  to over 1800. There are now 83 binary and millisecond pulsars
  associated with the disk of our Galaxy, and a further 140 pulsars in
  26 of the Galactic globular clusters. Recent highlights include
  the discovery of the young relativistic binary system PSR J1906+0746, 
  a rejuvination
  in globular cluster pulsar research including growing numbers
  of pulsars with masses in excess of $1.5\,M_{\odot}$, a precise measurement
  of relativistic spin precession in the double pulsar system
  and a Galactic millisecond pulsar in an eccentric ($e=0.44$) orbit
  around an unevolved companion.
\end{abstract}

\epubtkKeywords{pulsars}

\newpage


\section{Preamble}
\label{sec:preamble}

Pulsars -- rapidly rotating highly magnetised neutron stars -- have
many applications in physics and astronomy. Striking
examples include the confirmation of the existence of gravitational
radiation~\cite{tw82, tw89, nobpr1993}, the first extra-solar
planetary system~\cite{wf92, psrplanets} and the first detection of gas
in a globular cluster~\cite{fkl+01}. The diverse zoo of
radio pulsars currently known is summarized in Figure~\ref{fig:venn}.

\epubtkImage{venn.png}{
  \begin{figure}[htbp]
    \def\epsfsize#1#2{0.4#1}
    \centerline{\epsfbox{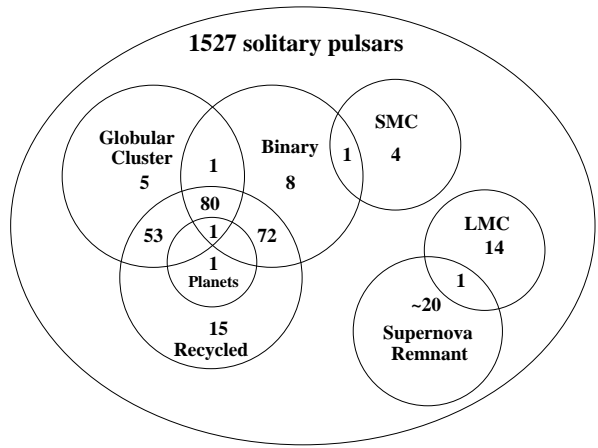}}
    \caption{Venn diagram showing the numbers and locations of the
      various types of radio pulsars known as of September 2008. The
      large and small Magellanic clouds are denoted by LMC and SMC.
     }
    \label{fig:venn}
  \end{figure}}

Pulsar research has proceeded at a rapid pace during the first three
versions of this article~\cite{lor98e, lor01, lor05}. Surveys with the
Parkes radio telescope~\cite{mbeampsr}, at Green
Bank~\cite{gbt}, Arecibo~\cite{arecibo} and the Giant Metre Wave Radio
Telescope~\cite{gmrt} have more than doubled the number of pulsars
known a decade ago. With new instrumentation coming online, and
new telescopes planned~\cite{fast,kat,ska}, 
these are exciting times for pulsar astronomy.

The aims of this review are to introduce the reader to the field and
to focus on some of the many applications of pulsar research in 
relativistic astrophysics. We begin in Section~\ref{sec:intro} with 
an overview of the pulsar phenomenon, a review of the key observed 
population properties, the origin and evolution of pulsars 
and the main search strategies. In
Section~\ref{sec:gal}, we review present understanding in pulsar
demography, discussing selection effects and their correction
techniques. This leads to empirical estimates of the total number of
normal and millisecond pulsars (see Section~\ref{sec:nmsppop}) and relativistic
binaries (see Section~\ref{sec:relpop}) in the Galaxy and has implications for
the detection of gravitational radiation by
current and planned telescopes. Our
review of pulsar timing in Section~\ref{sec:pultim} covers the basic techniques
(see Section~\ref{sec:tmodel}), timing stability (see Section~\ref{sec:tstab}), binary
pulsars (see Section~\ref{sec:tbin}), and 
using pulsars as sensitive detectors of long-period gravitational
waves (see Section~\ref{sec:gwdet}). We conclude with a brief
outlook to the future in Section~\ref{sec:future}. Up-to-date tables of
parameters of binary and millisecond pulsars are included in
Appendix~\ref{appendix}.

\newpage


\section{Pulsar Phenomenology}
\label{sec:intro}

Most of the basic observational facts about radio pulsars were
established shortly after their discovery~\cite{hbp68}
in 1967. Although there are still many open questions, the
basic model has long been established beyond all reasonable doubt,
i.e.\ pulsars are rapidly rotating, highly magnetised neutron stars
formed during the supernova explosions of massive stars. 


\subsection{The lighthouse model}
\label{sec:light}

Figure~\ref{fig:rotns} shows an animation depicting
the rotating neutron star, also known as the ``lighthouse'' model. 
As the neutron star spins, charged particles are accelerated along
magnetic field lines forming a beam (depicted by the gold
cones). The accelerating particles emit electromagnetic
radiation, most readily detected at radio frequencies as a sequence of
observed pulses. Each pulse is produced as the magnetic axis (and hence the radiation
beam) crosses the observer's line of sight each rotation. The
repetition period of the pulses is therefore simply the rotation
period of the neutron star. The moving ``tracker ball'' on the
pulse profile in the animation shows the relationship
between observed intensity and rotational phase of the neutron star.

\epubtkMovie{rotns.gif}{rotns.png}{
  \begin{figure}[htbp]
    \def\epsfsize#1#2{0.7#1}
    \centerline{\epsfbox{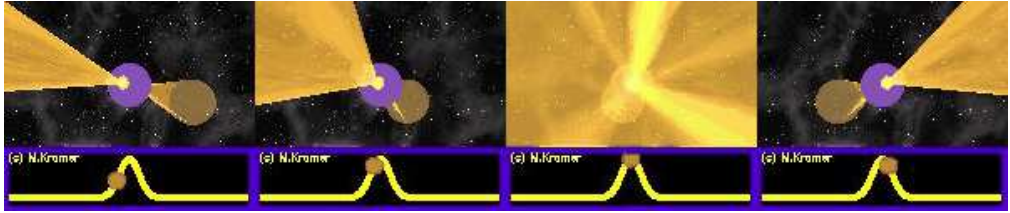}}
    \caption{The rotating neutron star (or ``lighthouse'') model for
    pulsar emission. Animation designed by Michael Kramer.}
    \label{fig:rotns}
  \end{figure}}

Neutron stars are essentially large celestial flywheels with moments
of inertia $\sim 10^{38} \mathrm{\ kg\ m}^2$. The rotating neutron star
model~\cite{pac68, gol68} predicts a gradual slowdown and hence an
increase in the pulse
period as the outgoing radiation carries away rotational kinetic
energy. This idea gained strong support when a period increase
of 36.5~ns per day was measured for the pulsar B0531+21\epubtkFootnote{Like
most astronomical sources, pulsars are
named after their position in the sky. Those pulsars discovered prior
to the mid 1990s are named based on the Besselian equatorial coordinate
system (B1950) and have a B prefix followed by their right ascension
and declination. More recently discovered pulsars follow the Julian
(J2000) epoch. Pulsars in globular clusters, where the positional
designation is not unique, have additional characters to distinguish
them. For example PSR~J1748$-$2446ad in the globular cluster
Terzan~5~\cite{hrs06}.} in the Crab
nebula~\cite{rc69b}, which implied that a rotating neutron star with a
large magnetic field must be the dominant energy supply for the
nebula~\cite{gol69}.


\subsection{Pulse periods and slowdown rates}
\label{sec:nms}

The pulsar
catalogue~\cite{mhth05,psrcat} contains up-to-date parameters for 1775
pulsars. Most of these are ``normal'' with pulse periods $P \sim 0.5 \mathrm{\ s}$
which increase secularly at rates $\dot{P} \sim 10^{-15} \mathrm{\ s/s}$.
A growing fraction are ``millisecond pulsars'',
with $1.4 \mathrm{\ ms} \lesssim P \lesssim 30 \mathrm{\ ms}$
and $\dot{P} \lesssim 10^{-19} \mathrm{\ s/s}$. As shown in the ``$P \mbox{--} \dot{P}$ diagram'' in
Figure~\ref{fig:ppdot}, normal and millisecond pulsars are distinct
populations. 

\epubtkImage{ppdot.png}{
  \begin{figure}[htbp]
    \def\epsfsize#1#2{0.6#1}
    \centerline{\epsfbox{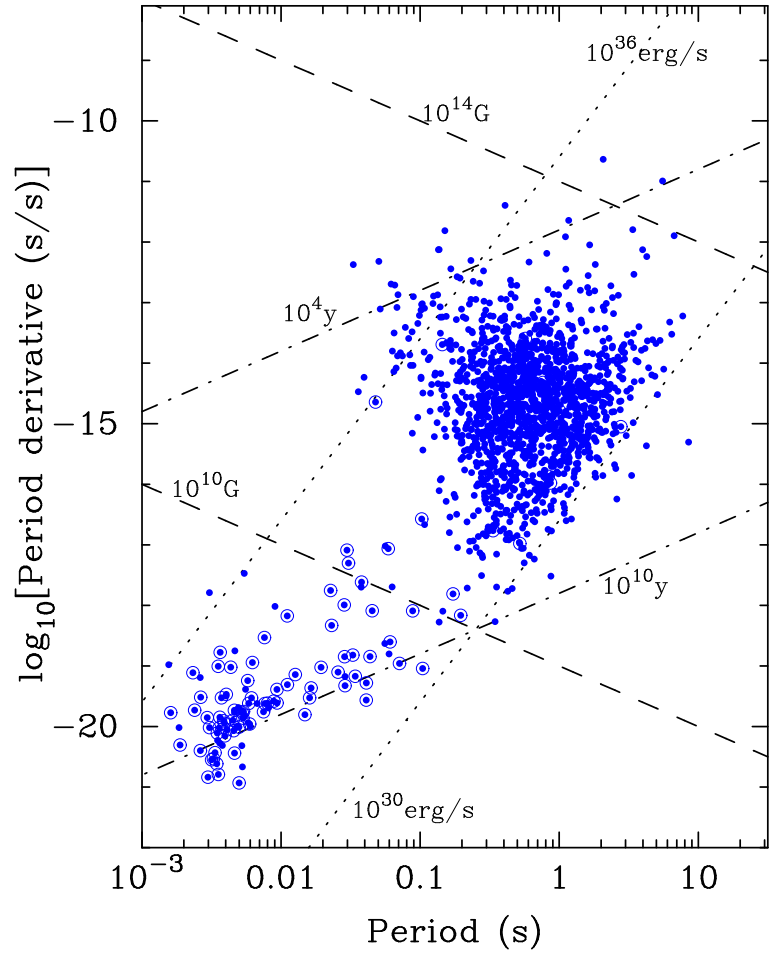}}
    \caption{The $P \mbox{--} \dot{P}$ diagram showing the current
      sample of radio pulsars. Binary pulsars are highlighted by open
      circles. Lines of constant magnetic field (dashed),
      characteristic age (dash-dotted) and spin-down energy loss rate
      (dotted) are also shown.}
    \label{fig:ppdot}
  \end{figure}}

The differences in $P$ and $\dot{P}$ imply fundamentally
different magnetic field strengths and ages. Treating the pulsar as a
rotating magnetic dipole, one may show~\cite{lk05} that the surface
magnetic field strength $B \propto (P \dot{P})^{1/2}$ and the
characteristic age $\tau_\mathrm{c} = P/(2\dot{P})$.
Lines of constant $B$ and $\tau_\mathrm{c}$ are drawn on
Figure~\ref{fig:ppdot}, from which we infer typical values of
10\super{12}~G and 10\super{7}~yr for the normal pulsars and
10\super{8}~G and 10\super{9}~yr for the millisecond
pulsars. For the rate of loss of kinetic energy, sometimes called the
spin-down luminosity, we have $\dot{E} \propto \dot{P}/P^3$. The lines
of constant $\dot{E}$ shown on Figure~\ref{fig:ppdot} indicate that the
most energetic objects are the very young normal pulsars and the most
rapidly spinning millisecond pulsars. 

The most rapidly rotating neutron star currently known,
J1748$-$2446ad,with a spin rate of 716~Hz, resides in the globular
cluster Terzan~5~\cite{hrs06}. As discussed by Lattimer \&
Prakash~\cite{lp07}, the limiting (non-rotating) radius of a
$1.4\,M_{\odot}$ neutron star with this period is 14.3~km. If a
precise measurement for the pulsar mass can be made through future
timing measurements (Section~\ref{sec:pultim}), then this pulsar could
be a very useful probe of the equation of state of super dense matter.

While the hunt for more rapidly rotating pulsars and even
``sub-millisecond pulsars'' continues, and most neutron star equations
of state allow higher spin rates than 716~Hz, it has been
suggested~\cite{bil98} that the dearth of pulsars with $P<1.5
\mathrm{\ ms}$ is caused by gravitational wave emission from
Rossby-mode instabilities~\cite{ajks00}.  
The most rapidly rotating pulsars~\cite{hrs06} are
predominantly members of eclipsing binary systems which could hamper
their detection in radio surveys. Independent constraints on the
limiting spin frequencies of neutron stars come from studies
of millisecond X-ray binaries~\cite{cmm03} which are not thought
to be selection-effect limited~\cite{cha05}. This analysis does
not predict a significant population of neutron stars with spin
rates in excess of 730~Hz.


\subsection{Pulse profiles}
\label{sec:profs}

Pulsars are weak radio sources. Measured intensities
at 1.4~GHz vary between
$5 \mu\mathrm{Jy}$
and $1 \mathrm{\ Jy}$ ($1 \mathrm{\ Jy} \equiv 10^{-26} \mathrm{\ W\
  m}^{-2} \mathrm{\ Hz}^{-1}$). As a result, even with a large radio telescope, the
coherent addition of many hundreds or even thousands of pulses is
usually required in order to produce a detectable ``integrated
profile''. Remarkably, although the individual pulses vary
dramatically, the integrated profile at a given observing
frequency is very stable and can be thought of as a fingerprint of the
neutron star's emission beam. Profile stability is of key importance
in pulsar timing measurements discussed in Section~\ref{sec:pultim}.

The selection of integrated profiles in Figure~\ref{fig:profs} shows a
rich diversity in morphology including two
examples of ``interpulses'' -- a secondary pulse separated by about
180 degrees from the main pulse. One interpretation for
this phenomenon is that the two pulses originate from opposite
magnetic poles of the neutron star (see however~\cite{ml77}). Since
this is an unlikely viewing angle, we would expect interpulses to be a
rare phenomenon. Indeed, the fraction of known
pulsars in which interpulses are observed in their pulse profiles is
only a few percent~\cite{kgm04}. Recent work~\cite{wj08} on the statistics
of interpulses among the known population show that the interpulse
fraction depends inversely on pulse period. The origin of this
effect could be due to alignment between the spin and
magnetic axis of neutron stars on timescales of order
10\super{7}~yr~\cite{wj08}. If we take this result at face value, and
assume that all millisecond pulsars descend from normal pulsars
(Section~\ref{sec:evolution}), then the implication is that
millisecond pulsars should be preferentially aligned
rotators. However, their appears to be no strong evidence in favour of
this expectation based on the pulse profile morphology of millisecond
pulsars~\cite{kxl98}.

\epubtkImage{profs.png}{
  \begin{figure}[htbp]
    \def\epsfsize#1#2{0.55#1}
    \centerline{\epsfbox{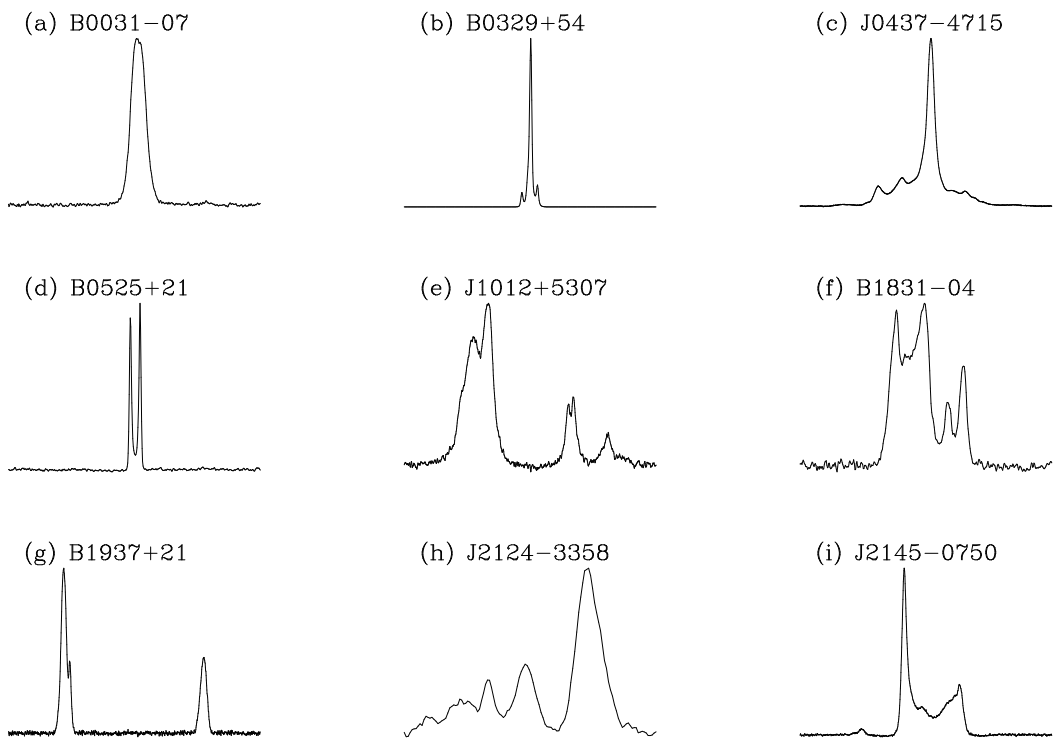}}
    \caption{A variety of integrated pulse profiles taken from the
      available literature. References: Panels~a, b, d, f~\cite{gl98},
      Panel~c~\cite{bjb97}, Panels~e, g, i~\cite{kxl98},
      Panel~h~\cite{bbm97}. Each profile represents 360 degrees of
      rotational phase. These profiles are freely available from an
      online database~\cite{epndb}.}
    \label{fig:profs}
  \end{figure}}

Two contrasting phenomenological models have been proposed
to explain the observed pulse shapes. The ``core and cone''
model~\cite{ran83} depicts the beam as a core surrounded by a series
of nested cones. Alternatively, the ``patchy beam''
model~\cite{lm88, hm01} has the beam populated by a series of
randomly-distributed emitting regions. Recent work suggests
that the observational data can be better explained by a hybrid
empirical model depicted in Figure~\ref{fig:cartoon} which employs
patchy beams in a core and cone structure~\cite{kj07}.

\epubtkImage{cartoon.png}{
  \begin{figure}[htbp]
    \def\epsfsize#1#2{0.6#1}
    \centerline{\epsfbox{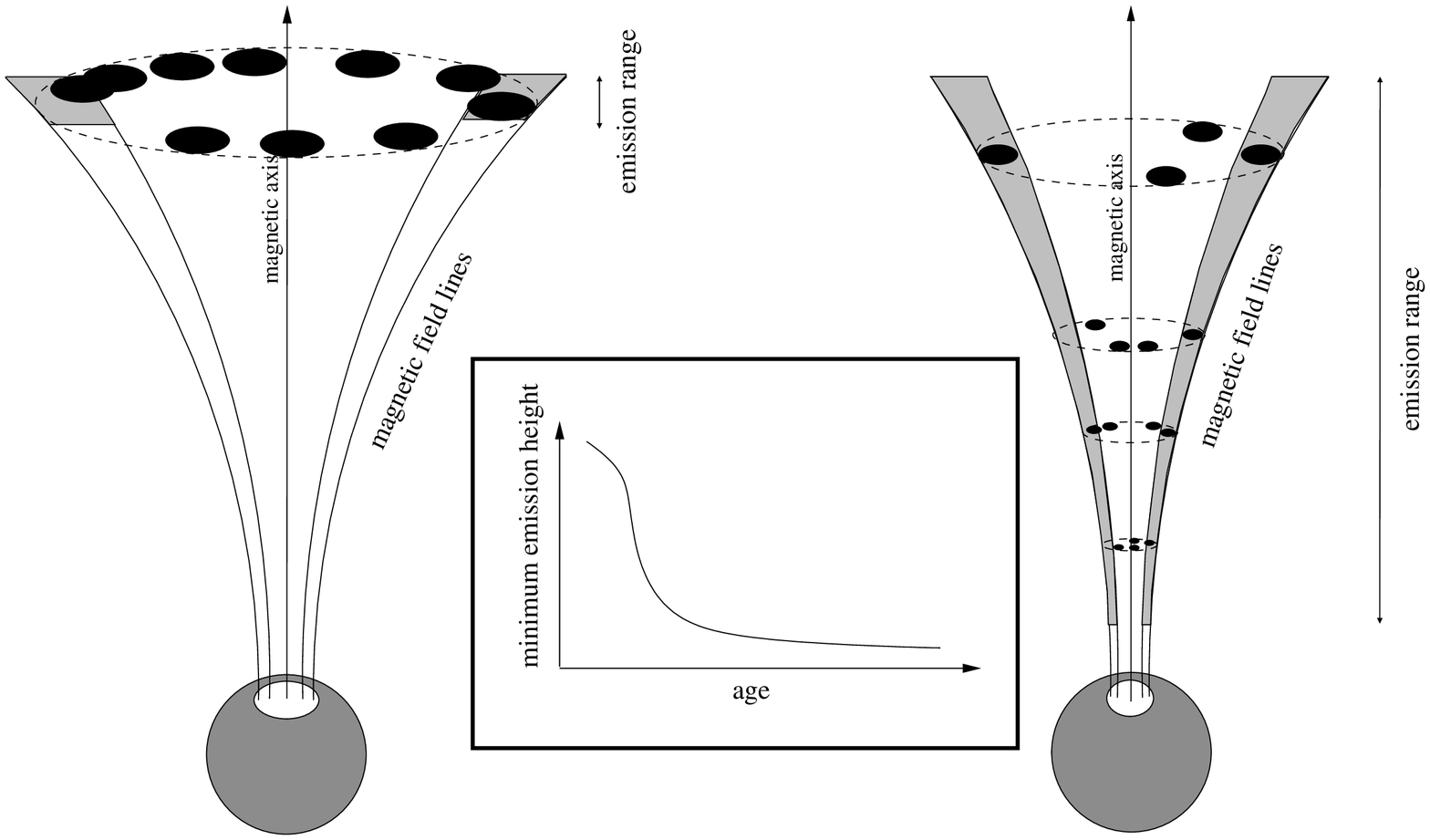}}
    \caption{A recent phenomenological model for pulse shape morphology.
      The neutron star is depicted by the grey sphere and only a single 
      magnetic pole is shown for clarity.
      Left: a young pulsar with emission from a patchy conal ring
      at high altitudes from the surface of the neutron star.
      Right: an older pulsar in which emission emanates from a
      series of patchy rings over a range of altitudes. Centre: schematic
      representation of the change in emission height with pulsar age.
      Figure designed by Aris Karastergiou and Simon Johnston~\cite{kj07}.}
    \label{fig:cartoon}
  \end{figure}}

A key feature of this new model is that the emission height of
young pulsars is radically different from that of the older
population. Monte Carlo simulations~\cite{kj07} using this phenomenological
model appear to be very successful at explaining the rich
diversity of pulse shapes. Further work in this area is
necessary to understand the origin of this model, and
improve our understanding of the shape and evolution of
pulsar beams and fraction of sky they cover. This is of key importance
to the results of population studies reviewed in Section~\ref{sec:corsamp}.


\subsection{The pulsar distance scale}
\label{sec:dist}

Quantitative estimates of the distance to each pulsar can be
made from the measurement of \emph{pulse dispersion} -- the delay in
pulse arrival times across a finite bandwidth. Dispersion occurs
because the group velocity of the pulsed radiation through the ionised
component of the interstellar medium is frequency dependent. As shown
in Figure~\ref{fig:dispersion}, pulses
emitted at lower radio frequencies travel slower through the
interstellar medium, arriving later than those emitted at higher
frequencies. 

\epubtkImage{dispersion.png}{
  \begin{figure}[htbp]
    \def\epsfsize#1#2{0.6#1}
    \centerline{\epsfbox{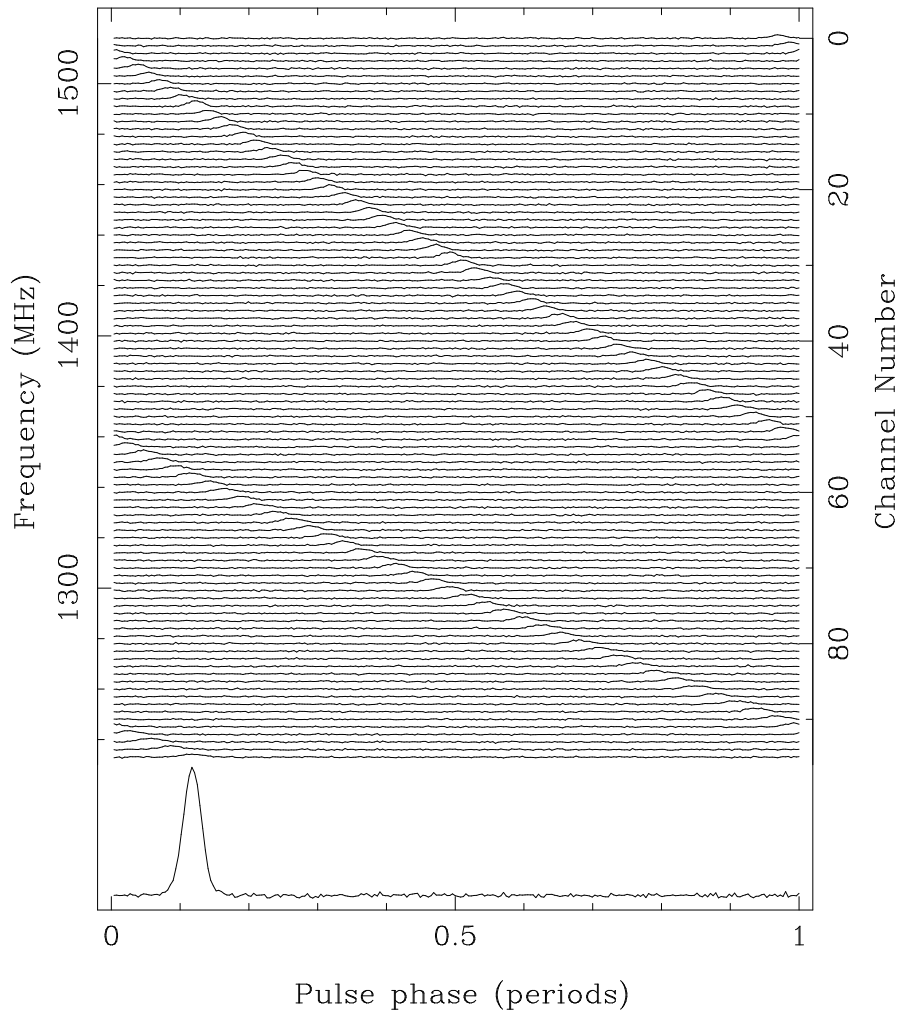}}
    \caption{Pulse dispersion shown in this Parkes observation of the
      128~ms pulsar B1356--60. The dispersion measure is
      295~cm\super{-3}~pc. The quadratic frequency dependence of the
      dispersion delay is clearly visible. Figure provided by Andrew
      Lyne.}
    \label{fig:dispersion}
  \end{figure}
}

Quantitatively, the delay $\Delta t$ in arrival times between a high
frequency $\nu_\mathrm{hi}$ and a low frequency $\nu_\mathrm{lo}$
pulse can be shown~\cite{lk05} to be
\begin{equation}
  \Delta t = 4.15 \mathrm{\ ms} \times
  \left[ \left( \frac{\nu_\mathrm{lo}}{\mathrm{GHz}} \right)^{-2} - 
  \left( \frac{\nu_\mathrm{hi}}{\mathrm{GHz}} \right)^{-2} \right] \times
  \left( \frac{\mathrm{DM}}{\mathrm{cm}^{-3}\mathrm{\ pc}} \right),
  \label{equ:defdt}
\end{equation}
where the dispersion measure
\begin{equation}
  \mathrm{DM} = \int_0^d \!\!\! n_\mathrm{e} \, dl,
  \label{equ:defdm}
\end{equation}
is the integrated column density of electrons, $n_\mathrm{e}$,
out to the pulsar
at a distance $d$. This equation may be solved for $d$ given a
measurement of DM and a model of the free electron distribution
calibrated from the 100 or so pulsars with independent distance
estimates and measurements of scattering for lines of sight towards
various Galactic and extragalactic sources~\cite{tc93, wei96}. A
recent model of this kind, known as NE2001~\cite{cl02a, cl02b},
provides distance estimates with an average uncertainty of $\sim 30$\%.

Because the electron density models are only as good as the scope of
their input data allow, one should be mindful of systematic
uncertainties. For example, studies of the Parkes multibeam pulsar
distribution~\cite{kbm03,lfl06} suggest that the NE2001 model
underestimates the distances of pulsars close to the Galactic
plane. This suspicion has been dramatically confirmed recently by an
extensive analysis~\cite{gmcm08} of pulsar DMs and measurements of
Galactic H$\alpha$ emission. This work shows that the distribution of
Galactic free electrons to be exponential in form with a scale height
of $1830^{+120}_{-250}$~pc.  This value is a factor of two higher than
previously thought. A revised version of the NE2001 model which takes
into account these and other developments is currently in
preparation~\cite{cor08}.


\subsection{Pulsars in binary systems}
\label{sec:bincomps}

As can be inferred from Figure~\ref{fig:venn}, only a few percent of
all known pulsars in the Galactic disk are members of binary systems.
Timing measurements (see Section~\ref{sec:pultim}) place useful constraints on
the masses of the companions which, often supplemented by observations at
other wavelengths, tell us a great deal about their nature. The
present sample of orbiting companions are either white dwarfs, main
sequence stars or other neutron stars. Two notable hybrid systems
are the ``planet pulsars'' B1257+12 and B1620$-$26. PSR B1257+12 is a
6.2-ms pulsar accompanied by at least three terrestrial-mass
bodies~\cite{wf92, psrplanets, wdk00} while B1620$-$26, an 11-ms pulsar
in the globular cluster M4, is part of a triple system with 
a $1 \mbox{\,--\,} 2\,M_\mathrm{Jupiter}$ planet~\cite{tat93, bfs93, tacl99, srh03}
orbiting a neutron star--white dwarf stellar binary system.
The current limits from pulsar timing 
favour a roughly 45-yr mildly-eccentric ($e \sim 0.16$) orbit with semi-major
axis $\sim$~25~AU. Despite several tentative claims over the years,
no other convincing cases for planetary companions to pulsars exist.
Orbiting companions are much more common around millisecond
pulsars ($\sim$~80\% of the observed sample) than around the normal
pulsars ($\lesssim$~1\%). In general, binary systems
with low-mass companions ($\lesssim 0.7\,M_\odot$
-- predominantly white dwarfs) have essentially circular orbits:
$10^{-5} \lesssim e \lesssim 0.01$. Binary pulsars with high-mass companions
($\gtrsim 1\,M_\odot$ -- massive white dwarfs, other neutron stars 
or main sequence stars) tend to have more eccentric orbits, 
$0.15 \lesssim e \lesssim 0.9$.


\subsection{Evolution of normal and millisecond pulsars}
\label{sec:evolution}

A simplified version of the presently favoured
model~\cite{bk74, fv75, sb76, acrs82} to explain the formation of the
various types of systems observed is shown in Figure~\ref{fig:bevol}.
Starting with a binary star system, a neutron star is formed during
the supernova explosion of the initially more massive star. From the
virial theorem, in the absence of any other factors,
the binary will be disrupted if more than half the
total pre-supernova mass is ejected from the system during the
(assumed symmetric) 
explosion~\cite{hil83, bv91}. In practice, the fraction of surviving
binaries is also affected by the magnitude and direction of any impulsive
``kick'' velocity the neutron star receives at
birth from a slightly asymmetric explosion~\cite{hil83, bai89}. 
Binaries that disrupt produce a
high-velocity isolated neutron star and an OB runaway
star~\cite{bla61}. The high probability of disruption explains
qualitatively why so few normal pulsars have companions. Those
that survive will likely have high orbital eccentricities due
to the violent conditions in the supernova explosion.
There are currently four known normal radio pulsars with massive main
sequence companions in eccentric orbits which are examples of
binary systems which survived the supernova 
explosion~\cite{jml92, kjb94, sml01, sml03, lyn05}. Over the
next 10\super{7-8}~yr after the explosion, the neutron star may be
observable as a normal radio pulsar spinning down to a period $\gtrsim$
several seconds. After this time, the energy output of the star
diminishes to a point where it no longer produces significant amounts
of radio emission.

\epubtkImage{bevol.png}{
  \begin{figure}[htbp]
    \def\epsfsize#1#2{0.21#1}
    \centerline{\epsfbox{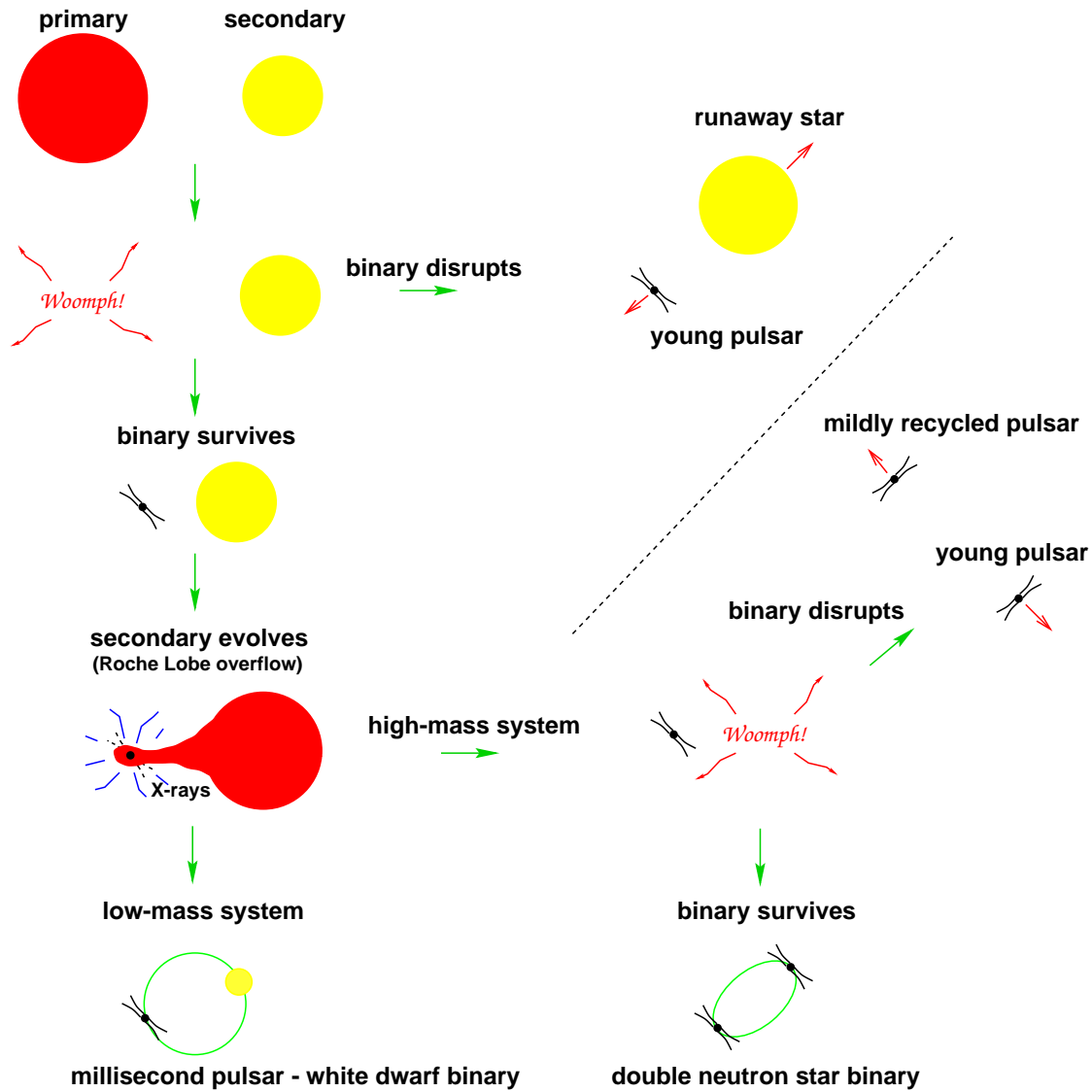}}
    \caption{Cartoon showing various evolutionary scenarios
      involving binary pulsars.}
    \label{fig:bevol}
  \end{figure}}

For those few binaries that remain bound, and in which the companion
is sufficiently massive to evolve into a giant and
overflow its Roche lobe, the old spun-down neutron star can gain a new
lease of life as a pulsar by accreting matter and angular momentum at
the expense of the orbital angular momentum of the binary
system~\cite{acrs82}. The term ``recycled pulsar'' is often used to
describe such objects. During this accretion phase, the X-rays
produced by the frictional heating of the infalling
matter onto the neutron star mean that such a system is expected to be
visible as an X-ray binary. Two classes of X-ray binaries relevant to
binary and millisecond pulsars exist: neutron stars with high-mass or
low-mass companions. Detailed reviews of the X-ray binary
population, including systems likely to contain black holes,
can be found elsewhere~\cite{bv91,rm06}. 

\subsubsection{High-mass systems}

In a high-mass X-ray binary, the companion is massive enough that it
also explodes as a supernova, producing a second neutron star. If
the binary system is lucky enough to survive the explosion, the result
is a double neutron star binary. Nine such
systems are now known, the original example being PSR
B1913+16~\cite{ht75a} -- a 59-ms radio pulsar which orbits its
companion every 7.75~hr~\cite{tw82, tw89}. In this formation scenario,
PSR~B1913+16 is an example of the older, first-born, neutron star that
has subsequently accreted matter from its companion.

For many years, no clear example was known where the second-born
neutron star was observed as a pulsar. The discovery of the double pulsar
J0737$-$3039~\cite{bdp03, lbk04}, where a 22.7-ms recycled pulsar ``A''
orbits a 2.77-s normal pulsar ``B'' every 2.4~hr, has now provided a
dramatic confirmation of this evolutionary model in which we identify
A and B as the first and second-born neutron stars respectively.
Just how many more observable double pulsar systems exist in our Galaxy
is not clear. While the population of double neutron star systems
in general is reasonably well understood (see Section~\ref{sec:nsns}),
a number of effects conspire to reduce the detectability of 
double pulsar systems. First, the lifetime of the second born pulsar 
cannot be prolonged by accretion and spin-up and hence 
is likely to be less than
one tenth that of the recycled pulsar. Second, the radio beam
of the longer period second-born pulsar
is likely to be much smaller 
than its spun-up partner making it harder to
detect (see Section~\ref{sec:beaming}). Finally, as discussed
in Section~\ref{sec:tbin}, the PSR~J0737--3039 system is viewed
almost perfectly edge-on to the line of sight and there is
strong evidence~\cite{mkl04} that the wind of the more rapidly spinning A
pulsar is impinging on B's magnetosphere which affects its radio
emission. In summary, the prospects of ever finding
more than a few 0737-like systems appears to be rather low. 

A notable recent addition to the sample of double neutron
star binaries is PSR~J1906+0746, a
144-ms pulsar in a highly relativistic 4-hr orbit with an eccentricity
of 0.085~\cite{lsf06}. Timing observations show the pulsar to be
young with a characteristic age of only 10\super{5}~yr. Measurements of
the relativistic periastron advance and gravitational redshift
(see Section~\ref{sec:tbin})
constrain the masses of the pulsar and its companion to be
$1.25\,M_{\odot}$ and $1.37\,M_{\odot}$ respectively~\cite{kas08}.
We note, however, that the pulsar exhibits significant amounts of timing
noise (see Section~\ref{sec:tstab}), making the analysis of the
system far from trivial. Taken at face value, these parameters suggest that
the companion is also a neutron star, and the system appears to be
a younger version of the double pulsar. Despite intensive searches,
no pulsations from the companion star (expected to be a recycled radio
pulsar) have so far been observed~\cite{lsf06}. This could be due
to unfavourable beaming and/or intrinsic radio faintness. Continued searches
for radio pulsations from these companions are strongly encouraged,
however, as geodetic precession may make their radio beams visible
in future years.

\subsubsection{Low-mass systems}

The companion in a low-mass X-ray binary evolves and transfers
matter onto the neutron star on a much longer timescale, spinning it
up to periods as short as a few ms~\cite{acrs82}. Tidal forces
during the accretion process serve to circularize the orbit.
At the end of the
spin-up phase, the secondary sheds its outer layers to become a white
dwarf in orbit around a rapidly spinning millisecond pulsar. This
model has gained strong support in recent years from the discoveries
of quasi-periodic kHz oscillations in a number of low-mass X-ray
binaries~\cite{wz97}, as well as Doppler-shifted 2.49-ms X-ray
pulsations from the transient X-ray burster SAX
J1808.4$-$3658~\cite{wv98, cm98}. Seven other ``X-ray millisecond
pulsars'' are now known with spin rates and orbital periods ranging
between 185\,--\,600~Hz and 40~min\,--\,4.3~hr
respectively~\cite{wij05, mkv05a, mkv05b, kmd+07}.

Numerous examples of these systems in their post X-ray phase are now
seen as the millisecond pulsar--white dwarf binary systems.
Presently, 20 of these systems have compelling optical identifications
of the white dwarf companion, and upper limits or tentative detections
have been found in about 30 others~\cite{vbjj05}. Comparisons between
the cooling ages of the white dwarfs and the millisecond pulsars
confirm the age of these systems and suggest that the accretion rate
during the spin-up phase was well below the Eddington limit~\cite{hp98}.

\epubtkImage{pbe.png}{
  \begin{figure}[htbp]
    \def\epsfsize#1#2{0.65#1}
    \centerline{\epsfbox{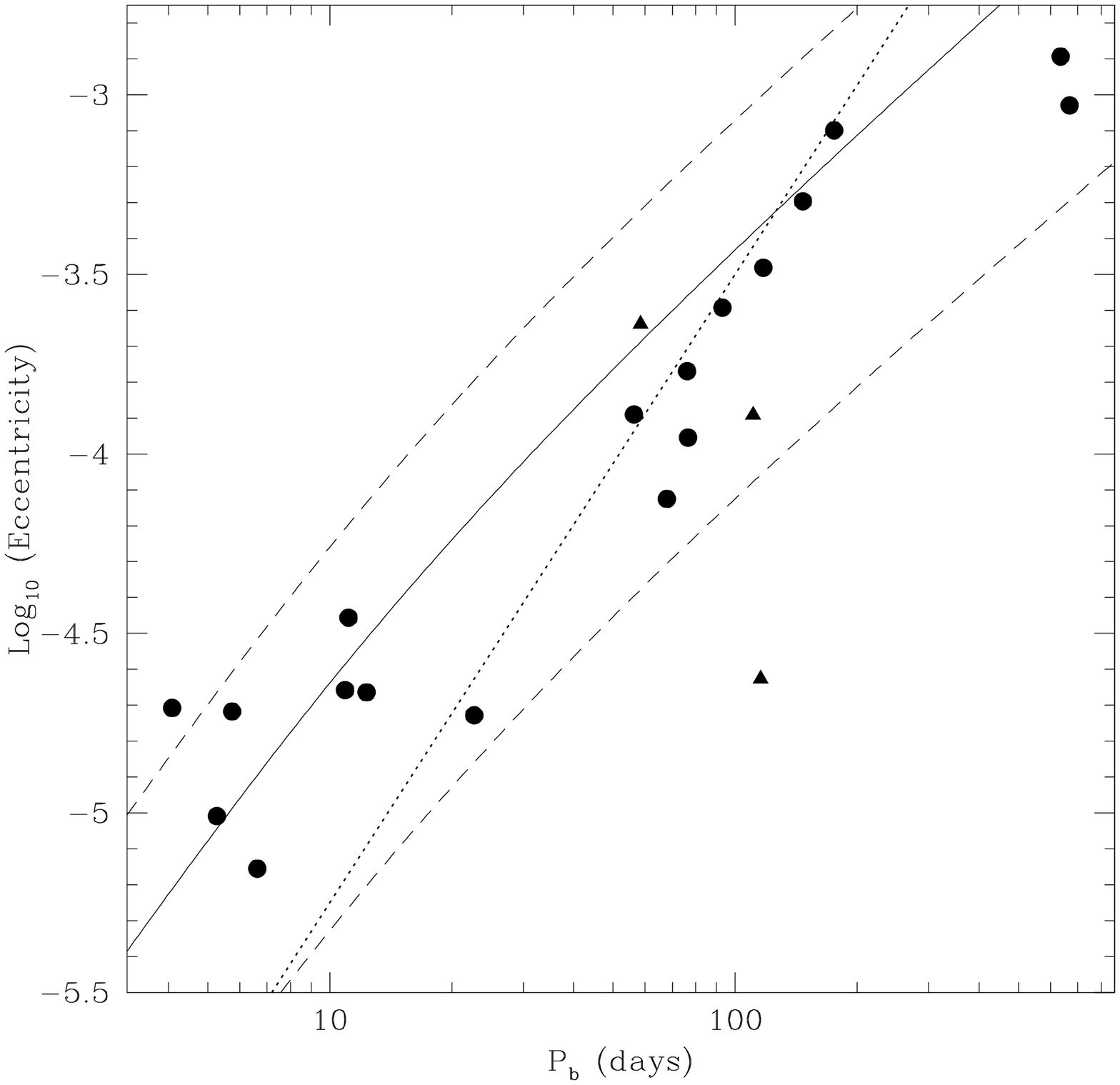}}
    \caption{Eccentricity versus orbital period for a sample of 21
      low-mass binary pulsars which are not in globular clusters, with
      the triangles denoting three recently discovered
      systems~\cite{sfl05}. The solid line shows the median of the
      predicted relationship between orbital period and
      eccentricity~\cite{phi92}. Dashed lines show 95\% the confidence
      limit about this relationship. The dotted line shows
      $P_\mathrm{b} \propto e^2$. Figure provided by Ingrid
      Stairs~\cite{sfl05} using an adaptation of the orbital
      period-eccentricity relationship tabulated by Fernando Camilo.}
    \label{fig:pbe}
  \end{figure}}

Further support for the above evolutionary scenarios comes from two
correlations in the observed sample of low-mass binary pulsars.
Firstly, as seen in Figure~\ref{fig:pbe}, there is a strong correlation
between orbital period and eccentricity. The data are in very good
agreement with a theoretical relationship which predicts a relic
orbital eccentricity due to convective eddy currents in the accretion
process~\cite{phi92}.
Secondly, as shown in Panel~b of Figure~\ref{fig:porbmwd}, where companion masses
have been measured accurately, through radio
timing~(see Section~\ref{sec:tbin}) and/or through optical
observations~\cite{vbjj05}, they are in good agreement with a relation
between companion mass and orbital period predicted by binary
evolution theory~\cite{ts99}. A word of caution is required in using
these models to make predictions, however. When confronted with a
larger ensemble of binary pulsars using statistical arguments to
constrain the companion masses (see Panel~a of Figure~\ref{fig:porbmwd}), current
models have problems in explaining the full range of orbital periods on this
diagram~\cite{sfl05}.

\epubtkImage{porbmwd.png}{
  \begin{figure}[htbp]
    \def\epsfsize#1#2{0.38#1}
    \centerline{\epsfbox{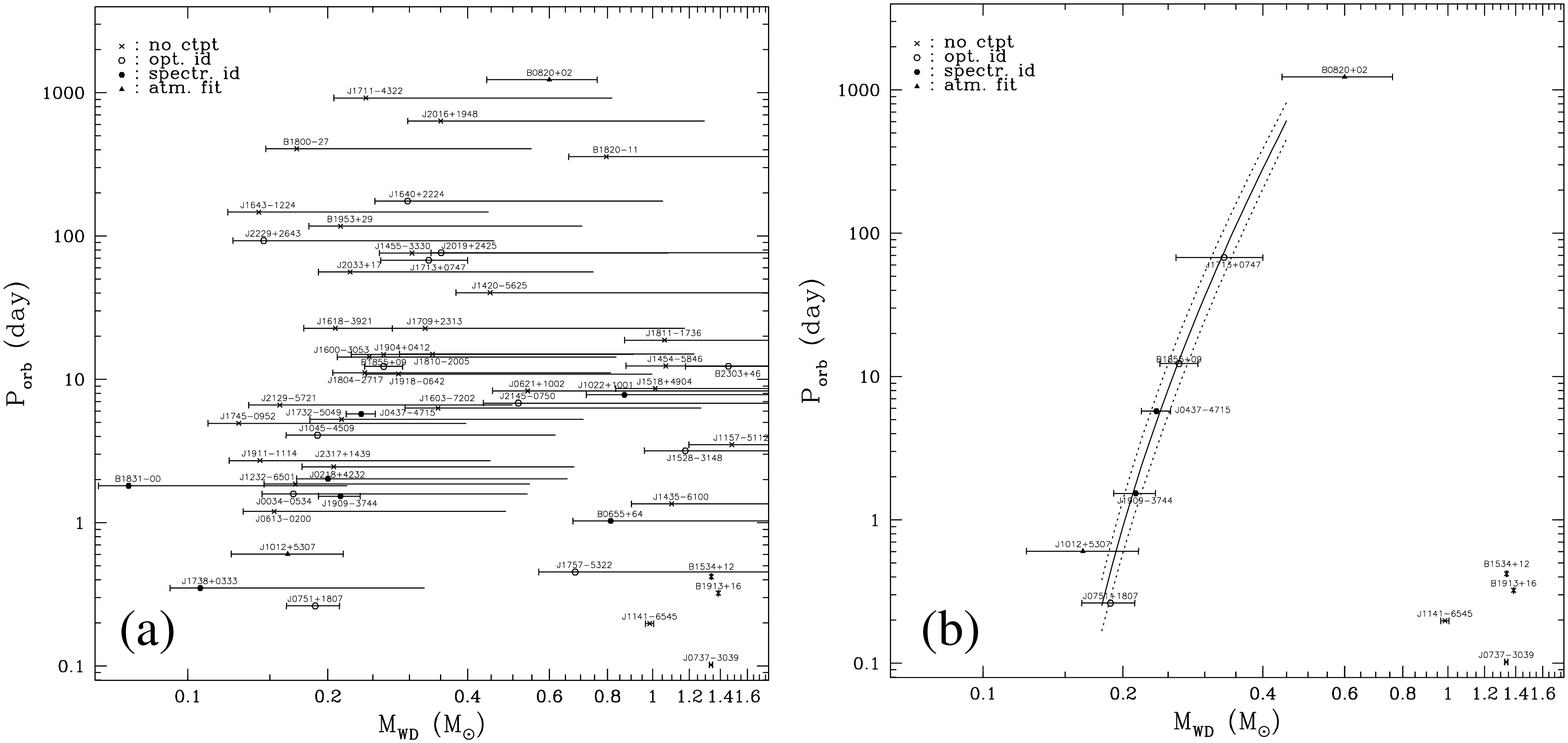}}
    \caption{Orbital period versus companion mass for binary pulsars
      showing the whole sample where, in the absence of mass
      determinations, statistical arguments based on a random
      distribution of orbital inclination angles (see
      Section~\ref{sec:tbin}) have been used to constrain the masses as
      shown (Panel~a), and only those with well determined companion
      masses (Panel~b). The dashed lines show the uncertainties in the
      predicted relation~\cite{ts99}. This relationship indicates that
      as these systems finished a period of stable mass transfer due to
      Roche-lobe overflow, the size and hence period of the orbit was
      determined by the mass of the evolved secondary star. Figure
      provided by Marten van Kerkwijk~\cite{vbjj05}.}
    \label{fig:porbmwd}
  \end{figure}}

\subsection{Intermediate mass binary pulsars}

The range of white dwarf masses observed is becoming broader. Since
this article originally appeared~\cite{lor98e}, the number of
``intermediate-mass binary pulsars''~\cite{cam96c} has grown
significantly~\cite{clm01}. These systems are distinct to the
millisecond pulsar--white dwarf binaries in several ways:
\begin{enumerate}
\item The spin period of the radio pulsar is generally longer
  (9\,--\,200~ms).
\item The mass of the white dwarf is larger (typically
  $\gtrsim 0.5\,M_\odot$).
\item The orbit, while still essentially circular, is often
  significantly more eccentric ($e \gtrsim 10^{-3}$).
\item The binary parameters do not necessarily follow the mass--period or
  eccentricity--period relationships.
\end{enumerate}
It is not presently clear whether these systems
originated from low- or high-mass X-ray binaries. It was suggested by
van den Heuvel~\cite{vdh94} that they have more in common with
high-mass systems. Subsequently, Li proposed~\cite{li02}
that a thermal-viscous instability in the accretion disk of a low-mass
X-ray binary could truncate the accretion phase and produce a more
slowly spinning neutron star.


\subsection{Isolated recycled pulsars}

The scenarios outlined qualitatively above represent a reasonable
understanding of binary evolution. There are, however, a number of
pulsars with spin properties that suggest a phase of recycling took
place but have no orbiting companions. While the existence of such
systems in globular clusters are more readily explained by the high
probability of stellar interactions compared to the disk~\cite{sp95},
it is somewhat surprising to find them in the Galactic disk. Out of a
total of 72 Galactic millisecond pulsars, 16 are isolated
(see Table~\ref{tab:imsps}). Although it has been proposed that these
millisecond pulsars have ablated their companion via their strong
relativistic winds~\cite{krst88} as may be happening in the
PSR~B1957+20 system~\cite{fst88}, it is not clear whether the
energetics or timescales for this process are
feasible~\cite{le91}. 

There are four further ``anomalous'' isolated pulsars with periods in
the range 28\,--\,60~ms~\cite{cnt93, lma04}. When placed on the
$P \mbox{--} \dot{P}$ diagram, these objects populate the region occupied by the
double neutron star binaries. The most natural explanation for their
existence, therefore, is that they are ``failed double neutron star
binaries'' which disrupted during the supernova explosion of the
secondary~\cite{cnt93}.
A simple calculation~\cite{lma04}, suggested that
for every double neutron star we should see of order ten such isolated
objects. Recent work~\cite{blr08} has investigated
why so few are observed.
Using the
most recent population synthesis models to follow the evolution of
binary systems~\cite{bkr08}, it appears that the discrepancy may not
be as significant as previously supposed.  In particular, the space
velocity distribution of surviving binary systems is narrower than for
the isolated objects that were during the second supernova
explosion. The isolated systems occupy a larger volume of
the Galaxy than the surviving binaries and are harder to detect. When
this selection effect is accounted for~\cite{blr08}, the relative
sample sizes appear to be consistent with the disruption
hypothesis.

\subsection{A new class of millisecond pulsars?}
\label{sec:1903}

One of the most remarkable recent discoveries is the binary pulsar
J1903+0327~\cite{crl08}.  Found in an on-going multibeam survey with
the Arecibo telescope~\cite{palfa,cfl06}, this 2.15-ms pulsar is
distinct from all other millisecond pulsars in that its 95-day orbit
has an eccentricity of 0.43! In addition, timing measurements of the
relativistic periastron advance and Shapiro delay in this system (see
Section \ref{sec:tbin}), show the mass of the pulsar to be $1.74 \pm
0.04\,M_{\odot}$ and the companion star to be $1.051 \pm
0.015\,M_{\odot}$.  Optical observations show a possible counterpart
which is consistent with a $1\,M_{\odot}$ star.  While similar systems
have been observed in globular clusters (e.g.~PSR~J0514$-$4002A in
NGC~1851\cite{frg07}), presumably a result of exchange interactions, the
standard recycling hypothesis outlined in Section \ref{sec:evolution}
cannot account for pulsars like J1903+0327 in the Galactic disk.

How could such an eccentric binary millisecond pulsar system form? One
possibility is that the binary system was produced in an exchange
interaction in a globular cluster and subsequently ejected, or the
cluster has since disrupted. Statistical estimates~\cite{crl08} of
the likelihood of both these channels are in the range 1--10\%,
implying that a globular cluster origin cannot be ruled out.

Another possibility is that the pulsar is a member of an hierarchical
triple system with a one solar mass white dwarf in the 95-day orbit,
and a main sequence star in a much wider and highly inclined orbit
which has so far not been revealed by timing. The origin of the high
eccentricity is through perturbations from the outer star, the
so-called Kozai mechanism~\cite{koz62}. Formation estimates based on
observational data on stellar multiplicity~\cite{pbh99} find that
around 4\% of all binary millisecond pulsars are expected to be triple
systems~\cite{crl08}. The existence of a single triple system among the
current sample of millisecond pulsars appears to be consistent with
this hypothesis.

If future observations of the proposed optical counterpart confirm it
as the binary companion through spectral line measurements of orbital
Doppler shifts, the above triple-system scenario will be ruled
out. Such an observation would favour a hybrid scenario suggested by van den
Heuvel~\cite{van08} in which the white dwarf and pulsar merge due to
gravitational radiation losses. Tidal disruption of the white dwarf in
the inspiral would produce an accretion disk and induce an
eccentricity in the orbit of the outer star leaving behind an
eccentric binary system. This idea could
naturally account for the high pulsar mass observed in this
system which could arise from accretion of a white-dwarf debris disk
following coalescence. Alternatively, as suggested by Champion et
al.~\cite{crl08}, the millisecond pulsar might have ablated the
white dwarf companion in a triple system leaving only the unevolved
companion in an elliptical orbit.


\subsection{Pulsar velocities}
\label{sec:pvel}

Pulsars have long been known to have space velocities at least an order
of magnitude larger than those of their main sequence progenitors, which
have typical values between 10 and 50~km~s\super{-1}.
Proper motions for over 250 pulsars have now been
measured largely by radio timing and interferometric
techniques~\cite{las82, bmk90b, fgl92, hla93, hllk05, zhw05}. These
data imply a broad velocity spectrum ranging from 0 to over
1000~km~s\super{-1}~\cite{ll94}, with the current record holder being
PSR~B1508+55 \cite{cvb05}, with a proper motion and parallax
measurement implying a transverse velocity of
$1083^{+103}_{-90}\mathrm{\ km\ s}^{-1}$. As Figure~\ref{fig:migrate}
illustrates, high-velocity pulsars born close to the Galactic plane
quickly migrate to higher Galactic latitudes.  Given such a broad
velocity spectrum, as many as half of all pulsars will eventually
escape the Galactic gravitational potential~\cite{ll94, cc98}.

\epubtkMovie{init.gif}{init.png}{
  \begin{figure}[htbp]
    \def\epsfsize#1#2{0.8#1}
    \centerline{\epsfbox{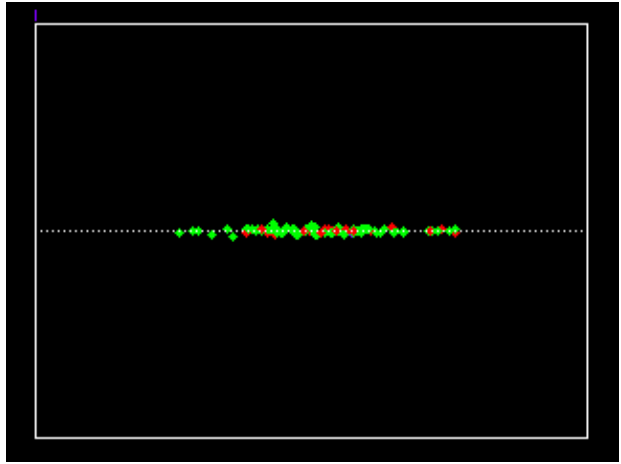}}
    \caption{A simulation following the motion of
      100 pulsars in a model gravitational potential of our Galaxy for
      200~Myr. The view is edge-on, i.e.\ the horizontal axis
      represents the Galactic plane (30~kpc across) while the vertical
      axis represents $\pm$10~kpc from the plane. This snapshot shows
      the initial configuration of young neutron stars.}
    \label{fig:migrate}
  \end{figure}}

Such large velocities are perhaps not surprising, given the violent
conditions under which neutron stars are
formed. If the explosion is
only slightly asymmetric, an impulsive ``kick'' velocity of up to
1000~km~s\super{-1} can be imparted to the neutron star~\cite{shk70}. In
addition, if the neutron star progenitor was a member of a binary
system prior to the explosion, the pre-supernova orbital velocity will
also contribute to the resulting speed of the newly-formed pulsar.
The relative contributions of these two factors
to the overall pulsar birth velocity distribution is currently not
well understood.

The distribution of pulsar velocities has 
a high velocity component due to the normal pulsars \cite{ll94,hllk05,fk06}, 
and a lower velocity component from binary and millisecond pulsars 
\cite{lor95,cc97,lml98,hllk05}. One reason for this dichotomy
appears to be that, in order to survive and subsequently form recycled
pulsars through the accretion process outlined above, the binary
systems contain only those neutron stars with lower birth velocities.
In addition, the surviving neutron star has to pull the companion
along with it, thus slowing the system down.

Further insights into pulsar kicks from analyses of proper 
motion and polarization data~\cite{jhv05,jkk07,ran07} find
strong evidence for an alignment between the spin axis
and the velocity vector at birth. These data have recently been combined
with modeling of pulsar-wind nebulae~\cite{nr07}, where
strong evidence is found for a model in which the
natal impulse is provided by an anisotropic flux of neutrinos from the 
proto-neutron star on timescales of a few seconds.


\subsection{Pulsar searches}
\label{sec:wheretolook}

The radio sky is being repeatedly searched for new pulsars in a variety
of ways. In the following, we outline the major search strategies that
are optimized for binary and millisecond pulsars.


\subsubsection{All-sky searches}
\label{sec:allsky}

The oldest radio pulsars form a relaxed population of stars
oscillating in the Galactic gravitational potential~\cite{hp97}. The
scale height for such a population is at least 500~pc~\cite{lor05}, about 10 times
that of the massive stars which populate the Galactic plane. Since the
typical ages of millisecond pulsars are several Gyr or more, we
expect, from our vantage point in the Galaxy, to be in the middle of
an essentially isotropic population of nearby sources. All-sky
searches for millisecond pulsars at high Galactic latitudes have been
very effective in probing this population. 

Motivated by the discovery of two recycled pulsars at high latitudes
with the Arecibo telescope~\cite{wol91a,wf92}, surveys carried out at
Arecibo, Parkes, Jodrell Bank and Green Bank by others in the
1990s~\cite{cnt93,cnt96,cnst96,mld96,lml98} discovered around 30
further objects.  Although further searching of this kind has been
carried out at Arecibo in the past decade~\cite{lxf05}, much of the
recent efforts have been concentrated along the plane of our Galaxy
and in globular clusters discussed below. Very recently, however, 
12,000 square degrees of sky was surveyed using a new 350-MHz receiver on the
Green Bank Telescope~\cite{350drift,boy08}. 
Processing of these data is currently
underway, with two millisecond pulsars found with around 10\% of the
dataset analysed.


\subsubsection{Searches close to the plane of our Galaxy}
\label{sec:plane}

Young pulsars are most likely to be found near to their places of
birth close to the Galactic plane. This was
the target region of the main Parkes multibeam survey and has
so far resulted in the discovery of 783
pulsars~\cite{mlc01, mhl02, kbm03, hfs04, fsk04, lfl06, kei08, bjd06}, almost half the number
currently known! Such a large haul inevitably results in a number of
interesting individual objects such as the relativistic binary 
pulsar J1141$-$6545~\cite{klm00, obv02, bokh03, hbo05, bbv08}, a
young pulsar orbiting an $\sim 11\,M_{\odot}$ star (probably a main
sequence B-star~\cite{sml01, sml03}), a young pulsar
in a $\sim$~5~yr-eccentric orbit ($e = 0.955$; the most
eccentric found so far) 
around a $10 \mbox{\,--\,} 20\,M_\odot$ companion~\cite{lyn05, lfl06},
several intermediate-mass binary pulsars~\cite{clm01}, and two double
neutron star binaries~\cite{lcm00, fkl05}. Further analyses of this
rich data set are now in progress and will ensure yet more discoveries
in the near future. 

Motivated by the successes at Parkes, a multibeam survey is now in
progress with the Arecibo telescope~\cite{palfa} and the Effelsberg
radio telescope.  The Arecibo survey has so far
discovered 46 pulsars~\cite{palfa,cfl06} with notable
finds including a highly relativistic
binary~\cite{lsf06} and an eccentric millisecond pulsar
binary~\cite{crl08}.  Hundreds more pulsars could be found in this 
survey over the next five years. A significant fraction of this yield
are expected to be distant millisecond pulsars in the disk of our
Galaxy. With the advent of sensitive low-noise receivers at lower observing
frequencies, surveys of the Galactic plane are being carried out
with the GMRT~\cite{jmk08}, Green Bank~\cite{hrk08} and 
Westerbork~\cite{rbe06}. At the time of writing, no new millisecond
pulsars have been found in these searches, though significant amounts
of data remain to be fully processed.


\subsubsection{Searches at intermediate and high Galactic latitudes}
\label{sec:intlat}

To probe more deeply into the population of millisecond and recycled
pulsars than possible at high Galactic latitudes, the Parkes multibeam
system was also used to survey intermediate
latitudes~\cite{edw00, eb01}. Among the 69 new pulsars found in the
survey, 8 are relatively distant recycled objects. Two of the new recycled
pulsars from this survey~\cite{eb01} are mildly relativistic neutron
star-white dwarf binaries. An analysis of the full results from this
survey should significantly improve our knowledge on the Galaxy-wide
population and birth-rate of millisecond pulsars. 
Arecibo surveys at intermediate latitudes also continue to find new
pulsars, such as the long-period binaries J2016+1948 and 
J0407+1607~\cite{naf03, lxf05}, and
the likely double neutron star system J1829+2456~\cite{clm04}.

Although the density of pulsars decreases with increasing Galactic
latitude, discoveries away from the plane provide strong constraints 
on the scale height of the millisecond pulsar population. Two recent
surveys with the Parkes multibeam system~\cite{bjd06,jbo07} have
resulted in a number of interesting discoveries. Pulsars at high
latitudes are especially important for the millisecond pulsar
timing array (Section~\ref{sec:array}) which benefits from widely
separated pulsars on the sky to search for correlations in the
cosmic gravitational wave background on a variety of angular scales.


\subsubsection{Targeted searches of globular clusters}
\label{sec:globs}

Globular clusters have long been known to be breeding grounds for
millisecond and binary pulsars~\cite{cr05}. The main reason for this
is the high stellar density and consequently high rate of stellar interaction
in globular clusters relative to most of
the rest of the Galaxy. As a result, low-mass X-ray binaries are
almost 10 times more abundant in clusters than in the Galactic
disk. In addition, exchange interactions between binary and multiple
systems in the cluster can result in the formation of exotic binary
systems~\cite{sp95}. To date, searches have revealed 140 pulsars in
26 globular clusters~\cite{gcpsrs}.  
Early highlights include the
double neutron star binary in M15~\cite{pakw91} and a low-mass binary
system with a 95-min orbital period in 47~Tucanae~\cite{clf00}, one of 23
millisecond pulsars currently known in this cluster alone~\cite{clf00, lcf03}.

On-going surveys of clusters continue to yield new
surprises~\cite{rgh01, dlm01}, with no less than 70 discoveries in
the past five years~\cite{ran08}. Among these is the most eccentric binary pulsar in
a globular cluster so far -- J0514$-$4002 is a 4.99~ms pulsar in a highly eccentric
($e=0.89$) binary system in the globular cluster NGC~1851~\cite{fgr04}. The cluster
with the most pulsars is now Terzan~5 which boasts 33~\cite{rhs05, gcpsrs}, 30 of
which were found with the Green Bank Telescope~\cite{gbt}.
The spin periods and orbital parameters of the new
pulsars reveal that, as a population, they are significantly different
to the pulsars of 47~Tucanae which have periods in the range
2\,--\,8~ms~\cite{lcf03}. The spin periods of the new pulsars span a much
broader range (1.4\,--\,80~ms) including the first, third and fourth
shortest spin periods of all pulsars currently known. The binary
pulsars include six systems with eccentric orbits and likely white
dwarf companions. No such systems
are known in 47~Tucanae. The difference between the two pulsar
populations may reflect the different evolutionary states and physical
conditions of the two clusters. In particular, the central stellar
density of Terzan~5 is about twice that of 47~Tucanae, suggesting that
the increased rate of stellar interactions might disrupt the recycling
process for the neutron stars in some binary systems and induce larger
eccentricities in others.


\subsubsection{Targeted searches of other regions}
\label{sec:othertargets}

While globular clusters are the richest targets for finding millisecond
pulsars, other regions of interest have been searched. Recently, a search
of error boxes from unidentified sources from the Energetic Gamma-Ray
Experiment Telescope (EGRET) revealed three new binary pulsars
J1614$-$2318, J1614$-$2230 and J1744$-$3922~\cite{hrr05,crh06a,ran08a}.
None of these pulsars is likely to be energetic enough to be
associated with their target EGRET sources~\cite{hrr05}.
While convincing EGRET associations with several young pulsars 
are now known~\cite{kbm03}, it is not clear whether millisecond pulsars are
relevant to the energetics of these enigmatic sources~\cite{cml05}. 
Despite this lack of success, it is quite possible that the
recent launches of the AGILE~\cite{agile} and GLAST~\cite{glast} gamma-ray 
observatories will provide further opportunities for follow-up.

Other targets of interest are X-ray point sources found with
the Chandra~\cite{chandra} and XMM-Newton~\cite{xmm} observatories
and TeV sources found with HESS~\cite{hess}. The X-ray sources
have been particularly fruitful targets for young pulsars, with
a number of discoveries of extremely faint objects~\cite{cam03}.
Although not directly relevant to the topic of this review, these
searches are revolutionizing our picture of the young neutron
star population and should provide valuable insights into the beaming
fraction and birthrate of these pulsars.


\subsubsection{Extragalactic searches}

The only radio pulsars known outside of the Galactic field and its
globular cluster systems are the 19 currently known in the Large and
Small Magellanic Clouds~\cite{mhah83,ckm01,mfl06}.  The lack of
millisecond pulsars in the sample so far is most likely due to the
limited sensitivity of the searches and large distance to the
clouds. Further surveys in the Magellanic clouds are warranted.
Surveys of more distant galaxies have so far been fruitless. Current
instrumentation is only sensitive to giant isolated pulsars of the
kind observed from the Crab~\cite{hkwe03} and the millisecond
pulsars~\cite{kbmo06}.  While surveys for such events are
on-going~\cite{klmr08}, detections of weaker periodic sources are
likely to require the enhanced sensitivity of the next generation 
radio telescopes.


\subsubsection{Surveys with new telescopes}
\label{sec:futuresurveys}

All surveys that have so far been conducted, or will be carried out in
the next few years, will ultimately be surpassed by the next
generation of radio telescopes. The Allen Telescope Array in 
California~\cite{ata} is now beginning operations and could allow
large-area coverage of the 1--10 GHz sky for pulsars and transients.
In Europe, the low-frequency
array~\cite{lofar,fvd07} is set to discover hundreds of faint nearby
pulsars~\cite{vs08} in the next five years. While the Square Kilometre
Array~\cite{ska} is not expected to be completed until 2020, a number
of pathfinder instruments are now under development.  In China,
the Five hundred meter Aperture Spherical Telescope~\cite{fast} is
scheduled for completion in 2013 and will provide significant advances
for pulsar research~\cite{nwz06}. The Australian Square Kilometre
Array Pathfinder, will have some applications as a pulsar 
instrument~\cite{jef07}. Very exciting wide-field search
capabilities will be offered by the South African 
MeerKAT array of 80 dishes set to begin operations in 2012~\cite{kat}.


\subsection{Going further}

Two books, {\it Pulsar
Astronomy}~\cite{ls05} and {\it Handbook of Pulsar
Astronomy}~\cite{lk05}, cover the literature and techniques and provide
excellent further reading. The morphological properties of pulsars
have recently been comprehensively discussed in recent
reviews~\cite{smi03, sw04}. Those seeking
a more theoretical viewpoint are advised to read {\it The Theory of
Neutron Star Magnetospheres}~\cite{mic91} and {\it The Physics of the
Pulsar Magnetosphere}~\cite{bgi93}. Our summary of evolutionary
aspects serves merely as a primer to the vast body of literature
available. Further insights can be found from more detailed
reviews~\cite{bv91, pk94, sta04}.

Pulsar resources available on the {\it Internet} are continually
becoming more extensive and useful. A good starting point for
pulsar-surfers is the {\it Handbook of Pulsar Astronomy} 
website~\cite{handbook}, as well as the
{\it Pulsar Astronomy} wiki
\cite{pulsarastronomy} and the Cool Pulsars site~\cite{coolpulsars}.

\newpage


\section{Pulsar Statistics and Demography}
\label{sec:gal}

The observed pulsar sample is heavily biased towards the brighter
objects that are the easiest to detect. What we observe represents
only the tip of the iceberg of a much larger underlying
population~\cite{go70}. The bias is well demonstrated by the
projection of pulsars onto the Galactic
plane shown in Figure~\ref{fig:incomplete}. The clustering of sources
around the Sun seen in the left panel is clearly at variance with the
distribution of other stellar populations which show a radial
distribution symmetric about the Galactic centre. Also shown in
Figure~\ref{fig:incomplete} is the cumulative number of pulsars as a
function of the projected distance from the Sun compared to the
expected distribution for a simple model population with no selection
effects. The observed number distribution becomes strongly deficient
beyond a few kpc.

\epubtkImage{xy.png}{
  \begin{figure}[htbp]
    \def\epsfsize#1#2{0.95#1}
    \centerline{\epsfbox{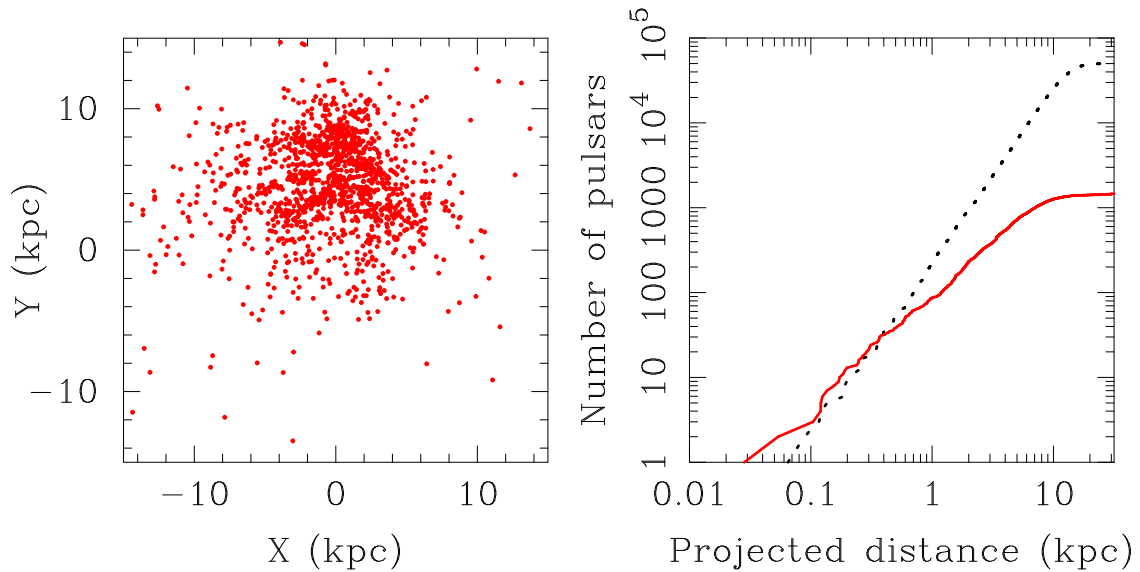}}
    \caption{Left panel: The current sample of all known radio
      pulsars projected onto the Galactic plane. The Galactic centre
      is at the origin and the Sun is at (0, 8.5)~kpc.
      Note the spiral-arm structure seen in the distribution which
      is now required by the most recent Galactic electron density
      model~\cite{cl02a, cl02b}. Right panel: Cumulative number of
      pulsars as a function of projected distance from the Sun. The
      solid line shows the observed sample while the dotted line shows
      a model population free from selection effects.}
    \label{fig:incomplete}
  \end{figure}}


\subsection{Selection effects in pulsar searches}
\label{sec:selfx}


\subsubsection{The inverse square law and survey thresholds}
\label{sec:invsq}

The most prominent selection effect is the inverse square law, i.e.\
for a given intrinsic luminosity\epubtkFootnote{Pulsar astronomers usually
define the luminosity $L \equiv S d^2$, where $S$ is the mean flux density
at 400~MHz (a standard observing frequency) and $d$ is the distance
derived from the DM (see Section~\ref{sec:dist}). Since this ignores any
assumptions about beaming or geometrical factors, it is sometimes
referred to as a ``pseudoluminosity''~\cite{acc02}.}, the observed
flux density varies inversely with the distance squared. This
results in the observed sample being dominated by nearby and/or high
luminosity
objects. Beyond distances of a few kpc from the Sun, the apparent
flux density falls below the detection thresholds $S_\mathrm{min}$ of most
surveys. Following~\cite{dss84}, we express this threshold as follows:
\begin{equation}
  S_\mathrm{min} =
  \frac{\mathrm{S/N}_\mathrm{min}}{\eta \sqrt{n_\mathrm{pol}}}
  \left( \frac{T_\mathrm{rec} + T_\mathrm{sky}}{\mathrm{K}} \right) 
  \left( \frac{G}{\mathrm{K\ Jy}^{-1}} \right)^{-1}
  \left( \frac{\Delta \nu}{\mathrm{MHz}} \right)^{-1/2}  
  \left( \frac{t_\mathrm{int}}{\mathrm{s}} \right)^{-1/2} 
  \left( \frac{W}{P-W} \right)^{1/2} \mathrm{\ mJy},
  \label{equ:defsmin}
\end{equation}
where $\mathrm{S/N}_\mathrm{min}$ is the threshold signal-to-noise
ratio, $\eta$ is a generic fudge factor ($\lesssim 1$) which accounts
for losses in sensitivity (e.g., due to sampling and digitization noise),
$n_\mathrm{pol}$ is the number of polarizations recorded (either 1 or 2),
$T_\mathrm{rec}$ and $T_\mathrm{sky}$ are the receiver and sky noise
temperatures, $G$ is the gain of the antenna, $\Delta \nu$ is the
observing bandwidth, $t_\mathrm{int}$ is the integration time, $W$ is the
detected pulse width and $P$ is the pulse period.


\subsubsection{Interstellar pulse dispersion and multipath scattering}
\label{sec:dispandscatt}

It follows from Equation~(\ref{equ:defsmin}) that the minimum flux
density increases as $W/(P-W)$ and hence $W$
increases. Also note that if $W \gtrsim P$, the
pulsed signal is smeared into the background emission and is no longer
detectable, regardless of how luminous the source may be. The detected
pulse width $W$ may be broader than the intrinsic value largely as a
result of pulse dispersion and multipath scattering by free electrons in the
interstellar medium. The dispersive smearing scales as $\Delta
\nu/\nu^3$, where $\nu$ is the observing frequency. This can largely
be removed by dividing the pass-band into a number of channels and
applying successively longer time delays to higher frequency channels
\emph{before} summing over all channels to produce a sharp profile.
This process is known as incoherent dedispersion.

The smearing across the individual frequency channels, however, still
remains and becomes significant at high dispersions when searching for
short-period pulsars. Multipath scattering from electron density
irregularities results in a one-sided broadening of the pulse profile due to the delay in
arrival times. A simple scattering model is shown in
Figure~\ref{fig:scatt} in which the scattering electrons are assumed to
lie in a thin screen between the pulsar and the observer~\cite{sch68}.
The timescale of this effect varies roughly as $\nu^{-4}$, which can
not currently be removed by instrumental means.

\epubtkImage{scatt.png}{
  \begin{figure}[htbp]
    \def\epsfsize#1#2{0.32#1}
    \centerline{\epsfbox{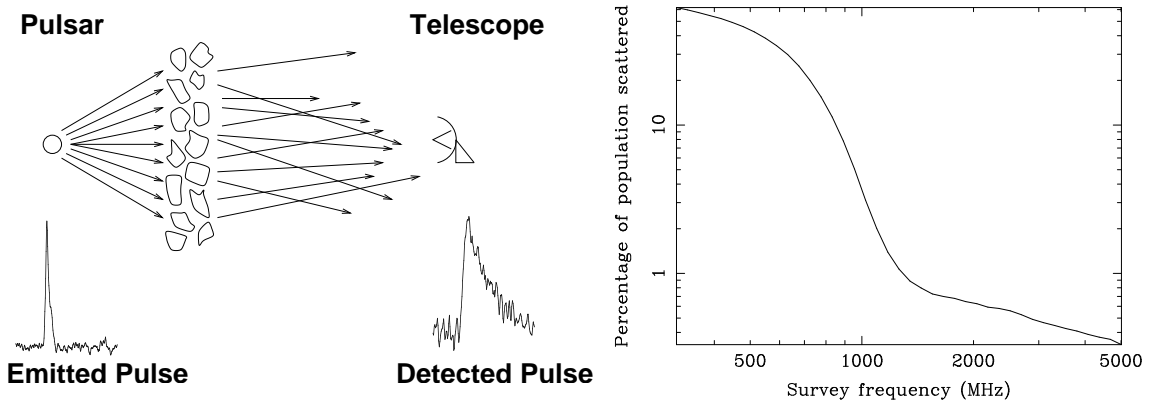}}
    \caption{Left panel: Pulse scattering caused by irregularities
      in the interstellar medium. The different path lengths and
      travel times of the scattered rays result in a ``scattering
      tail'' in the observed pulse profile which lowers its
      signal-to-noise ratio. Right panel: A simulation showing the
      percentage of Galactic pulsars that are likely to be
      undetectable due to scattering as a function of observing
      frequency. Low-frequency ($\lesssim$~1~GHz) surveys
      clearly miss a large percentage of the population due to this
      effect.}
    \label{fig:scatt}
  \end{figure}}

Dispersion and scattering are most severe for distant pulsars in
the inner Galaxy where the number of free electrons along the line of
sight becomes large. The strong frequency dependence of both effects
means that they are considerably less of a problem for surveys at
observing frequencies $\gtrsim$~1.4~GHz~\cite{clj92, jlm92} compared 
to the 400-MHz search frequency used in early surveys.
An added bonus for such
observations is the reduction in $T_\mathrm{sky}$ which
scales with frequency as approximately
$\nu^{-2.8}$~\cite{lmop87}. Pulsar intensities also have an inverse
frequency dependence, with the average scaling being $\nu^{-1.6}$~\cite{lylg95}, so that flux
densities are roughly an order of magnitude lower at 1.4~GHz compared
to 400~MHz. Fortunately, this can be at least partially compensated for by the 
use of larger receiver bandwidths at higher radio frequencies. For
example, the 1.4-GHz system at Parkes has a bandwidth of 288~MHz~\cite{lcm00}
compared to the 430-MHz system, where nominally
32~MHz is available~\cite{mld96}.


\subsubsection{Orbital acceleration}
\label{sec:accn}

Standard pulsar searches use Fourier techniques~\cite{lk05} to search
for \emph{a priori} unknown periodic signals and usually assume that
the apparent pulse period remains constant throughout the
observation. For searches with integration times much greater than a
few minutes, this assumption is only valid for solitary pulsars or
binary systems with orbital periods longer than
about a day. For shorter-period binary systems, the Doppler-shifting
of the period results in a spreading of the signal power over a number
of frequency bins in the Fourier domain, leading to a reduction in
S/N~\cite{jk91}. An observer will perceive the frequency of a pulsar
to shift by an amount $aT/(Pc)$, where $a$ is the (assumed constant)
line-of-sight acceleration during the observation of length $T$, $P$
is the (constant) pulsar period in its rest frame and $c$ is the speed
of light. Given that the width of a frequency bin in the Fourier
domain is $1/T$, we see that the signal will drift into more than one
spectral bin if $aT^2/(Pc)>1$. Survey sensitivities to
rapidly-spinning pulsars in tight orbits are therefore significantly
compromised when the integration times are large.

\epubtkImage{1913acsearch.png}{
  \begin{figure}[htbp]
    \def\epsfsize#1#2{0.6#1}
    \centerline{\epsfbox{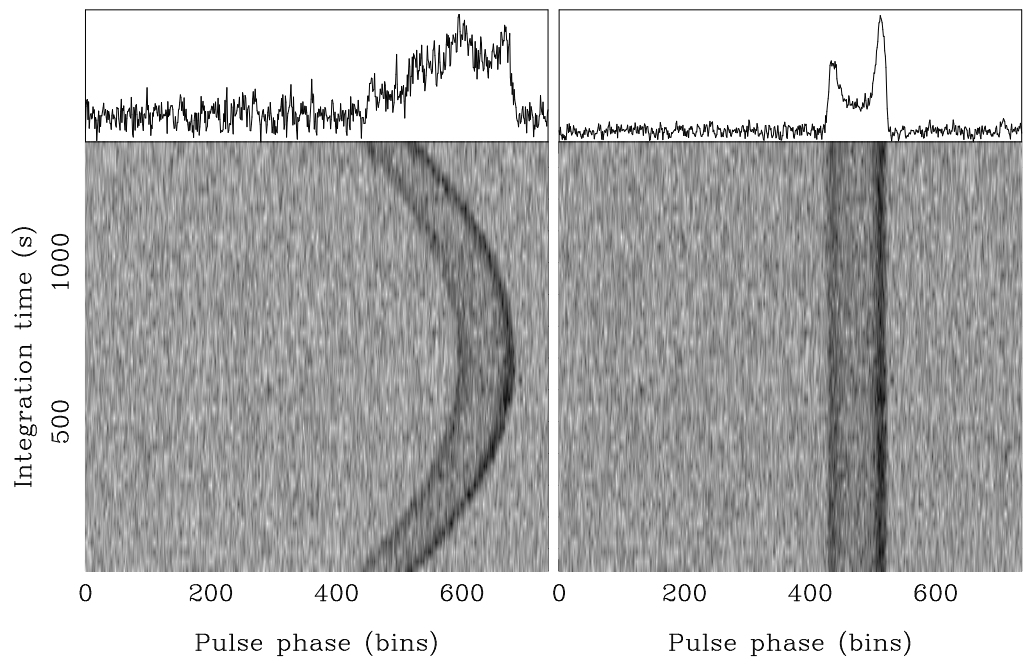}}
    \caption{Left panel: A 22.5-min Arecibo observation of the
      binary pulsar B1913+16. The assumption that the pulsar has a
      constant period during this time is clearly inappropriate given
      the quadratic drifting in phase of the pulse during the
      observation (linear grey scale plot). Right panel: The same
      observation after applying an acceleration search. This shows
      the effective recovery of the pulse shape and a significant
      improvement in the signal-to-noise ratio.}
    \label{fig:1913acc}
  \end{figure}}

As an example of this effect, as seen in the time domain,
Figure~\ref{fig:1913acc} shows a 22.5-min search mode observation of the
binary pulsar~B1913+16~\cite{ht75a, tw82, tw89}. Although this
observation covers only about 5\% of the orbit (7.75~hr), the severe effects
of the Doppler smearing on the pulse signal are very apparent. While
the standard search code nominally detects the pulsar with $\mathrm{S/N}=9.5$
for this observation, it is clear that this value is significantly
reduced due to the Doppler shifting of the
pulse period seen in the individual sub-integrations.

It is clearly desirable to employ a technique to recover the loss in
sensitivity due to Doppler smearing. One such technique, the so-called
``acceleration search''~\cite{mk84}, assumes the pulsar has a constant
acceleration during the observation. Each time series can then be
re-sampled to refer it to the frame of an inertial observer using the
Doppler formula to relate a time interval $\tau$ in the pulsar frame
to that in the observed frame at time $t$, as $\tau(t) \propto ( 1 +
at/c )$. Searching over a range of accelerations is desirable to find
the time series for which the trial acceleration most closely matches
the true value. In the ideal case, a time series is produced with a
signal of constant period for which full sensitivity is recovered (see
right panel of Figure~\ref{fig:1913acc}). This technique was first
used to find PSR B2127+11C~\cite{agk90}, a double neutron star binary
in M15 which has parameters similar to B1913+16. Its
application to 47~Tucanae~\cite{clf00} resulted in the discovery of
nine binary millisecond pulsars, including one in a 96-min orbit
around a low-mass ($0.15\,M_{\odot}$) companion. This is
currently the shortest binary period for any known radio pulsar.
The majority of binary millisecond pulsars with orbital periods
less than a day found in recent globular cluster searches would
not have been discovered without the use of acceleration searches.

For intermediate orbital periods, in the range 30~min\,--\,several hours,
another promising technique is the dynamic power spectrum search
shown in Figure~\ref{fig:dps}. 
Here the time series is split into a number of smaller contiguous
segments which are Fourier-transformed separately. The individual
spectra are displayed as a two-dimensional (frequency versus time)
image. Orbitally modulated pulsar signals appear as sinusoidal signals
in this plane as shown in Figure~\ref{fig:dps}.

\epubtkImage{dps.png}{
  \begin{figure}[htbp]
    \def\epsfsize#1#2{0.9#1}
    \centerline{\epsfbox{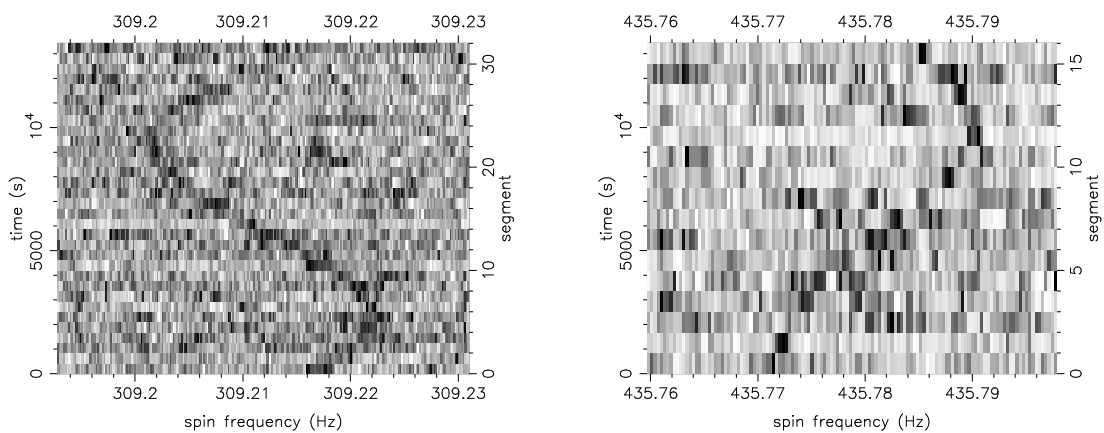}}
    \caption{Dynamic power spectra showing two recent pulsar
      discoveries in the globular cluster M62 showing fluctuation
      frequency as a function of time. Figure provided by Adam
      Chandler.}
    \label{fig:dps}
  \end{figure}}

This technique has been used by various groups where spectra are
inspected visually~\cite{lmbm00}. Much of
the human intervention can be removed using a hierarchical scheme for
selecting significant events~\cite{cha03}. This approach was recently
applied to a search of the globular cluster M62 resulting in the
discovery of three new pulsars. One of the new discoveries -- M62F, a
faint 2.3-ms pulsar in a 4.8-hr orbit -- was detectable only using the
dynamic power spectrum technique.

For the shortest orbital periods, the assumption of a constant
acceleration during the observation clearly breaks down. In this case,
a particularly efficient algorithm has been
developed~\cite{cor95, ran01, jrs02, rce03} which is optimised to finding
binaries with periods so short that many orbits can take place during
an observation. This ``phase modulation'' technique exploits the fact
that the Fourier components are modulated by the orbit to create a
family of periodic sidebands around the nominal spin frequency of the
pulsar. While this technique has so far not resulted in any new
discoveries, the existence of short period binaries in
47~Tucanae~\cite{clf00}, Terzan~5~\cite{rhs05} and the 11-min X-ray
binary X1820$-$303 in NGC~6624~\cite{spw87}, suggests that there are more
ultra-compact radio binary pulsars that await discovery.


\subsection{Correcting the observed pulsar sample}
\label{sec:corsamp}

In the following, we review common techniques to account for many of 
the aforementioned selection effects and form a less biased 
picture of the true pulsar population.


\subsubsection{Scale factor determination}
\label{sec:sfacts}

A very useful approach~\cite{pb81, vn81}, is to
define a scaling factor $\xi$ as the ratio of the total Galactic
volume weighted by pulsar density to the volume in which a pulsar
is detectable:
\begin{equation}
  \xi(P, L) = \frac{\int \int_\mathrm{G} \Sigma(R, z) R \, dR \, dz}
  {\int \int_{P, L} \Sigma(R, z) R \, dR \, dz}.
  \label{equ:sfac}
\end{equation}
Here, $\Sigma(R, z)$ is the assumed pulsar space density distribution
in terms of galactocentric radius $R$ and height above the Galactic
plane $z$. Note that $\xi$ is primarily a function of period $P$ and
luminosity $L$ such that short-period/low-luminosity pulsars have
smaller detectable volumes and therefore higher $\xi$ values than
their long-period/high-luminosity counterparts. 

In practice, $\xi$ is calculated for each pulsar separately using a Monte Carlo
simulation to model the volume of the Galaxy probed by the major
surveys~\cite{nar87}. For a sample of $N_\mathrm{obs}$ observed pulsars
above a minimum luminosity $L_\mathrm{min}$, the total number of
pulsars in the Galaxy 
\begin{equation}
  N_\mathrm{G} = \sum_{i=1}^{N_\mathrm{obs}} \frac{\xi_i}{f_i},
  \label{equ:ngal}
\end{equation}
where $f$ is the model-dependent ``beaming fraction'' discussed below
in Section~\ref{sec:beaming}. Note that this estimate 
applies to those pulsars with
luminosities $\gtrsim L_\mathrm{min}$. Monte Carlo simulations have shown
this method to be reliable, as long as $N_\mathrm{obs}$ is reasonably
large~\cite{lbdh93}.


\subsubsection{The small-number bias}
\label{sec:smallnumber}

For small samples of observationally-selected objects, the detected
sources are likely to be those with larger-than-average luminosities.
The sum of the scale factors~(\ref{equ:ngal}), therefore,
will tend to underestimate the true size of the population. This
``small-number bias'' was first pointed out~\cite{kal00, knst01} for
the sample of double neutron star binaries where we know of only five
systems relevant for calculations of the merging rate
(see Section~\ref{sec:nsns}). Only when $N_\mathrm{obs} \gtrsim 10$ does
the sum of the scale factors become a good indicator of the true population size.

\epubtkImage{smallnumber.png}{
  \begin{figure}[htbp]
    \def\epsfsize#1#2{0.9#1}
    \centerline{\epsfbox{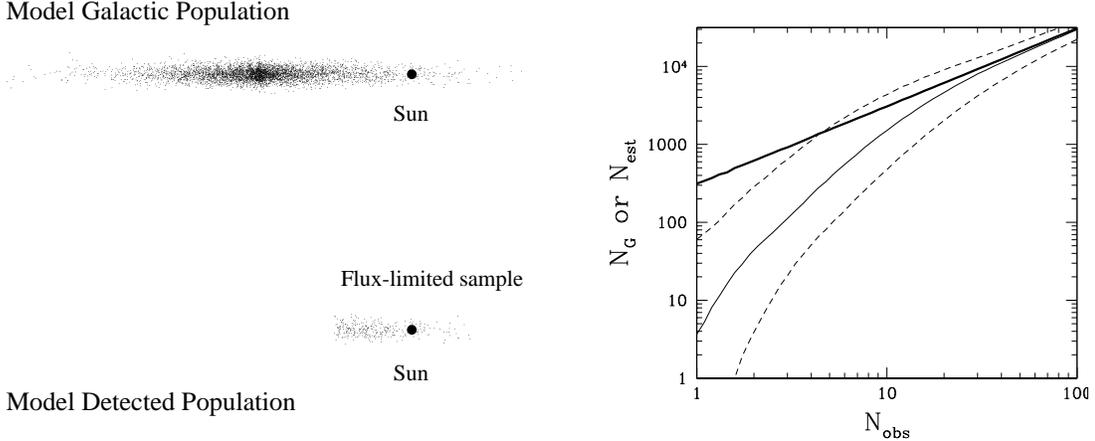}}
    \caption{Small-number bias of the scale factor estimates
      derived from a synthetic population of sources where the true
      number of sources is known. Left panel: An edge-on view of a
      model Galactic source population. Right panel: The thick line
      shows $N_\mathrm{G}$, the true number of objects in the model
      Galaxy, plotted against $N_\mathrm{observed}$, the number
      detected by a flux-limited survey. The thin solid line shows
      $N_\mathrm{est}$, the median sum of the scale factors, as a
      function of $N_\mathrm{obs}$ from a large number of Monte Carlo
      trials. Dashed lines show 25 and 75\% percentiles of the
      $N_\mathrm{est}$ distribution.}
    \label{fig:smallnumber}
  \end{figure}}

Despite a limited sample size, it has been
demonstrated~\cite{kkl03} that rigorous confidence
intervals of $N_\mathrm{G}$ can be derived using Bayesian techniques.  Monte Carlo simulations
verify that the simulated number of detected objects
$N_\mathrm{detected}$
closely follows a Poisson distribution and that $N_\mathrm{detected} = \alpha
N_\mathrm{G}$, where $\alpha$ is a constant. By varying the
value of $N_\mathrm{G}$ in the simulations, the mean of
this Poisson distribution can be measured.
The Bayesian analysis~\cite{kkl03}
finds, for a single object, the probability density function
of the total population is
\begin{equation}
  P(N_\mathrm{G}) = \alpha^2 N_\mathrm{G} \exp(-\alpha N_\mathrm{G}).
\end{equation}
Adopting the necessary assumptions required in the Monte
Carlo population about the underlying pulsar distribution,
this technique can be used to place interesting constraints
on the size and, as we shall see later, birth rate of the
underlying population.


\subsubsection{The beaming correction}
\label{sec:beaming}

The ``beaming fraction'' $f$ in Equation~(\ref{equ:ngal}) is the
fraction of $4\pi$ steradians swept out by a pulsar's radio beam
during one rotation. Thus $f$ is the probability that the beam cuts
the line-of-sight of an arbitrarily positioned observer. A na\"{\i}ve
estimate for $f$ of roughly 20\% assumes a circular beam of width $\sim 10^{\circ}$
and a randomly distributed inclination angle between the spin and
magnetic axes~\cite{tm77}. Observational evidence summarised in Figure~\ref{fig:bfracts} suggests that
shorter period pulsars have wider beams and therefore larger beaming
fractions than their long-period
counterparts~\cite{nv83, lm88, big90b, tm98}.

\epubtkImage{beam.png}{
  \begin{figure}[htbp]
    \def\epsfsize#1#2{0.5#1}
    \centerline{\epsfbox{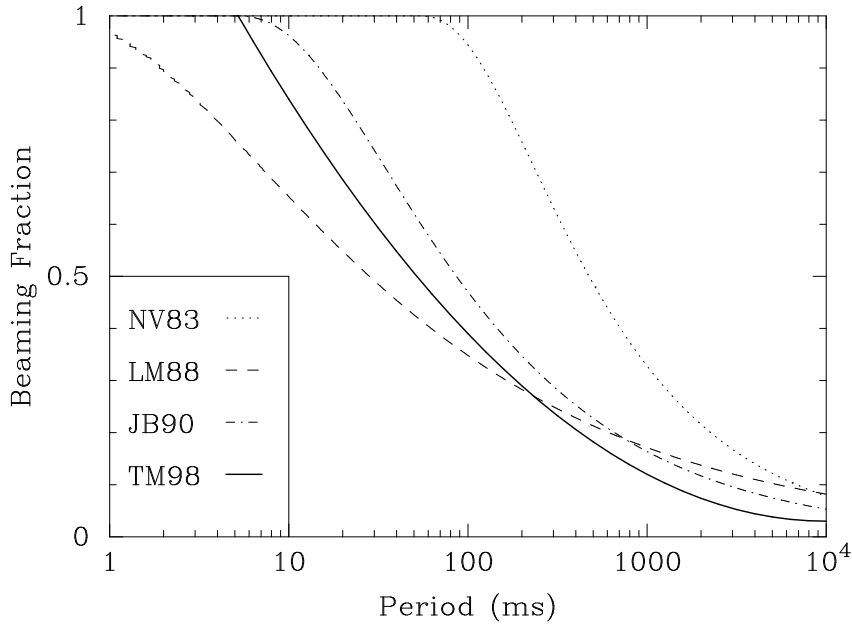}}
    \caption{Beaming fraction plotted against pulse period for
      four different beaming models: Narayan \& Vivekanand 1983 (NV83)
      \cite{nv83}, Lyne \& Manchester 1988 (LM88) \cite{lm88}, Biggs
      1990 (JDB90) \cite{big90b} and Tauris \& Manchester 1998 (TM98)
      \cite{tm98}.}
    \label{fig:bfracts}
  \end{figure}}

When most of these beaming models were originally proposed, the sample of
millisecond pulsars was $\lesssim 5 $ and hence their predictions about
the beaming fractions of short-period pulsars relied largely on
extrapolations from the normal pulsars. An analysis of a large
sample of millisecond pulsar profiles~\cite{kxl98} 
suggests that their beaming fraction lies between 50 and 100\%. 
Independent constraints on $f$ for millisecond pulsars come
from deep {\it Chandra} observations of the globular cluster
47~Tucanae~\cite{gch02} and radio pulsar surveys~\cite{clf00}
which suggest that $f>0.4$ and likely close to unity~\cite{hge05}.
The large beaming fraction and narrow pulses often observed
strongly suggests a fan beam model for millisecond pulsars~\cite{mic91}.


\subsection{The population of normal and millisecond pulsars}
\label{sec:nmsppop}


\subsubsection{Luminosity distributions and local number estimates}
\label{sec:lumfuns}

Based on a number of all-sky surveys carried out in the 1990s, the
scale factor approach has been used to derive the characteristics of
the true normal and millisecond pulsar populations and is based on the
sample of pulsars within 1.5~kpc of the Sun~\cite{lml98}. Within this
region, the selection effects are well understood and easier to
quantify than in the rest of the Galaxy. These calculations
should therefore give a reliable \emph{local pulsar population} estimate.

\epubtkImage{lumfuns.png}{
  \begin{figure}[htbp]
    \def\epsfsize#1#2{0.32#1}
    \centerline{\epsfbox{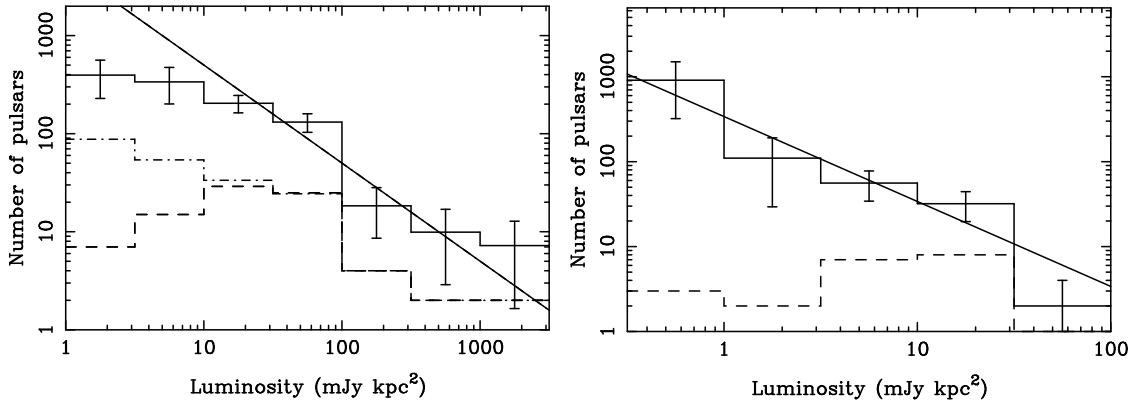}}
    \caption{Left panel: The corrected luminosity distribution
      (solid histogram with error bars) for normal pulsars. The
      corrected distribution \emph{before} the beaming model has been
      applied is shown by the dot-dashed line. Right panel: The
      corresponding distribution for millisecond pulsars. In both cases,
      the observed distribution is shown by the dashed line and the
      thick solid line is a power law with a slope of $-1$. The
      difference between the observed and corrected distributions
      highlights the severe under-sampling of low-luminosity pulsars.}
    \label{fig:lumfuns}
  \end{figure}}

The 430-MHz luminosity distributions obtained from this analysis are shown in
Figure~\ref{fig:lumfuns}. For the normal pulsars, integrating the
corrected distribution above $1 \mathrm{\ mJy\ kpc}^2$ and dividing by $\pi
\times (1.5)^2 \mathrm{\ kpc}^2$ yields a local surface density, assuming
a beaming model~\cite{big90b}, of $156 \pm 31 \mathrm{pulsars\
  kpc}^{-2}$ for 430-MHz luminosities above $1 \mathrm{\ mJy\ kpc}^2$. The same
analysis for the millisecond pulsars, assuming a mean beaming fraction
of 75\%~\cite{kxl98}, leads to a local surface density of $38 \pm 16
\mathrm{\ pulsars\ kpc}^{-2}$ also for 430-MHz luminosities above $1 \mathrm{\ mJy\ kpc}^2$.


\subsubsection{Galactic population and birth-rates}
\label{sec:psrpop}

Integrating the local surface densities of pulsars over the whole
Galaxy requires a knowledge of the presently rather uncertain
Galactocentric radial distribution~\cite{joh94, lor04}. One approach is
to assume that pulsars have a radial distribution similar to that of
other stellar populations and to scale the local number density with 
this distribution in order to estimate the total Galactic population. 
The corresponding local-to-Galactic scaling is
1000$\pm$250~kpc\super{2}~\cite{rv89}. This implies a population of
$\sim$~160,000 active normal pulsars and $\sim$~40,000
millisecond pulsars in the Galaxy. 

Based on these estimates, we are in a position to deduce the
corresponding rate of formation or birth-rate required to sustain
the observed population. From the
$P \mbox{--} \dot{P}$ diagram in Figure~\ref{fig:ppdot}, we infer a typical
lifetime for normal pulsars of $\sim$~10\super{7}~yr, corresponding to a
Galactic birth rate of $\sim$~1 per 60~yr -- consistent with the rate
of supernovae~\cite{vt91}. As noted in Section~\ref{sec:nms}, the
millisecond pulsars are much older, with ages close to that of the
Universe $\tau_\mathrm{u}$ (we assume here $\tau_\mathrm{u}$~=~13.8~Gyr~\cite{yjb05}).
Taking the maximum age of the millisecond pulsars
to be $\tau_\mathrm{u}$, we infer a mean birth rate of at least 1 per
345,000~yr. This is consistent, within the uncertainties, of other studies of the millisecond pulsar population~\cite{fw07, sgh07} and
with the birth-rate of low-mass X-ray binaries~\cite{lnl95, kh06}.


\subsection{The population of relativistic binaries}
\label{sec:relpop}

Although no radio pulsar has so far been observed in orbit around a
black hole companion, we now know of several double neutron star and
neutron star--white dwarf binaries which will merge due to
gravitational wave emission within a reasonable timescale. The
current sample of objects is shown as a function of orbital period and
eccentricity in Figure~\ref{fig:mplane}. Isochrones showing various
coalescence times $\tau_\mathrm{g}$ are calculated using the
expression
\begin{equation}
  \tau_\mathrm{g} \simeq 9.83 \times 10^6 \mathrm{\ yr}
  \left( \frac{P_\mathrm{b}}{\mathrm{hr}} \right)^{8/3} 
  \left( \frac{m_1 + m_2}{M_\odot} \right)^{-2/3} 
  \left( \frac{\mu}{M_\odot} \right)^{-1}
  \left( 1 - e^2 \right)^{7/2},
  \label{equ:tgw}
\end{equation}
where $m_1$ and $m_2$ are the masses of the two stars, 
$\mu=m_1m_2/(m_1+m_2)$ is the ``reduced mass'', 
$P_\mathrm{b}$ is the binary period and $e$ is
the eccentricity. This formula is a good analytic approximation
(within a few percent) to the numerical solution of the exact
equations for $\tau_\mathrm{g}$~\cite{pet64, pm63}. 

\epubtkImage{mplane.png}{
  \begin{figure}[htbp]
    \def\epsfsize#1#2{0.5#1}
    \centerline{\epsfbox{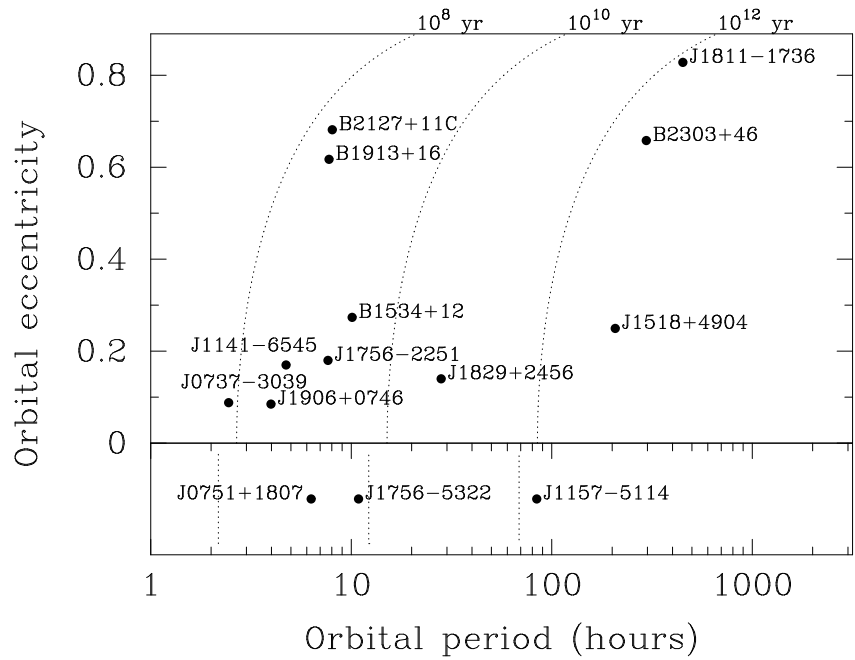}}
    \caption{The relativistic binary merging plane. Top: Orbital
      eccentricity versus period for eccentric binary systems
      involving neutron stars. Bottom: Orbital period distribution for
      the massive white dwarf--pulsar binaries. Isocrones show
      coalescence times assuming neutron stars of
      $1.4\,M_{\odot}$ and white dwarfs of $0.3\,M_{\odot}$.}
    \label{fig:mplane}
  \end{figure}}

In addition to tests of strong-field gravity through observations of
relativistic binary systems (see Section~\ref{sec:tbin}), estimates of their Galactic
population and merger rate are of great interest as one of the prime
sources for current gravitational wave detectors such as
GEO600~\cite{geo600}, LIGO~\cite{ligo}, VIRGO~\cite{virgo} and
TAMA~\cite{tama}. In the following, we review empirical
determinations of the population sizes and merging rates of binaries
where at least one component is visible as a radio pulsar.


\subsubsection{Double neutron star binaries}
\label{sec:nsns}

As discussed in Section~\ref{sec:nms}, double neutron star (DNS) binaries
are expected to be rare. This is certainly the case; as
summarized in Table~\ref{tab:nsns},
only around ten DNS binaries are
currently known. Although we only
see both neutron stars as pulsars in J0737$-$3039~\cite{lbk04}, we are
``certain'' of the identification in five other systems from
precise component mass measurements from pulsar timing
observations (see Section~\ref{sec:tbin}). The
other systems
listed in Table~\ref{tab:nsns}
have eccentric
orbits, mass functions and periastron advance measurements that are
\emph{consistent} with a DNS identification, but for which there is
presently not sufficient component mass information to confirm their nature.
One further DNS candidate, the 95-ms pulsar J1753$-$2243 (see
Table~\ref{tab:ebpsrs}), has recently been
discovered~\cite{kei08}. Although the mass function for this pulsar is
lower than the DNS systems listed in Table~\ref{tab:nsns}, a neutron
star companion cannot be ruled out in this case. Further observations
should soon clarify the nature of this system. 
We note, however, that the 13.6-day
orbital period of this system means that it will not contribute
to gravitational wave inspiral rate calculations discussed below.

\begin{table}[htbp]
  \caption[Known and likely DNS binaries.]{Known and likely DNS
  binaries. Listed are the pulse period $P$, orbital period
  $P_\mathrm{b}$, orbital eccentricity $e$, characteristic age
  $\tau_\mathrm{c}$, and expected binary coalescence timescale
  $\tau_\mathrm{g}$ due to gravitational wave emission calculated from
  Equation~(\ref{equ:tgw}). To distinguish between definite and
  candidate DNS systems, we also list whether the masses of both
  components have been determined via the measurement of two or more
  post-Keplerian parameters as described in Section~\ref{sec:tbin}.}
  \label{tab:nsns}
  \vskip 4mm
  \centering
  \begin{tabular}{l|rrrrr}
    \hline \hline
    \vsp & J0737$-$3039 & J1518+4904 & B1534+12 & J1756$-$2251 & J1811$-$1736\\ [0.2 em]
    \hline
    $P$ [ms] & 22.7/2770 & 40.9 & 37.9 & 28.5 & 104.2\\
    $P_\mathrm{b}$ [d] & 0.102 & 8.6 & 0.4 & 0.32 & 18.8 \\
    $e$ & 0.088 & 0.25 & 0.27 & 0.18 & 0.83 \\
    $\log_{10}(\tau_\mathrm{c}$/[$\mathrm{yr}$]) & 8.3/7.7 & 10.3 & 8.4 & 8.6 & 9.0\\
    $\log_{10}(\tau_\mathrm{g}$/[$\mathrm{yr}$]) & 7.9 & 12.4 & 9.4 & 10.2 & 13.0\\
    Masses measured? & Yes & No & Yes & Yes & Yes  \\
    \hline \hline
    \vsp & B1820$-$11 & J1829+2456 & J1906+0746 & B1913+16 & B2127+11C \\ [0.2 em]
    \hline
    $P$ [ms] & 279.8 & 41.0 & 144.1 & 59.0 & 30.5 \\
    $P_\mathrm{b}$ [d] & 357.8 & 1.18 & 0.17 & 0.3 & 0.3 \\
    $e$ & 0.79 & 0.14 & 0.085 & 0.62 & 0.68 \\
    $\log_{10}(\tau_\mathrm{c}$/[$\mathrm{yr}$])& 6.5 & 10.1 & 5.1 & 8.0 & 
8.0 \\
    $\log_{10}(\tau_\mathrm{g}$/[$\mathrm{yr}$])& 15.8 & 10.8 & 8.5  & 8.5 & 8.3 \\
    Masses measured? & No & No & Yes & Yes & Yes \\
    \hline \hline
  \end{tabular}
\end{table}

Despite the uncertainties in identifying DNS binaries, for the
purposes of determining the Galactic merger rate, the
systems
for
which
$\tau_\mathrm{g}$ is less than $\tau_\mathrm{u}$
(i.e.\ PSRs J0737$-$3039, B1534+12, J1756$-$2251, J1906+0746, B1913+16 and
B2127+11C) are primarily of interest. Of these PSR B2127+11C is in the
process of being ejected from the globular cluster
M15~\cite{pakw91, ps91} and is thought to make only a negligible
contribution to the merger rate~\cite{phi91}. The general approach
with the remaining systems is to derive scale factors for each
object, construct the probability density function of their total
population (as outlined in Section~\ref{sec:sfacts}) and then divide these
by a reasonable estimate for the lifetime. Getting such estimates is, however, difficult. It has been
proposed~\cite{knst01} that the \emph{observable lifetimes} for these
systems are determined by the timescale on which the current orbital
period is reduced by a factor of two~\cite{acw99}. Below this point,
the orbital smearing selection effect discussed in Section~\ref{sec:accn}
will render the binary undetectable by current surveys. More
recent work~\cite{cb05} has suggested that a significant population
of highly eccentric binary systems could easily evade detection
due to their short lifetimes before gravitational wave inspiral.
If this selection effect is significant, then the merger rate estimates 
quoted below could easily be underestimated by a factor of a few.

The results of the most recent DNS merger rate estimates of this
kind~\cite{kim08}
are summarised in the left panel of
Figure~\ref{fig:rates}. The combined Galactic merger rate, dominated by
the double pulsar and J1906+0746 is found to be $118_{-79}^{+174} \mathrm{\ Myr}^{-1}$, where
the uncertainties reflect the 95\% confidence level using the
techniques summarised in Section~\ref{sec:smallnumber}. Extrapolating this
number to include DNS binaries detectable by LIGO in other
galaxies~\cite{phi91}, the expected event rate is
$49^{+73}_{-33}\times 10^{-3} \mathrm{\ yr}^{-1}$ for initial LIGO and
$265^{+390}_{-178}\mathrm{\ yr}^{-1}$ for advanced LIGO. Future prospects for
detecting gravitational wave emission from binary neutron star
inspirals are therefore very encouraging, especially if the population of highly eccentric systems is significant~\cite{cb05}. Since much of the
uncertainty in the rate estimates is due to our ignorance of the
underlying distribution of double neutron star systems, future
gravitational wave detection could ultimately constrain the properties of this
exciting binary species.

\epubtkImage{rates.png}{
  \begin{figure}[htbp]
    \def\epsfsize#1#2{0.37#1}
    \centerline{\epsfbox{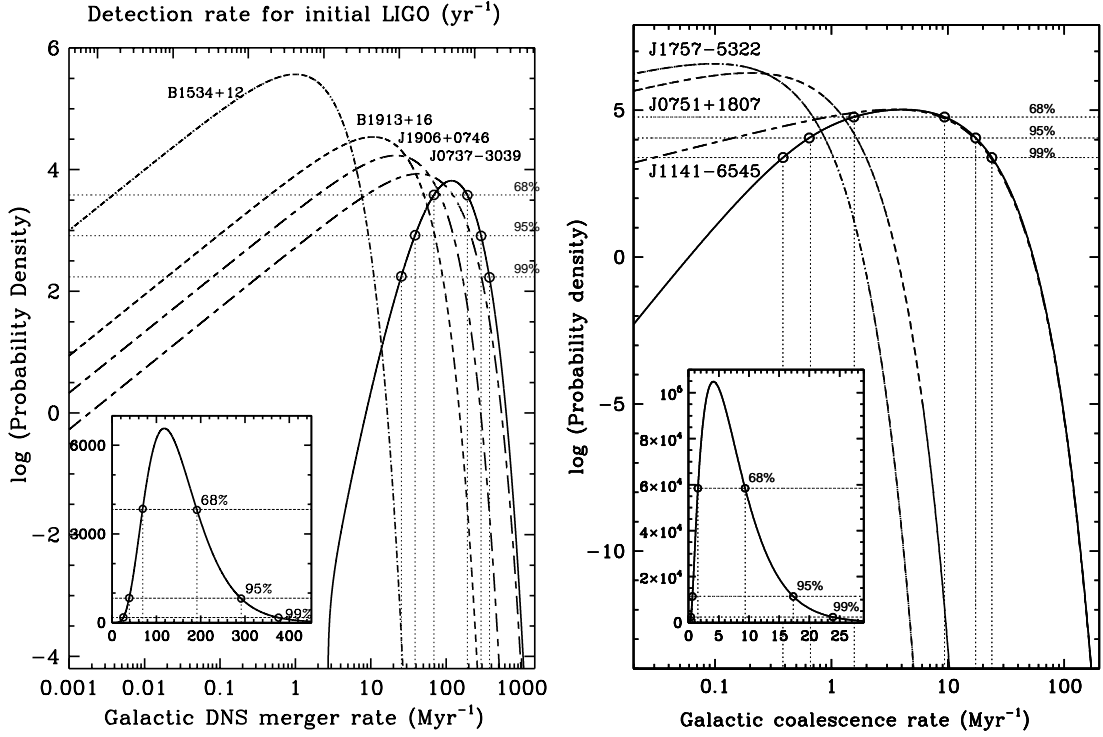}}
    \caption{The current best empirical estimates of the
      coalescence rates of relativistic binaries involving neutron
      stars. The individual contributions from each known binary
      system are shown as dashed lines, while the solid line shows the
      total probability density function on a logarithmic and (inset)
      linear scale. The left panel shows the most recent analysis for DNS
      binaries~\cite{kim08}, while the right panel shows the
      equivalent results for NS--WD binaries~\cite{kklw04}.
      Figures provided by Chunglee Kim.}
    \label{fig:rates}
  \end{figure}}

Although the double pulsar system J0737$-$3039 will not be important
for ground-based detectors until its final coalescence in another
85~Myr, it may be a useful calibration source for the future
space-based detector LISA~\cite{lisa}. It is calculated~\cite{kkl04}
that a 1-yr observation with LISA would
detect (albeit with $\mathrm{S/N} \sim 2$) the continuous emission at a
frequency of 0.2~mHz based on the current orbital parameters. Although
there is the prospect of using LISA to detect similar systems
through their continuous emission, current calculations~\cite{kkl04}
suggest that significant ($\mathrm{S/N}>5$) detections are not likely. Despite
these limitations, it is likely that LISA observations will be able to
place independent constraints on the Galactic DNS binary population
after several years of operation.


\subsubsection{White dwarf--neutron star binaries}
\label{sec:nswd}

Although the population of white dwarf--neutron star (WDNS) binaries
in general is substantial, the fraction which will merge due to
gravitational wave emission is small. Like the DNS binaries, the
observed WDNS sample suffers from small-number statistics. From
Figure~\ref{fig:mplane}, we note that only three WDNS systems
are currently known that will merge within $\tau_\mathrm{u}$, PSRs
J0751+1807~\cite{lzc95}, J1757$-$5322~\cite{eb01} and 
J1141$-$6545~\cite{klm00}. Applying the same techniques as used
for the DNS population, the merging rate contributions of the
three systems can be calculated~\cite{kklw04} and are shown
in Figure~\ref{fig:rates}. The combined Galactic coalescence rate is
$4_{-3}^{+5} \mathrm{\ Myr}^{-1}$ (at 68\% confidence interval). 
This result is not corrected for beaming and therefore
should be regarded as a lower
limit on the total event rate. Although
the orbital frequencies of these objects at coalescence are too 
low to be detected by LIGO, they do fall within the band planned
for LISA~\cite{lisa}. Unfortunately, an extrapolation of the
Galactic event rate out to distances at which such events would be detectable
by LISA does not suggest that these systems will be a major source
of detection~\cite{kklw04}. Similar conclusions were reached by
considering the statistics of low-mass X-ray binaries~\cite{coo04}.

\subsection{Going further}

Good starting points for further reading can be found in other review articles~\cite{pk94,kim08}.
Our coverage of compact object coalescence rates has concentrated on empirical
methods.  Two software packages are freely available which allow the user to
synthesize and search radio pulsar populations~\cite{gppg,psrpop,rl08}.
An alternative approach is to undertake a full-blown Monte Carlo
simulation of the most likely evolutionary scenarios described in
Section~\ref{sec:evolution}. In this scheme, a
population of primordial binaries is synthesized with a number of
underlying distribution functions: primary mass, binary mass ratio,
orbital period distribution etc. The evolution of both stars is then
followed to give a predicted sample of binary systems of all the
various types. Although selection effects are not always taken into account
in this approach, the final census is usually normalized to the
star formation rate.

Numerous population syntheses (most often to populations of binaries
where one or both members are NSs) can be
found in the literature~\cite{dc87, rom92, ty93, pv96, lpp96, bkb02, bkr08}.  A group
in Moscow has made a web interface to their code~\cite{scenario}.
Although extremely instructive, the uncertain assumptions about
initial conditions, the physics of mass transfer and the kicks applied
to the compact object at birth result in a wide range of predicted
event rates which are currently broader than the empirical
methods~\cite{kkl04b, kklw04, kkl05}. Ultimately, the detection
statistics from the gravitational wave detectors could provide far
tighter constraints on the DNS merging rate than the pulsar surveys
from which these predictions are made.  Very
recently~\cite{okkb08} the results from empirical population
constraints and full-blown binary population synthesis codes have been
combined to constrain a variety of input parameters and physical
conditions.  The results of this work are promising, with stringent
constraints being placed on the kick distributions, mass-loss fraction
during mass transfer and common envelope assumptions.

\newpage


\section{Principles and Applications of Pulsar Timing}
\label{sec:pultim}

Pulsars are excellent
celestial clocks. The period of the first pulsar~\cite{hbp68} was
found to be stable to one part in 10\super{7} over a few months. Following
the discovery of the millisecond pulsar B1937+21~\cite{bkh82} it was
demonstrated that its period could be measured to one part in
10\super{13} or better~\cite{dtwb85}. This unrivaled stability leads to a
host of applications including uses as time keepers, probes of relativistic
gravity and natural gravitational wave detectors.


\subsection{Observing basics}
\label{sec:timobs}

Each pulsar is typically observed at least once or twice per month
over the course of a year to establish its basic
properties. Figure~\ref{fig:timing} summarises the essential steps
involved in a ``time-of-arrival'' (TOA) measurement. Pulses from the
neutron star traverse the interstellar medium before being received at
the radio telescope where they are dedispersed and added to form a
mean pulse profile.

\epubtkImage{timing.png}{
  \begin{figure}[htbp]
    \def\epsfsize#1#2{0.6#1}
    \centerline{\epsfbox{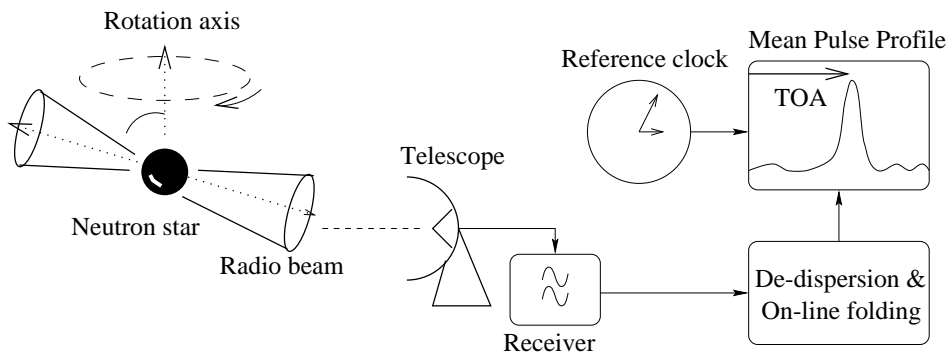}}
    \caption{Schematic showing the main stages involved in pulsar
      timing observations.}
    \label{fig:timing}
  \end{figure}}

During the observation, the data regularly receive a time stamp,
usually based on a caesium time standard or hydrogen maser at the
observatory plus a signal from the Global Positioning System of
satellites (GPS; see~\cite{gps}). The TOA is defined as the arrival time
of some fiducial point on the integrated profile with respect to either
the start or the midpoint of the observation. Since the profile has a
stable form at any given observing frequency (see Section~\ref{sec:profs}), the
TOA can be accurately determined by cross-correlation of the observed
profile with a high S/N ``template'' profile obtained from
the addition of many observations at the particular observing frequency.

Successful pulsar timing requires optimal TOA precision which
largely depends on the signal-to-noise ratio (S/N) of the 
pulse profile. Since the TOA uncertainty $\epsilon_\mathrm{TOA}$
is roughly the pulse width divided by the S/N, using 
Equation~(\ref{equ:defsmin}) we may write the fractional error as
\begin{equation}
  \frac{\epsilon_\mathrm{TOA}}{P} \simeq 
  \left( \frac{S_\mathrm{psr}}{\mathrm{mJy}} \right)^{-1}
  \left( \frac{T_\mathrm{rec} + T_\mathrm{sky}}{\mathrm{K}} \right)
  \left( \frac{G}{\mathrm{K\ Jy}^{-1}} \right)^{-1}
  \left( \frac{\Delta \nu}{\mathrm{MHz}} \right)^{-1/2}
  \left( \frac{t_\mathrm{int}}{\mathrm{s}} \right)^{-1/2}
  \left( \frac{W}{P} \right)^{3/2}\!\!\!\!\!\!\!.
  \label{equ:defsnr}
\end{equation}
Here, $S_\mathrm{psr}$ is the flux density of the pulsar, $T_\mathrm{rec}$
and $T_\mathrm{sky}$ are the receiver and sky noise temperatures, $G$ is
the antenna gain, $\Delta \nu$ is the observing bandwidth,
$t_\mathrm{int}$ is the integration time, $W$ is the pulse width and $P$ is the
pulse period (we assume $W \ll P$). Optimal results are thus obtained
for observations of short period pulsars with large flux densities and
small duty cycles (i.e.~small $W/P$) using large telescopes with low-noise
receivers and large observing bandwidths.

One of the main problems of employing large bandwidths is pulse
dispersion. As discussed in Section~\ref{sec:dist}, 
pulses emitted at lower radio frequencies travel
slower and arrive later than those emitted at higher
frequencies. This process has the effect of ``stretching'' the pulse
across a finite receiver bandwidth, increasing $W$
and therefore increasing $\epsilon_\mathrm{TOA}$.
For normal pulsars, dispersion can largely be compensated for
by the incoherent dedispersion process outlined in Section~\ref{sec:selfx}.

The short periods of millisecond pulsars
offer the ultimate in timing precision. In order to fully exploit this,
a better method of dispersion removal is required. Technical
difficulties in building devices with very narrow channel bandwidths
require another dispersion removal technique. In the process of
coherent dedispersion~\cite{han71, lk05} the incoming signals are
dedispersed over the whole bandwidth using a filter which has the
inverse transfer function to that of the interstellar medium.
The signal processing can be done online either using high speed devices such as field
programmable gate arrays~\cite{casper,pbc05} or completely in 
software~\cite{sta98, sst00}.
Off-line data reduction, while disk-space limited,
allows for more flexible analysis schemes~\cite{bai03}.

\epubtkImage{crab.png}{
  \begin{figure}[htbp]
    \def\epsfsize#1#2{0.5#1}
    \centerline{\epsfbox{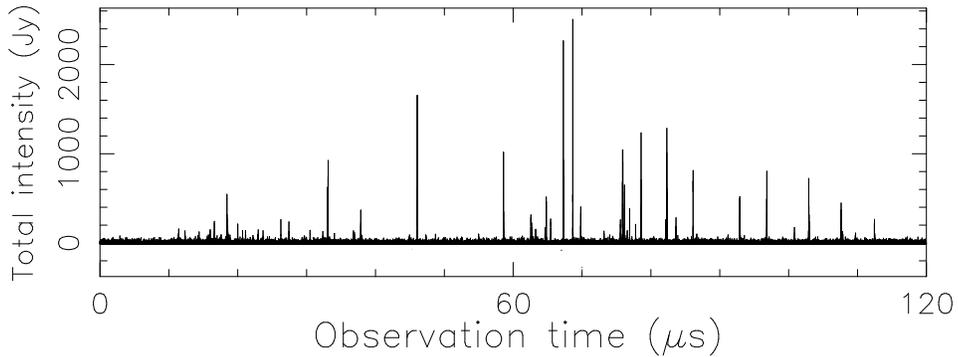}}
    \caption{A 120~$\mu$s window centred on a coherently-dedispersed
    giant pulse from the Crab pulsar showing high-intensity nanosecond
    bursts. Figure provided by Tim Hankins~\cite{hkwe03}.}
    \label{fig:crab}
  \end{figure}}

The maximum time resolution obtainable via coherent dedispersion is the
inverse of the total receiver bandwidth. The current state of the art
is the detection~\cite{hkwe03} of features on nanosecond timescales in
pulses from the 33-ms pulsar B0531+21 in the Crab nebula shown in
Figure~\ref{fig:crab}. Simple light travel-time arguments can be made
to show that, in the absence of relativistic beaming
effects~\cite{gm04}, these incredibly bright  bursts originate from
regions less than 1~m in size.


\subsection{The timing model}
\label{sec:tmodel}

To model the rotational behaviour of the neutron
star we ideally require TOAs measured by an inertial observer. An observatory
located on Earth experiences accelerations with respect to the neutron
star due to the Earth's rotation and orbital
motion around the Sun and is therefore not in
an inertial frame. To a very good approximation, the solar system
centre-of-mass (barycentre) can be regarded as an
inertial frame. It is now standard practice~\cite{hun71} to transform the
observed TOAs to this frame using a planetary ephemeris such as the
JPL DE405~\cite{sn96}. The transformation
between barycentric ($\cal T$) and observed ($t$) takes the form
\begin{equation}
  {\cal T} - t =
  \frac{\underline{r} . \hat{\underline{s}}}{c} +
  \frac{(\underline{r} . \hat{\underline{s}})^2 - |\underline{r}|^2}{2cd} +
  \Delta t_\mathrm{rel} - \Delta t_\mathrm{DM}.
  \label{equ:bary}
\end{equation}
Here $\underline{r}$ is the position of the observatory with respect to the
barycentre, $\hat{\underline{s}}$ is a unit vector in the direction
of the pulsar at a distance $d$ and $c$ is the speed of
light. The first term on the right hand side of
Equation~(\ref{equ:bary}) is the light travel time from the observatory to the
solar system barycentre. Incoming pulses from all but the nearest pulsars
can be approximated by plane wavefronts. The second term, which
represents the delay due to spherical wavefronts, yields the
parallax and hence $d$. This has so far only been measured
for five nearby millisecond pulsars~\cite{sbm97, tsb99, lkd04, sns05,
lkn06}. The term $\Delta t_\mathrm{rel}$ represents the Einstein and
Shapiro corrections due to general
relativistic time delays in the solar system~\cite{bh86}. Since
measurements can be carried out at different observing frequencies
with different dispersive delays, TOAs are generally referred to
the equivalent time that would be observed at infinite frequency.
This transformation is the term $\Delta t_\mathrm{DM}$ (see also
Equation~(\ref{equ:defdt})).

Following the accumulation of a number of TOAs, a surprisingly 
simple model is usually sufficient to account for the TOAs during
the time span of the observations and to predict the arrival times of
subsequent pulses. The model is a Taylor expansion of the
rotational frequency $\Omega = 2 \pi/P$ about a model value
$\Omega_0$ at some reference epoch ${\cal T}_0$. 
The model pulse phase
\begin{equation}
  \phi({\cal T}) =
  \phi_0 + ({\cal T} - {\cal T}_0) \Omega_0 +
  \frac{1}{2} ({\cal T} - {\cal T}_0)^2 \dot{\Omega}_0 +
  \dots,
  \label{equ:phi}
\end{equation}
where ${\cal T}$ is the barycentric time and $\phi_0$ is the
pulse phase at ${\cal T}_0$. Based on this simple model, and
using initial estimates of the position, dispersion measure and pulse
period, a ``timing residual'' is calculated for each TOA as the
difference between the observed and predicted pulse phases.

A set of timing residuals for the nearby pulsar B1133+16 spanning
almost 10 years is shown in Figure~\ref{fig:1133}. Ideally, the
residuals should have a zero mean and be free from any systematic
trends (see Panel~a of Figure~\ref{fig:1133}). To reach this point, however, the model
needs to be refined in a bootstrap fashion. Early sets of residuals
will exhibit a number of trends indicating a systematic error in one
or more of the model parameters, or a parameter not incorporated into
the model.

\epubtkImage{1133.png}{
  \begin{figure}[htbp]
    \def\epsfsize#1#2{0.65#1}
    \centerline{\epsfbox{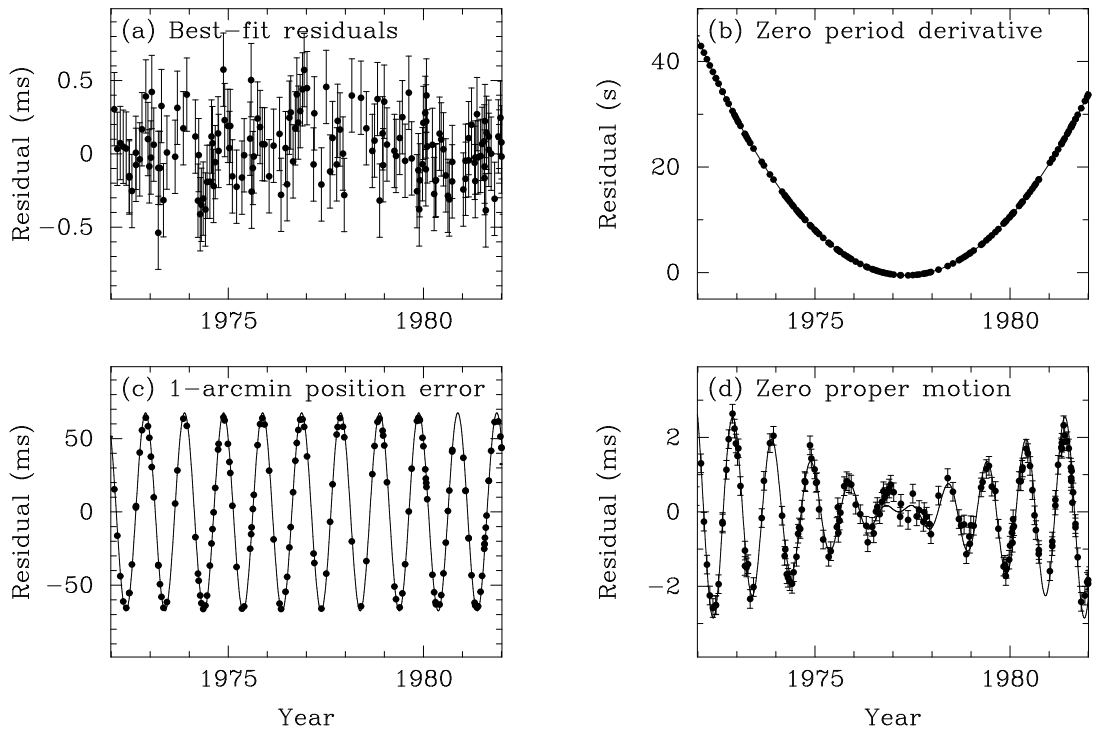}}
    \caption{Timing model residuals for PSR~B1133+16. Panel~a:
      Residuals obtained from the best-fitting model which includes
      period, period derivative, position and proper motion. Panel~b:
      Residuals obtained when the period derivative term is set to
      zero. Panel~c: Residuals showing the effect of a 1-arcmin
      position error. Panel~d: Residuals obtained neglecting the
      proper motion. The lines in Panels~b--d show the expected
      behaviour in the timing residuals for each effect. Data provided
      by Andrew Lyne.}
    \label{fig:1133}
  \end{figure}}

From Equation~(\ref{equ:phi}), an error in the assumed
$\Omega_0$ results in a linear slope with time. A parabolic
trend results from an error in $\dot{\Omega}_0$
(see Panel~b of Figure~\ref{fig:1133}). Additional effects will arise if the assumed
position of the pulsar (the unit vector $\hat{\underline{s}}$ in
Equation~(\ref{equ:bary})) used in the barycentric time calculation is
incorrect. A position error results in an annual sinusoid
(see Panel~c of Figure~\ref{fig:1133}). A proper motion produces an annual sinusoid
of linearly increasing magnitude (see Panel~d of Figure~\ref{fig:1133}).

After a number of iterations, and with the benefit of a modicum of
experience, it is possible to identify and account for each of these
various effects to produce a ``timing solution'' which is phase
coherent over the whole data span. The resulting model parameters
provide spin and astrometric information with a
precision which improves as the length of the data span
increases. For example, timing observations of the original
millisecond pulsar B1937+21, spanning almost 9~years
(exactly 165,711,423,279 rotations!), measure a period of
1.5578064688197945$\pm$0.0000000000000004~ms~\cite{ktr94, kas94}
defined at midnight UT on
December~5, 1988! Measurements of other parameters are no
less impressive, with astrometric errors of $\sim 3 \mathrm{\ \mu arcsec}$ being
presently possible for the bright millisecond pulsar J0437$-$4715~\cite{vbv08}.


\subsection{Timing stability}
\label{sec:tstab}

Ideally, after correctly applying a timing model,
we would expect a set of uncorrelated timing residuals
with a zero mean and a Gaussian scatter with a standard deviation
consistent with the measurement uncertainties. As can be
seen in Figure~\ref{fig:tnoise}, this is not always the
case; the residuals of many pulsars exhibit a quasi-periodic
wandering with time.

\epubtkImage{tnoise.png}{
  \begin{figure}[htbp]
    \def\epsfsize#1#2{0.5#1}
    \centerline{\epsfbox{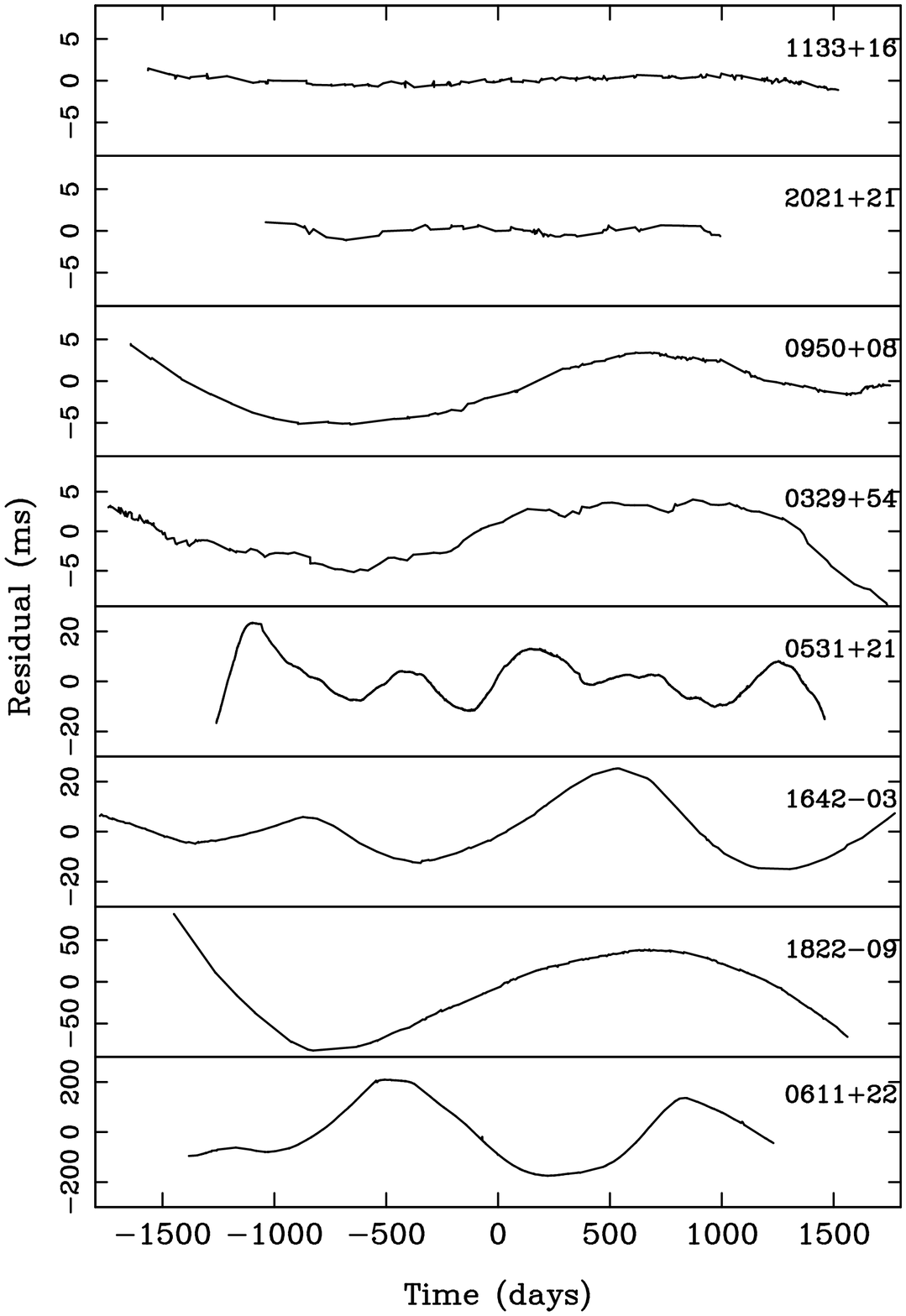}}
    \caption{Examples of timing residuals for a number of normal
      pulsars. Note the varying scale on the ordinate axis, the
      pulsars being ranked in increasing order of timing
      ``activity''. Data taken from the Jodrell Bank timing
      program~\cite{sl96, hlk04}. Figure provided by Andrew Lyne.}
    \label{fig:tnoise}
  \end{figure}}

Such ``timing noise'' is most prominent in the youngest of the normal
pulsars~\cite{mt74, ch80} and present at a lower level in the much older
millisecond pulsars~\cite{ktr94, antt94}. While the physical processes of
this phenomenon are not well understood, it seems likely that they
may be connected to superfluid processes and temperature changes
in the interior of the neutron star~\cite{anp86}, or to processes in the
magnetosphere~\cite{che87a, che87b}.

The relative dearth of timing noise for the older pulsars is a very
important finding. It implies that the measurement
precision presently depends primarily on the particular hardware constraints of
the observing system. Consequently, a large effort in hardware
development is now being made to improve the precision of these
observations using, in particular, coherent dedispersion outlined in
Section~\ref{sec:timobs}. Much progress in this area has been made by
groups at Princeton~\cite{pripsr}, Berkeley~\cite{bkypsr},
Jodrell Bank~\cite{jodpsr},
UBC~\cite{ubcpsr}, Swinburne~\cite{swinpsr} and ATNF~\cite{atnfpsr}.
From high quality observations spanning over a
decade~\cite{rt91a, rt91b, ktr94}, these groups have demonstrated that
the timing stability of millisecond pulsars over such timescales is
comparable to terrestrial atomic clocks.

\epubtkImage{sigmaz.png}{
  \begin{figure}[htbp]
    \def\epsfsize#1#2{0.5#1}
    \centerline{\epsfbox{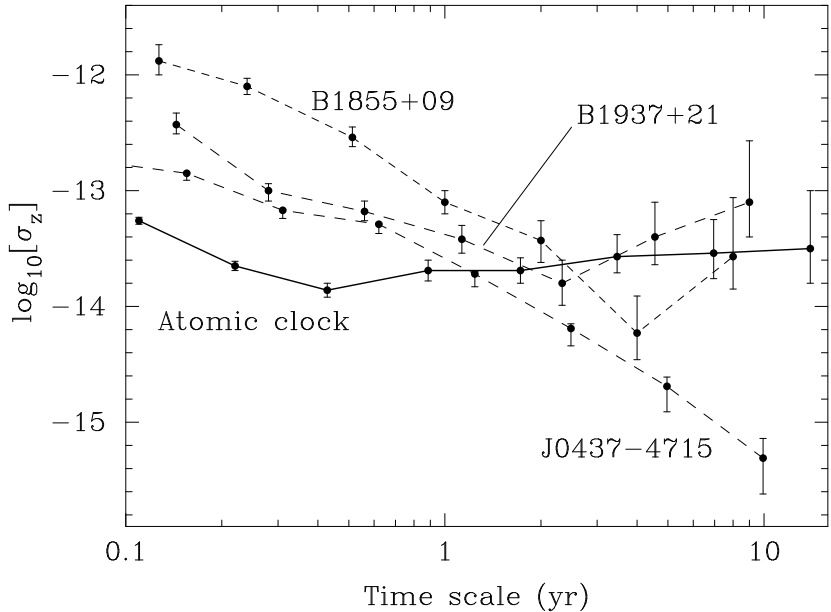}}
    \caption{The fractional stability of three millisecond pulsars
      compared to an atomic clock. Both PSRs B1855+09 and B1937+21 are
      comparable, or just slightly worse than, the atomic clock
      behaviour over timescales of a few years~\cite{mtem97}. More
      recent timing of the millisecond pulsar J0437$-$4715~\cite{vbv08}
      indicates that it is inherently a very stable clock. Data for
      the latter pulsar provided by Joris Verbiest.}
    \label{fig:sigmaz}
  \end{figure}}

This phenomenal stability is demonstrated in Figure~\ref{fig:sigmaz}
which shows $\sigma_z$~\cite{mtem97,sigmaz}, a parameter closely
resembling the Allan variance used by the clock community to estimate
the stability of atomic clocks~\cite{tay91, allan}. Both PSRs B1937+21
and B1855+09 seem to be limited by a power law component which
produces a minimum in $\sigma_z$ after 2~yr and 5~yr
respectively. This is most likely a result of a small amount of
intrinsic timing noise~\cite{ktr94}. The $\sigma_z$ based on timing
observations~\cite{vbv08} of the bright millisecond pulsar
J0437$-$4715 is now 1\,--\,2 orders of magnitude smaller than the other
two pulsars or the atomic clock!


\subsection{Timing binary pulsars}
\label{sec:tbin}

For binary pulsars, the timing model introduced in
Section~\ref{sec:tmodel} needs to be extended to incorporate the additional
motion of the pulsar as it orbits the common centre-of-mass of the
binary system. Describing the binary orbit using Kepler's laws to refer
the TOAs to the binary barycentre requires five additional model
parameters: the orbital period $P_\mathrm{b}$, projected semi-major
orbital axis $x$, orbital eccentricity
$e$, longitude of periastron $\omega$ and the epoch of periastron
passage $T_0$. This description, using five ``Keplerian
parameters'', is identical to that used for spectroscopic binary
stars. Analogous to the radial velocity curve in a spectroscopic
binary, for binary pulsars the orbit is described by the apparent
pulse period against time. An example of this is shown in Panel~a of
Figure~\ref{fig:orbits}. Alternatively, when radial accelerations can be
measured, the orbit can also be visualised in a plot of acceleration
versus period as shown in Panel~b of Figure~\ref{fig:orbits}. This method is particularly
useful for determining binary pulsar
orbits from sparsely sampled data~\cite{fkl01}.

\epubtkImage{orbitcurves.png}{
  \begin{figure}[htbp]
    \def\epsfsize#1#2{0.65#1}
    \centerline{\epsfbox{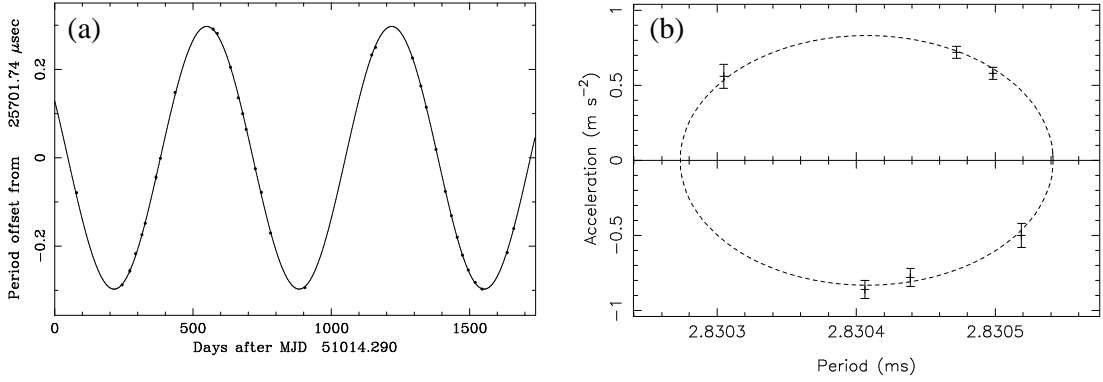}}
    \caption{Panel~a: Keplerian orbital fit to the 669-day binary
      pulsar J0407+1607~\cite{lxf05}. Panel~b: Orbital fit in the
      period-acceleration plane for the globular cluster pulsar
      47~Tuc~S~\cite{fkl01}.}
    \label{fig:orbits}
  \end{figure}}

Constraints on the masses of the pulsar $m_\mathrm{p}$ and the
orbiting companion $m_\mathrm{c}$ can be placed by combining $x$ and
$P_\mathrm{b}$ to obtain the mass function
\begin{equation}
  f_\mathrm{mass} = \frac{4\pi^2}{G} \frac{x^3}{P_\mathrm{b}^2} = 
  \frac{(m_\mathrm{c} \sin i)^3}{(m_\mathrm{p} + m_\mathrm{c})^2},
  \label{equ:massfn}
\end{equation}
where $G$ is Newton's gravitational constant and $i$ is the (initially
unknown) angle between the orbital plane and the plane of the sky
(i.e.\ an orbit viewed edge-on corresponds to $i=90^{\circ}$). In the
absence of further information, the standard practice is to consider a
random distribution of inclination angles. Since the probability that
$i$ is \emph{less} than some value $i_0$ is $p(<i_0) = 1 - \cos(i_0)$,
the 90\% confidence interval for $i$ is $26^{\circ} < i <
90^{\circ}$. For an assumed pulsar mass, the 90\% confidence interval
for $m_\mathrm{c}$ can be obtained by solving Equation~(\ref{equ:massfn})
for $i = 26^{\circ}$ and $90^{\circ}$.

Although most of the presently known binary pulsar systems can be
adequately timed using Kepler's laws, there are a number which require
an additional set of ``post-Keplerian'' (PK) parameters which have
a distinct functional form for a given relativistic theory of
gravity~\cite{dt92}. In general relativity (GR) the PK formalism gives the relativistic
advance of periastron
\begin{equation}
  \dot\omega = 3 \left( \frac{P_\mathrm{b}}{2\pi} \right)^{-5/3}
  (T_\odot M)^{2/3} (1 - e^2)^{-1},
  \label{equ:omdot}
\end{equation}
the time dilation and gravitational redshift parameter
\begin{equation}
  \gamma = e \left(\frac{P_\mathrm{b}}{2\pi}\right)^{1/3}
  T_\odot^{2/3}M^{-4/3} m_\mathrm{c} (m_\mathrm{p} + 2m_\mathrm{c}),
  \label{equ:gamma}
\end{equation}
the rate of orbital decay due to gravitational radiation
\begin{equation}
  \dot P_\mathrm{b} = -\frac{192\pi}{5}
  \left( \frac{P_\mathrm{b}}{2\pi} \right)^{-5/3}
  \left( 1 + \frac{73}{24} e^2 + \frac{37}{96} e^4 \right)
  \left( 1 - e^2 \right)^{-7/2}
  T_\odot^{5/3} m_\mathrm{p} m_\mathrm{c} M^{-1/3}
  \label{equ:pbdot}
\end{equation}
and the two Shapiro delay parameters
\begin{equation}
  r = T_\odot m_\mathrm{c}
  \label{equ:r}
\end{equation}
and
\begin{equation}
  s = x \left( \frac{P_\mathrm{b}}{2\pi} \right)^{-2/3}
  T_\odot^{-1/3} M^{2/3} m_\mathrm{c}^{-1}
  \label{equ:s}
\end{equation}
which describe the delay in the pulses around superior conjunction
where the pulsar radiation traverses the gravitational well of
its companion. In the above expressions, all masses are in solar units,
$M\equiv m_\mathrm{p}+m_\mathrm{c}$, $x\equiv
a_\mathrm{p} \sin i/c$,
$s\equiv\sin i$ and $T_\odot\equiv G M_\odot/c^3 = 4.925490947
\mathrm{\ \mu s}$.
Some combinations, or all, of the PK parameters have now been 
measured for a number of binary pulsar systems.
Further PK parameters due to aberration and relativistic
deformation~\cite{dd86} are not listed here but may soon
be important for the double pulsar~\cite{ksm06}.


\subsection{Testing general relativity}
\label{sec:gr}

The key point in the PK definitions introduced in the previous section is that,
given the precisely measured Keplerian parameters, the only two unknowns
are the masses of the pulsar and its companion,
$m_\mathrm{p}$ and $m_\mathrm{c}$. Hence, from a measurement of just
two PK parameters (e.g., $\dot{\omega}$ and $\gamma$) one can
solve for the two masses and, using 
Equation~(\ref{equ:massfn}), find the orbital inclination angle $i$.
If three (or more) PK parameters are measured, the system is
``overdetermined'' and can be used to test GR (or, more
generally, any other
theory of gravity) by comparing
the third PK parameter with the predicted value based on
the masses determined from the other two.

The first binary pulsar used to test GR in this way was PSR~B1913+16
discovered by Hulse \& Taylor in 1974~\cite{ht75a}. Measurements of
three PK parameters ($\dot{\omega}$, $\gamma$ and
$\dot{P_\mathrm{b}}$) were obtained from long-term timing observations
at Arecibo~\cite{tw82, tw89, wt05}. The measurement of orbital decay,
which corresponds to a shrinkage of about $3.2 \mathrm{\ mm}$ per
orbit, is seen most dramatically as the gradually increasing shift in
orbital phase for periastron passages with respect to a non-decaying
orbit shown in Figure~\ref{fig:1913}. This figure includes recent
Arecibo data taken since the upgrade of the telescope in the mid
1990s. 

\epubtkImage{1913.png}{
  \begin{figure}[htbp]
    \def\epsfsize#1#2{0.5#1}
    \centerline{\epsfbox{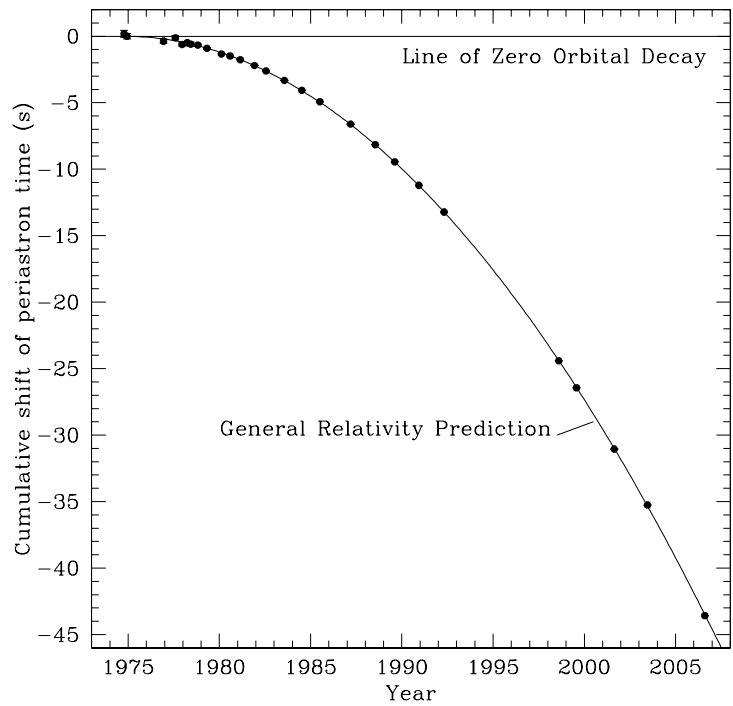}}
    \caption{Orbital decay in the binary pulsar B1913+16 system
      demonstrated as an increasing orbital phase shift for periastron
      passages with time. The GR prediction due entirely to the
      emission of gravitational radiation is shown by the
      parabola. Figure provided by Joel Weisberg.}
    \label{fig:1913}
  \end{figure}}

The measurement of orbital decay in the B1913+16 system, obtained from
observations spanning a 30-yr baseline~\cite{wt05},
is within 0.2\% of the GR prediction and provided the first indirect
evidence for the existence of gravitational waves. Hulse and Taylor
were awarded the 1993 Nobel Physics prize~\cite{nobpr1993, hul94,
tay94} in recognition of their discovery of this remarkable laboratory
for testing GR.  A similar, though less precise, test of GR from this
combination of PK parameters has recently been performed in the double
neutron star binary PSR~B2127+11C in the globular cluster
M15~\cite{jcj06}.  For this system, the measurement uncertainties
permit a test of GR to 3\% precision which is unlikely to improve
due to various kinematic contaminations including the acceleration
of the binary system in the cluster's gravitational potential.

Five PK parameters have been
measured for the double pulsar discussed in detail below and
for the double neutron star system PSR
B1534+12~\cite{sttw02} where the test of GR comes from measurements of
$\dot{\omega}$, $\gamma$ and $s$. In this system, the agreement
between theory and observation is within 0.7\%~\cite{sttw02}. This
test will improve in the future as the timing baseline extends and a
more significant measurement of $r$ can be made. 

Currently the best binary pulsar system for testing GR in the
strong-field regime is the double pulsar J0737$-$3039. In this system,
where two independent pulsar clocks can be timed, five PK parameters
of the 22.7-ms pulsar ``A'' have been measured as well as two
additional constraints from the measured mass function and projected
semi-major axis of the 2.7-s pulsar ``B''. A useful means of
summarising the limits so far is Figure~\ref{fig:0737m1m2} which shows
the allowed regions of parameter space in terms of the masses of the
two pulsars. The shaded regions are excluded by the requirement that
$\sin i < 1$. Further constraints are shown as pairs of lines
enclosing permitted regions as predicted by GR. The
measurement~\cite{ksm06} of $\dot{\omega}=16.899 \pm 0.001 \mathrm{\
  deg\ yr}^{-1}$ gives the total system mass $M=2.5871 \pm
0.0002\,M_\odot$.

\epubtkImage{0737m1m2.png}{
  \begin{figure}[htbp]
    \def\epsfsize#1#2{0.6#1}
    \centerline{\epsfbox{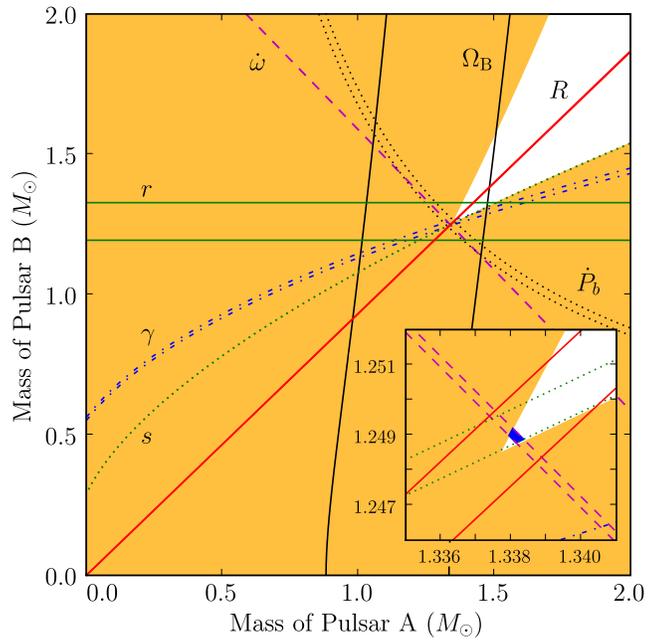}}
    \caption{`Mass--mass' diagram showing the observational
      constraints on the masses of the neutron stars in the double
      pulsar system J0737$-$3039. Inset is an enlarged view of the
      small square encompassing the intersection of the tightest
      constraints. Figure provided by Ren{\'e} Breton~\cite{bkk08}.}
    \label{fig:0737m1m2}
  \end{figure}}

The measurement of the projected semi-major axes of both
orbits gives the mass ratio $R=1.071 \pm 0.001$. The mass ratio
measurement is unique to the double pulsar system and rests on
the basic assumption that momentum is conserved. This constraint
should apply to any reasonable theory of gravity.
The intersection between the lines for $\dot{\omega}$ and $R$
yield the masses of A and B as $m_\mathrm{A}=1.3381\pm0.007\,M_\odot$
and $m_\mathrm{B}=1.2489\pm0.0007\,M_\odot$. From these values,
using Equations~(\ref{equ:gamma}--\ref{equ:s}) the expected values of
$\gamma$, $\dot{P}_\mathrm{b}$, $r$ and $s$ may be calculated and
compared with the observed values. These four tests of GR all agree
with the theory to within the uncertainties. Currently the tightest
constraint is the Shapiro delay parameter $s$ where the observed value
is in agreement with GR at the 0.05\% level~\cite{ksm06}.

Another unique feature of the double pulsar system is the interactions
between the two pulsars' radiation beams. Specifically, the signal
from A is eclipsed for 30~s each orbit by the magnetosphere of
B~\cite{lbk04,krb04,mll04} and the radio pulses from B are modulated
by the relativistic wind from A during two phases of the
orbit~\cite{mkl04}. These provide unique insights into plasma
physics~\cite{sm05} and, as shown shown in Figure~\ref{fig:0737m1m2},
a measurement of relativistic spin--orbit coupling in the binary
system and, hence, another constraint on GR.  By careful modeling of
the change in eclipse profiles of A over a four-year baseline, Breton
et al.~\cite{bkk08} have been able to fit a remarkably simple
model~\epubtkFootnote{The reader is encouraged to view the beautiful movies
explaining the eclipse model~\cite{breton}} and determine the
precession of B's spin axis about the orbital angular momentum vector.
This remarkable measurement agrees, within the 13\% measurement
uncertainty, to the GR prediction.

It took only two years for the double pulsar system to surpass the
tests of GR possible from three decades of monitoring PSR~B1913+16 and
over a decade of timing PSR~B1534+12.  On-going precision timing
measurements of the double pulsar system should soon provide even more
stringent and new tests of GR. Crucial to these measurements will be
the timing of the 2.7-s pulsar B, where the
observed profile is significantly affected by A's relativistic
wind~\cite{lbk04, mkl04}. A careful decoupling of these profile
variations is required to accurately measure TOAs for this pulsar and
determine the extent to which the theory-independent mass ratio $R$
can be measured. This task is compounded by the fact that pulsar
B's signal is now~\cite{bkk08} significantly weaker compared to 
the discovery observations five years ago~\cite{lbk04}. This appears
to be a direct result of its beam precessing out of our line of sight. 

The relativistic effects observed in the double pulsar system are so
large that corrections to higher post-Newtonian order may soon need to
be considered. For example, $\dot{\omega}$ may be measured precisely
enough to require terms of second post-Newtonian order to be included
in the computations~\cite{ds88}. In addition, in contrast to
Newtonian physics, GR predicts that the spins of the neutron stars
affect their orbital motion via spin-orbit coupling. This effect
would most clearly be visible as a contribution to the observed
$\dot{\omega}$ in a secular~\cite{bo75b} and periodic
fashion~\cite{wex95}. For the J0737$-$3039 system, the expected
contribution is about an order of magnitude larger than for PSR
B1913+16~\cite{lbk04}. As the exact value depends on the pulsars'
moment of inertia, a potential measurement of this effect would allow the
moment of inertia of a neutron star to be determined for the first
time~\cite{ds88}. Such a measurement would be invaluable for studies
of the neutron star equation of state and our understanding of
matter at extreme pressure and densities~\cite{lp07}.

Finally, in the neutron star--white dwarf binary J1141$-$6545, the
measurement of $\dot{P_b}$, $\dot{\omega}$, $\gamma$ and $s$ permit a
6\% test of GR~\cite{bbv08}. Despite this being nominally only a relatively
weak constraint for GR, the significantly different self gravities of
the neutron star and white dwarf permits constraints on two
coupling parameters between matter and the scalar gravitational
field. These parameters are non-existent in GR, which describes
gravity in terms of a tensor field, but are predicted by some
alternative theories of gravity which consider tensor and scalar field
components.  The constraints provided by PSR~J1141$-$6545 for the
strongly non-linear coupling parameter are currently the most
stringent to date and are set to improve dramatically over the next
five years of timing observations.


\subsection{Pulsar timing and neutron star masses}
\label{sec:masses}

Multiple PK parameters have now been measured for a number of binary
pulsars which provide very precise measurements of the neutron star
masses~\cite{tc99, sta04}. Figure~\ref{fig:masses} shows the distribution
of masses following updates to a recent compilation~\cite{sta04}. While the young
pulsars and the double neutron star binaries are consistent
with, or just below, the canonical $1.4\,M_\odot$, we note
that the millisecond pulsars in binary systems have, on average,
significantly larger masses. This provides strong support for their
formation through an extended period of accretion in the past,
as discussed in Section~\ref{sec:evolution}.

\epubtkImage{masses.png}{
  \begin{figure}[htbp]
    \def\epsfsize#1#2{0.6#1}
    \centerline{\epsfbox{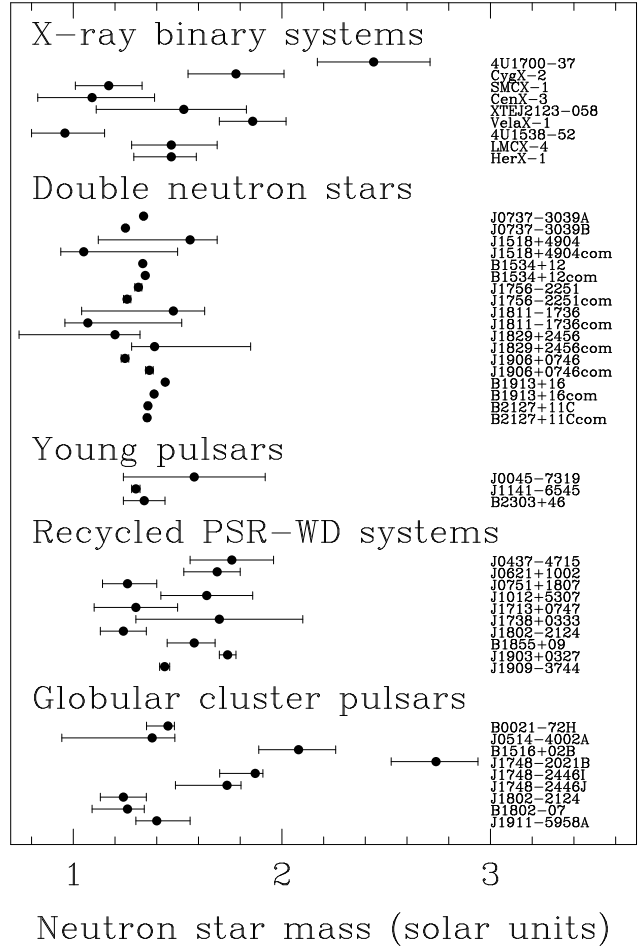}}
    \caption{Distribution of neutron star masses as inferred from
      timing observations of binary pulsars and X-ray binary systems.
      Figure adapted from an original version provided
      by Ingrid stairs. Additional information on globular cluster
      pulsar mass constraints provided by Paulo Freire. Error
      bars show one-sigma uncertainties on each mass determination.
    \label{fig:masses}}
  \end{figure}}

Also shown on this diagram are several eccentric binary systems
in globular clusters which have their masses constrained from
measurements of the relativistic advance of periastron and
the Keplerian mass function. In these cases, the condition $\sin i<1$
sets a lower limit on the companion mass
$m_\mathrm{c}>(f_\mathrm{mass} M^2)^{1/3}$ and a corresponding upper limit
on the pulsar mass. Probability density functions for both 
$m_\mathrm{p}$ and $m_\mathrm{c}$ can also be estimated in a statistical sense
by {\it assuming} a random distribution of orbital inclination angles.
An example of this is shown in Figure~\ref{fig:mass-massM5} for
the eccentric binary millisecond pulsar in M5~\cite{fwvh08}
where the nominal pulsar mass is $2.08 \pm 0.19\,M_{\odot}$!
This is a potentially outstanding result.
If confirmed by the measurement of
other relativistic parameters, these supermassive neutron stars
will have important constraints on the equation of state of superdense
matter.

\epubtkImage{mass-massM5.png}{
  \begin{figure}[htbp]
    \def\epsfsize#1#2{0.6#1}
    \centerline{\epsfbox{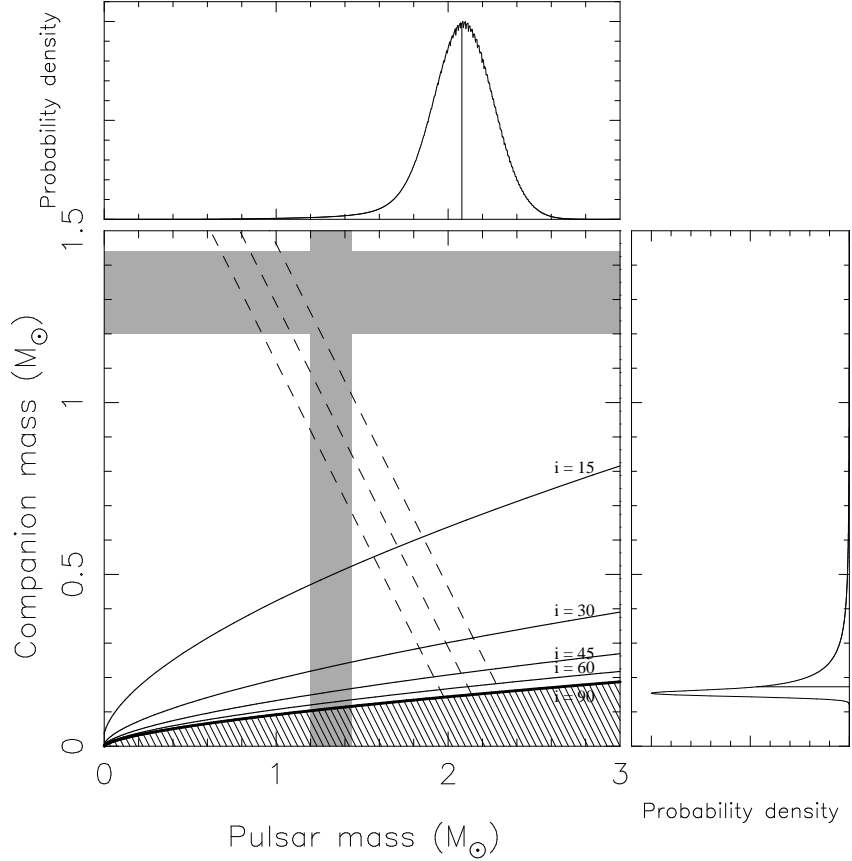}}
    \caption{Pulsar mass versus companion mass diagram showing
      mass constraints for the eccentric binary millisecond pulsar
      B1516+02B~\cite{fwvh08}. The allowed parameter space is bounded
      by the dashed lines which show the uncertainty on the total mass
      from the measurement of the relativistic advance of
      periastron. Masses within the hatched region are disallowed by
      the Keplerian mass function and the constraint that $\sin i <
      1$. Assuming a random distribution of orbital inclination angles
      allows the probability density functions of the pulsar and
      companion mass to be derived, as shown by the top and right hand
      panels. Figure provided by Paulo Freire.
    \label{fig:mass-massM5}}
  \end{figure}}

Currently the largest measurement of a radio pulsar mass through
multiple PK parameters is the eccentric millisecond pulsar binary
system J1903+0327~\cite{crl08} (see also Section~\ref{sec:1903}).
Thus, in addition to challenging models of millisecond pulsar formation,
this new discovery has important implications for fundamental physics.
When placed on the mass--radius diagram for neutron stars~\cite{lp07},
this $1.74\,M_{\odot}$ pulsar appears to be incompatible with at least
four equations of state for superdense matter.  Future timing
measurements are required to consolidate and verify this potentially
very exciting result.


\subsection{Pulsar timing and gravitational wave detection}
\label{sec:gwdet}

We have seen in the Section~\ref{sec:tbin} how pulsar timing can be used to
provide indirect evidence for the existence of gravitational waves
from coalescing stellar-mass binaries. In this final section, we look
at how pulsar timing might soon be used to detect gravitational
radiation directly. The idea to use pulsars as natural gravitational
wave detectors was first explored in the late
1970s~\cite{saz78, det79}. The basic concept is to treat the solar
system barycentre and a distant pulsar as opposite ends of an
imaginary arm in space. The pulsar acts as the reference clock at one
end of the arm sending out regular signals which are monitored by an
observer on the Earth over some timescale $T$. The effect of a
passing gravitational wave would be to perturb the local spacetime
metric and cause a change in the observed rotational frequency of the
pulsar. For regular pulsar timing observations with typical TOA
uncertainties of $\epsilon_\mathrm{TOA}$, this ``detector'' would be
sensitive to waves with dimensionless amplitudes
$h \gtrsim \epsilon_\mathrm{TOA}/T$ and frequencies
$f \sim 1/T$~\cite{bcr83, bnr84}.


\subsubsection{Probing the gravitational wave background}
\label{sec:uplim}

Many cosmological models predict that the Universe is presently filled
with a low-frequency
stochastic gravitational wave background (GWB) produced during
the big bang era~\cite{pee93}. A significant 
component~\cite{rr95, jb03} is the gravitational
radiation from the inspiral prior to supermassive black hole mergers.
In the ideal case, the change in the observed frequency
caused by the GWB should be detectable in the set of timing residuals
after the application of an appropriate model for the rotational,
astrometric and, where necessary, binary parameters of the pulsar. As
discussed in Section~\ref{sec:tmodel}, all other effects being negligible,
the rms scatter of these residuals $\sigma$ would be due to the
measurement uncertainties and intrinsic timing noise from the neutron star.

For a GWB with a flat energy spectrum in the frequency band $f \pm
f/2$ there is an additional contribution to the timing residuals
$\sigma_\mathrm{g}$~\cite{det79}. When $fT \gg 1$, the corresponding
wave energy density is
\begin{equation}
  \rho_\mathrm{g} = \frac{243 \pi^3 f^4 \sigma_\mathrm{g}^2}{208 G}.
\end{equation}
An upper limit to
$\rho_\mathrm{g}$ can be obtained from a set of timing residuals by assuming
the rms scatter is entirely due to this effect ($\sigma=\sigma_\mathrm{g}$).
These limits are commonly expressed as a fraction of the
energy density required to close the Universe
\begin{equation}
  \rho_\mathrm{c} = \frac{3 H_0^2}{8 \pi G} \simeq
  2 \times 10^{-29} h^2 \mathrm{\ g\ cm}^{-3},
\end{equation}
where the Hubble constant $H_0 = 100 \, h \mathrm{\ km\ s}^{-1} \mathrm{\ Mpc}$.

This technique was first applied~\cite{rt83} to a set of TOAs for PSR
B1237+25 obtained from regular observations over a period of 11 years
as part of the JPL pulsar timing programme~\cite{dr83}. This pulsar
was chosen on the basis of its relatively low level of timing activity
by comparison with the youngest pulsars, whose residuals are
ultimately plagued by timing noise (see Section~\ref{sec:tstab}). By ascribing
the rms scatter in the residuals ($\sigma = 240 \mathrm{\ ms}$) to the GWB, the
limit is $\rho_\mathrm{g}/\rho_\mathrm{c} \lesssim 4 \times 10^{-3} h^{-2}$ for a
centre frequency $f = 7 \mathrm{\ nHz}$.

This limit, already well below the energy density required to close
the Universe, was further reduced following the long-term timing
measurements of millisecond pulsars at Arecibo
(see Section~\ref{sec:tstab}). In the intervening period, more
elaborate techniques had been devised~\cite{bcr83, bnr84, srtr90} to
look for the likely signature of a GWB in the frequency spectrum of
the timing residuals and to address the possibility of ``fitting
out'' the signal in the TOAs. Following~\cite{bcr83} it is convenient
to define
\begin{equation}
  \Omega_\mathrm{g} = \frac{1}{\rho_\mathrm{c}}
  \frac{\mathrm{d}\log \rho_\mathrm{g}}{\mathrm{d}\log f},
\end{equation}
i.e.\ the energy density of the GWB per logarithmic
frequency interval relative to $\rho_\mathrm{c}$. With this definition, the
power spectrum of the GWB~\cite{hr84, bnr84} is
\begin{equation}
  {\cal P}(f) = \frac{G \rho_\mathrm{g}}{3\pi^3 f^4} =
  \frac{H_0^2 \Omega_\mathrm{g}}{8\pi^4 f^5} =
  1.34 \times 10^4 \Omega_\mathrm{g} h^2
  f^{-5}_{\mathrm{yr}^{-1}} \mathrm{\ \mu s}^2 \mathrm{\ yr},
\end{equation}
where $f_\mathrm{yr^{-1}}$ is frequency in cycles per year. The long-term
timing stability of B1937+21, discussed in Section~\ref{sec:tstab}, limits its use
for periods $\gtrsim$~2~yr. Using the more stable residuals for PSR B1855+09,
Kaspi et al.~\cite{ktr94} placed an upper limit of $\Omega_\mathrm{g}
h^2 < 1.1 \times 10^{-7}$.
This limit is difficult to reconcile with most
cosmic string models for galaxy formation~\cite{ca92, td96}.

For binary pulsars, the orbital period provides an additional clock
for measuring the effects of gravitational waves. In this case, the
range of frequencies is not limited by the time span of the
observations, allowing the detection of waves with periods as large as
the light travel time to the binary system~\cite{bcr83}. The most
stringent results presently available are based on the B1855+09 limit
$\Omega_\mathrm{g} h^2 < 2.7 \times 10^{-4}$ in the frequency range
$10^{-11} < f < 4.4 \times 10^{-9} \mathrm{\ Hz}$~\cite{kop97}.


\subsubsection{Constraints on massive black hole binaries}

In addition to probing the GWB, pulsar timing is beginning to place
interesting constraints on the existence of massive black hole
binaries. Arecibo data for PSRs~B1937+21 and J1713+0747 already make
the existence of an equal-mass black hole binary in Sagittarius~A*
unlikely~\cite{lb01}. More recently, timing data from B1855+09
have been used to virtually rule out the existence of a proposed
supermassive black hole as the explanation for the periodic motion
seen at the centre of the radio galaxy 3C66B~\cite{simy03}.

A simulation of the expected modulations of the timing
residuals for the putative binary system, with a total mass of 
$5.4\times 10^{10}\,M_\odot$, is shown along with the
observed timing residuals in Figure~\ref{fig:3c66b}. Although
the exact signature depends on the orientation and eccentricity
of the binary system, Monte Carlo simulations show that the
existence of such a massive black hole binary is ruled out
with at least 95\% confidence~\cite{jllw04}.

\epubtkImage{3c66b.png}{
  \begin{figure}[htbp]
    \def\epsfsize#1#2{0.65#1}
    \centerline{\epsfbox{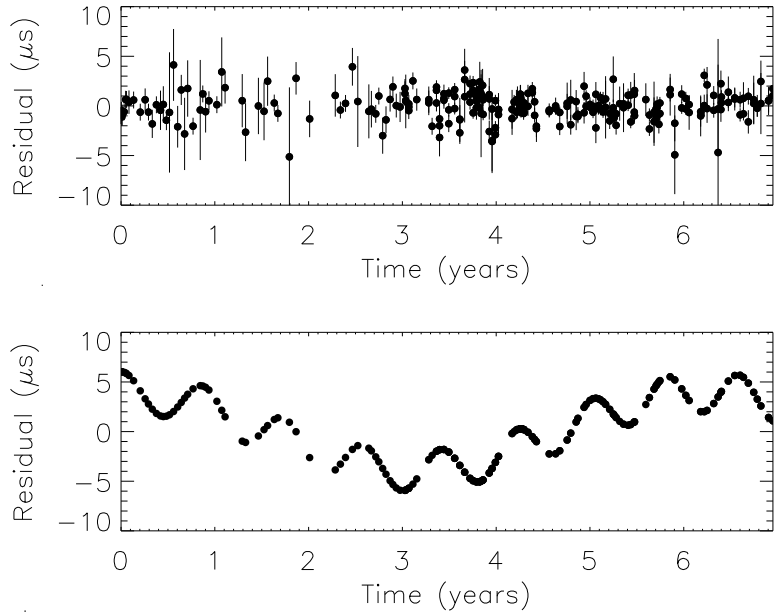}}
    \caption{Top panel: Observed timing residuals for
      PSR~B1855+09. Bottom panel: Simulated timing residuals induced
      from a putative black hole binary in 3C66B. Figure provided by
      Rick Jenet~\cite{jllw04}.}
    \label{fig:3c66b}
  \end{figure}}


\subsubsection{A millisecond pulsar timing array}
\label{sec:array}

A natural extension of the single-arm detector concept discussed above
is the idea of using timing data for a number of pulsars distributed
over the whole sky to detect gravitational waves~\cite{hd83}. Such a
``timing array'' would have the advantage over a single arm
in that, through a cross-correlation analysis of the residuals for
pairs of pulsars distributed over the sky, it should be possible to
separate the timing noise of each pulsar from the signature of the
GWB common to all pulsars in the array. To illustrate this, consider
the fractional frequency shift of the $i$th pulsar in an array
\begin{equation}
  \frac{\delta \nu_i}{\nu_i} = \alpha_i {\cal A}(t) + {\cal N}_i(t),
\end{equation}
where $\alpha_i$ is a geometric factor dependent on
the line-of-sight direction to the pulsar and the propagation
and polarisation vectors of the gravitational wave of dimensionless
amplitude ${\cal A}$. The timing noise intrinsic to the pulsar
is characterised by the function ${\cal N}_i$. The result of a
cross-correlation between pulsars $i$ and $j$ is then
\begin{equation}
  \alpha_i \alpha_j \langle {\cal A}^2 \rangle +
  \alpha_i \langle {\cal A}{\cal N}_j \rangle +
  \alpha_j \langle {\cal A}{\cal N}_i \rangle +
  \langle {\cal N}_i{\cal N}_j \rangle,
\end{equation}
where the bracketed terms indicate cross-correlations. Since the wave
function and the noise contributions from the two pulsars are
independent, the cross correlation tends to $\alpha_i
\alpha_j \langle{\cal A}^2\rangle$ as the number of residuals becomes
large. Summing the cross-correlation functions over a large number of
pulsar pairs provides additional information on this term as a
function of the angle on the sky~\cite{hel90} and allows
the separation of the effects of clock and 
ephemeris errors from the GWB~\cite{fb90}.

A recent analysis~\cite{jhv06}
applying the timing array concept to data for seven
millisecond pulsars 
has reduced the energy density
limit to $\Omega_\mathrm{g} h^2 < 1.9 \times
10^{-8}$ for a background of supermassive black hole sources.
The corresponding limits on the background of
relic gravitational waves and cosmic strings are
$2.0 \times 10^{-8}$ and
$1.9 \times 10^{-8}$ respectively.
These limits can be used to constrain the merger rate of
supermassive black hole binaries at high redshift, investigate
inflationary parameters and place limits on the tension
of currently proposed cosmic string scenarios.

The region of the gravitational wave
energy density spectrum probed by the current pulsar timing
array is shown in Figure~\ref{fig:gwspec} where it can be
seen that the pulsar timing regime is complementary to the
higher frequency bands of LISA and LIGO.

\epubtkImage{gwspec.png}{
  \begin{figure}[htbp]
    \def\epsfsize#1#2{0.7#1}
    \centerline{\epsfbox{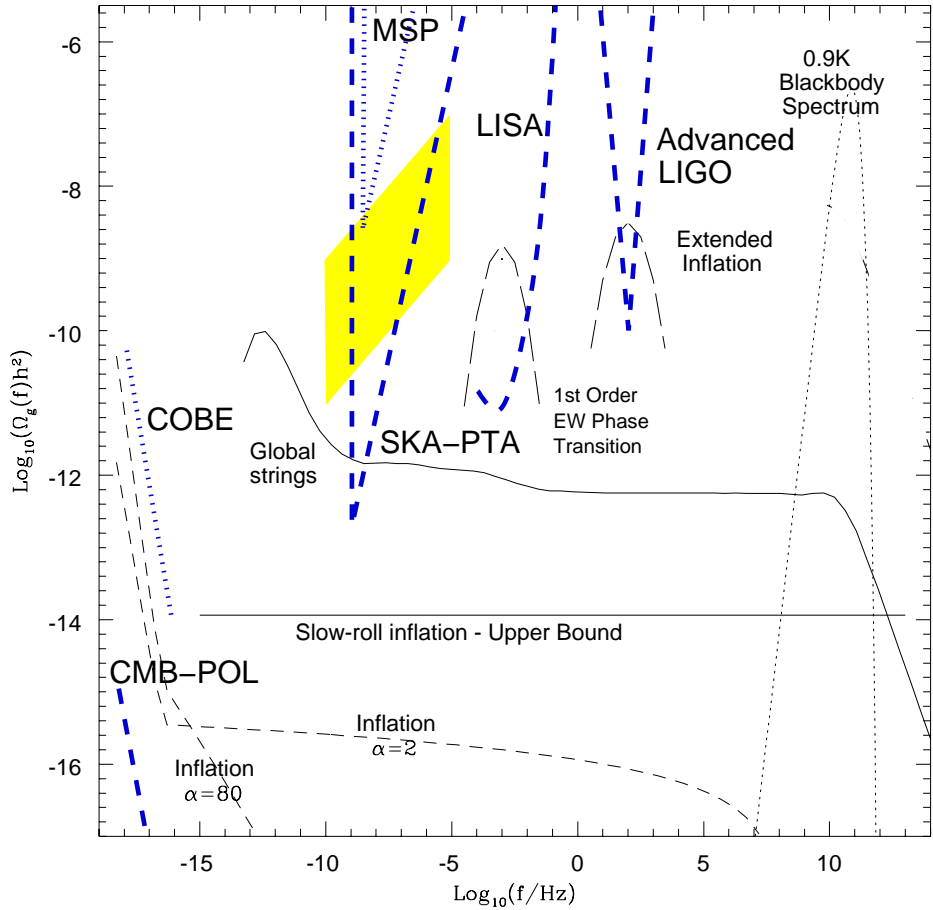}}
    \caption{Summary of the gravitational wave spectrum showing
      the location in phase space of the pulsar timing array (PTA) and
      its extension with the Square Kilometre Array (SKA). Figure
      updated by Michael Kramer~\cite{kbc04} from an original design
      by Richard Battye.}
    \label{fig:gwspec}
  \end{figure}}

A number of long-term timing projects are now underway to make a
large-scale pulsar timing array a reality. The Parkes pulsar timing
array \cite{ppta,hob05} observes twenty millisecond pulsars twice
a month. The European Pulsar Timing Array \cite{epta} uses the
Lovell, Westerbork, Effelsberg and Nancay radio telescopes to 
regularly observe a similar number. Finally, at Arecibo and Green Bank,
regular timing of a dozen or more millisecond pulsars is carried
out by a North American consortium~\cite{nanograv}. It is expected
that a combined analysis of all these efforts could
reach a limits of $\Omega_\mathrm{g} h^2 < 10^{-10}$ before 2010~\cite{hob05}.
Looking further ahead, the increase in sensitivity provided by
the Square Kilometre Array~\cite{ska, kbc04} should further
improve the limits of the spectrum probed by pulsar timing. 
As Figure~\ref{fig:gwspec} shows, the SKA could provide
up to two orders of magnitude improvement over current capabilities.


\subsection{Going further}
\label{sec:tfurther}

For further details on the techniques and prospects of pulsar timing,
a number of excellent reviews are
available~\cite{bh86, tay91, bel98, sta03, lk05}. Two audio files and
slides from lectures at the centennial meeting of the American
Physical Society are also available
on-line~\cite{backertalk, willtalk}. The tests of general relativity
discussed in Section~\ref{sec:tbin}, along with further effects, are
reviewed elsewhere in this journal by Will~\cite{wil06} and
Stairs~\cite{sta03}. A recent review by Kramer \& Stairs~\cite{ks08} provides a
detailed account of the double pulsar. Further discussions on the realities of using
pulsars as gravity wave detectors can be found in other review
articles~\cite{rom89, bac96, hob05}.

\newpage


\section{Summary and Future Prospects}
\label{sec:future}

The main aim of this article was to review some of the many
astrophysical applications provided by the present sample of binary
and millisecond radio pulsars that are relevant to gravitational physics. The topics covered here, along with the
bibliography and associated tables of observational parameters, should
be useful to those wishing to delve deeper into the vast body of literature
that exists.

Through an understanding of the Galactic population of radio pulsars
summarised in Section~\ref{sec:gal} it is possible to predict the detection
statistics of terrestrial gravitational wave detectors to nearby
rapidly spinning neutron stars (see Section~\ref{sec:nmsppop}), as well as
coalescing relativistic binaries at cosmic distances
(see Section~\ref{sec:relpop}). Continued improvements in gravitational wave
detector sensitivities should result in a number of interesting
developments and contributions in this area. These developments and
contributions might include the detection of presently known radio
pulsars, as well as a population of coalescing binary systems which
have not yet been detected as radio pulsars.

The phenomenal timing stability of radio pulsars leads naturally to a
large number of applications, including their use as laboratories for
relativistic gravity (see Section~\ref{sec:gr}), constraining the equation of state of superdense matter (see Section~\ref{sec:masses}) and as natural detectors
of gravitational radiation (see Section~\ref{sec:gwdet}). Long-term timing
experiments of the present sample of millisecond and binary pulsars
currently underway appear to have tremendous potential in these areas
and may detect the gravitational wave background (if it exists) 
within the next decade.

These applications will benefit greatly from the continuing discovery
of new systems by the present generation of radio pulsar searches
which continue to probe new areas of parameter space. Based on the
results presented in Section~\ref{sec:nmsppop}, it is clear that we are
aware of only a few per cent of the total active pulsar population in our
Galaxy. In all likelihood, then, we have not seen all of the pulsar
zoo. The current and
futures
surveys described in Section~\ref{sec:wheretolook} will ultimately
provide a far more complete census of the Galactic pulsar
population. Three possible ``holy
grails'' of pulsar astronomy which could soon be found are:
\begin{description}
\item[An X-ray/radio millisecond pulsar]
  While the link between millisecond pulsars and X-ray binaries
  appears to be compelling, despite intensive searches~\cite{bbp03},
  no radio pulsations have been detected in these binaries. This could
  be a result of free-free absorption of any radio waves by the thick
  accretion disk, or perhaps quenching of the accelerating potential in
  the neutron star magnetosphere by infalling matter. The recent
  discovery of radio pulsations from two magnetars~\cite{crh06,crhr07}
  demonstrates that radio emission can be found in seemingly unlikely
  sites. A millisecond pulsar in transition between an X-ray and radio
  phase would be another such remarkable discovery.
\item[A pulsar--black hole binary system:]
  Following the wide variety of science possible from the new double
  pulsar J0737$-$3039 (see Sections~\ref{sec:evolution}, \ref{sec:nsns}
  and~\ref{sec:tbin}), a radio pulsar with a black-hole companion would
  without doubt be a fantastic gravitational laboratory. Excellent places
  to look for such systems are globular clusters and the Galactic
  centre~\cite{pl04}.
\end{description}
\begin{description}
\item[A sub-millisecond pulsar:]
  The most rapidly spinning radio pulsar currently known is the 1.39-ms
  PSR~J1748$-$2446ad in
  Terzan~5~\cite{hrs06}. Do neutron stars with kHz
  spin rates exist? Searches now have sensitivity to such
  objects~\cite{bd97} and a discovery of even one would constrain the
  equation of state of matter at high densities. As mentioned in
  Section~\ref{sec:nms}, the R-mode instability may prevent neutron
  stars from reaching such high spin rates~\cite{bil98, cmm03}. Of course,
  this does not stop observers from looking for such objects! The
  current generation of surveys summarized in Section~\ref{sec:wheretolook}
  continues to be more and more sensitive to this putative source population.
\end{description}
The discovery of the eccentric millisecond binary pulsar
J1903+0327~\cite{crl08} (see Sections~\ref{sec:1903} and
\ref{sec:masses}) was completely unprecedented within the framework of
our theories for binary pulsar formation. It therefore serves as a
stark reminder that we should ``expect the unexpected''! Even though
the study of radio pulsars is now 40 years old, these experiences tell
us that many exciting discoveries lie ahead of us.

Pulsar astronomy remains an extremely active area of modern
astrophysics and the next decade will undoubtedly continue to produce
new results from currently known objects as well as new surprises. Although it is often stated that we are bound by available computational resources,
in my opinion, pulsar research is currently limited by a shortage of
researchers, and not necessarily computational resources. Keen students
more than ever are needed to help shape the future of this
exciting and continually evolving field.

\newpage


\section{Acknowledgements}

Many thanks to Chunglee Kim and Evan Kean who
read and provided very useful comments on this revised review,
and to Scott Ransom and Matthew Bailes for providing unpublished data 
given in Table~\ref{tab:bmsps}. I am indebted to a number of colleagues who, as noted in many
of the figure captions, kindly gave permission to use their figures in
this article. Frequent use was made of NASA's magnificent Astrophysics
Data System~\cite{nasaads}, the arXiv e-print service~\cite{lanl} and
of course the {\it Google} during the literature searches. My
research is partially supported by West Virginia EPSCoR through a Research
Challenge Grant.  Finally, I would like to thank the {\it Living Reviews}
editors for their support and patience during the completion of this
long-overdue update.

\newpage

\appendix


\section{Tables of Binary and Millisecond Pulsars}
\label{appendix}

\begin{table}[htbp]
  \caption[Parameters for the 20 isolated millisecond pulsars
    currently known in the Galactic disk.]{Parameters for the 20
    isolated millisecond pulsars currently known in the Galactic
    disk. Four of the pulsars (J0609+2130, J1038+0032, J1753$-$1914
    and J2235+1506) are thought to be the descendants of high-mass
    binary systems. Listed are the spin period $P$, the base-10
    logarithms of the characteristic age $\tau_\mathrm{c}$ and surface
    magnetic field strength $B$ (see Section~\ref{sec:nms}), the
    distance $d$ derived from the Cordes \& Lasio electron density
    model~\cite{cl02a, cl02b} or independently (when available), and
    the transverse speed $v_\mathrm{t}$ inferred from $d$ and a proper
    motion measurement (when available). Key publications for each
    pulsar are referenced to the bibliography.}
  \label{tab:imsps}
  \vskip 4mm
  \centering
  \begin{tabular}{l|rrrrrr}
    \toprule
    \vsp PSR &
    \multicolumn{1}{c}{$P$} &
    \multicolumn{1}{c}{$\log \tau_\mathrm{c}$} &
    \multicolumn{1}{c}{$\log B$} &
    \multicolumn{1}{c}{$d$} &
    \multicolumn{1}{c}{$v_\mathrm{t}$} &
    \multicolumn{1}{c}{Refs.} \\
    &
    \multicolumn{1}{c}{[ms]} &
    \multicolumn{1}{c}{[$\log(\mathrm{yr})$]} &
    \multicolumn{1}{c}{[$\log(\mathrm{G})$]} &
    \multicolumn{1}{c}{[kpc]} &
    \multicolumn{1}{c}{[km/s]} & \\ [0.4 em]
    \midrule
    J0030+0451  &  4.865 &  9.9 &  8.3 & 0.23  & $<$~65 & \cite{lzb01} \\
    J0609+2130  & 55.698 &  9.6 &  9.6 & 1.20  &     ? & \cite{lma04, lxf05} \\
    J0711$-$6830 &  5.491 & 10.4 &  8.2 & 1.04  &   139 & \cite{bjb97, tsb99} \\
    J1024$-$0719 &  5.162 &  9.7 &  8.5 & 0.35  &    45 & \cite{bjb97, tsb99} \\
    J1038+0032  & 28.852 &  9.8 &  9.1 & 2.4\z &     ? & \cite{bjd06} \\ [1.0 em]

    J1453+1902  &  5.792 &  9.9 &  8.4 & 0.95  &    46 & \cite{lmcs07}\\
    J1629$-$6902 &  6.000 & 10.0 &  8.4 & 1.36  &     ? & \cite{eb01b} \\ 
    J1721$-$2457 &  3.497 & 10.0 &  8.2 & 1.56  &     ? & \cite{eb01b} \\
    J1730$-$2304 &  8.123 &  9.9 &  8.6 & 0.51  &    53 & \cite{lnl95, tsb99} \\
    J1744$-$1134 &  4.075 &  9.9 &  8.3 & 0.17  &    20 & \cite{bjb97, tsb99} \\ [1.0 em]

    J1753$-$1914 & 62.955 &  8.7 & 10.1 & 2.7\z &     ? & \cite{lfl06} \\
    J1801$-$1417 &  3.625 & 10.0 &  8.1 & 1.80  &     ? & \cite{fsk04} \\
    J1843$-$1113 &  1.846 &  9.5 &  8.1 & 1.97  &     ? & \cite{hfs04} \\ 
    J1911+1347  &  4.625 &  9.6 &  8.5 & 1.61  &     ? & \cite{lfl06} \\
    B1937+21    &  1.558 &  8.4 &  8.6 & 9.65  &    22 & \cite{bkh82, ktr94} \\ [1.0 em]

    J1944+0907  &  5.185 &  9.7 &  8.5 & 1.8\z &     ? & \cite{mla05} \\
    J2010$-$1323 &  5.223 & 10.2 &  8.2 & 1.3\z &     ? & \cite{jbo07} \\
    J2124$-$3358 &  4.931 &  9.9 &  8.4 & 0.25  &    67 & \cite{bjb97, tsb99} \\
    J2235+1506  & 59.767 & 10.0 &  9.4 & 1.15  &    98 & \cite{cnt93} \\
    J2322+2057  &  4.808 & 10.2 &  8.2 & 0.78  &    89 &\cite{nt95, cnt96} \\
    \bottomrule
  \end{tabular}
\end{table}

\begin{landscape}

\begin{table}
  \caption[Parameters for the 19 eccentric ($e>0.05$) binary pulsars
  currently known that are not in globular clusters.]{Parameters for
  the 19 eccentric ($e>0.05$) binary pulsars currently known that are
  not in globular clusters. In addition to the parameters listed in
  Table~\ref{tab:imsps}, we also give the binary period
  $P_\mathrm{b}$, the projected semi-major axis of the orbit $x$ in
  units of light seconds, the orbital eccentricity $e$, and the
  companion mass $m_2$ evaluated from the mass function assuming a
  pulsar mass of $1.4\,M_\odot$ and an inclination angle of 60 degrees
  (see Section~\ref{sec:tbin}) or (when known) from independent
  measurements.}
  \label{tab:ebpsrs}
  \vskip 4mm
  \centering
  \begin{tabular}{l|rrrrrrrrrr}
    \toprule
    \vsp PSR &
    \multicolumn{1}{c}{$P$} &
    \multicolumn{1}{c}{$\log \tau_\mathrm{c}$} &
    \multicolumn{1}{c}{$\log B$} &
    \multicolumn{1}{c}{$d$} &
    \multicolumn{1}{c}{$v_\mathrm{t}$} &
    \multicolumn{1}{c}{$P_\mathrm{b}$} &
    \multicolumn{1}{c}{$x$} &
    \multicolumn{1}{c}{$e$} &
    \multicolumn{1}{c}{$m_2$} &
    \multicolumn{1}{c}{Refs.} \\
    &
    \multicolumn{1}{c}{[ms]} &
    \multicolumn{1}{c}{[$\log(\mathrm{yr})$]} &
    \multicolumn{1}{c}{[$\log(\mathrm{G})$]} &
    \multicolumn{1}{c}{[kpc]} &
    \multicolumn{1}{c}{[km/s]} &
    \multicolumn{1}{c}{[days]} &
    \multicolumn{1}{c}{[s]} & &
    \multicolumn{1}{c}{[$M_\odot$]} & \\ [0.4 em]
    \midrule
    J0045$-$7319  &   926.276 &  6.5 & 12.3 &  57.00 &   ? &   51.2\z &  174.3\z &  0.81\z &    8.8\z & \cite{bbs95} \\
    J0737$-$3039A &    22.699 &  8.3 &  9.8 &   0.57 &  10 &    0.1\z &    1.4\z &   0.088 &     1.25 & \cite{bdp03, lbk04, ksm06} \\
    J0737$-$3039B & 2773.46\z &  7.7 & 12.2 &   0.57 &  10 &    0.1\z &    1.5\z &   0.088 &     1.34 & \cite{bdp03, lbk04, ksm06} \\
    J1141$-$6545  &   393.898 &  6.2 & 12.1 &   3.20 & 115 &     0.20 &     1.86 &  0.17\z &   1.0\z  & \cite{klm00, obv02} \\
    B1259$-$63    &    47.762 &  5.5 & 11.5 &   4.60 &   ? & 1237\dzz & 1296\dzz &  0.87\z &   10.0\z & \cite{jlm92, jml92} \\ [1.0 em]
    J1518+4904   &    40.935 & 10.3 &  9.0 &   0.70 &  27 &     8.63 &    20.04 &  0.25\z &    1.3\z & \cite{nst96, nst99} \\ 
    B1534+12     &     7.904 &  8.4 & 10.0 &   0.68 &  80 &     0.42 &     3.73 &  0.27\z &    1.3\z & \cite{wol91a, sac98} \\
    J1638$-$4715  &  763.933  &  6.4 & 12.3 &   6.79 &   ? & 1940\dzz & 2382\dzz &  0.955  &    8.1\z & \cite{lyn05, lfl06} \\
    J1740$-$3052  &   570.309 &  5.5 & 12.6 & 10.8\z &   ? &  231\dzz &  756.9\z &  0.58\z &   16\dzz & \cite{sml01, sml03} \\
    J1753$-$2243  &    95.138 &  8.4 & 10.4 &  3.0\z &   ? &    13.64 &   18.1\z &  0.30\z &    0.6\z & \cite{kei08} \\ [1.0 em]
    J1756$-$2251  &    28.462 &  8.6 &  9.7 &  2.5\z &   ? &     0.32 &    2.8\z &  0.18\z &    1.2\z & \cite{fkl05} \\ 
    J1811$-$1736  &   104.182 &  8.9 & 10.1 &   5.94 &   ? &    18.77 &    34.78 &  0.83\z &    0.7\z & \cite{lcm00} \\   
    B1820$-$11    &   279.828 &  6.5 & 11.8 &   6.26 &   ? &  357.8\z &  200.7\z &  0.79\z &    0.7\z & \cite{lm89, pv91} \\
    J1822$-$0848  &   834.839 &  8.0 & 11.5 &   4.20 &   ? &  286.8\z &    97.73 &  0.059  &    0.4\z & \cite{lfl06} \\
    J1829+2456   &    41.010 & 10.1 &  9.2 &   0.75 &   ? &     1.18 &     7.24 &  0.14\z &    1.3\z & \cite{clm04, cha05} \\ [1.0 em]
    J1903+0327   &     2.150 &  9.3 &  8.3 &   6.40 &   ? &    95.17 &   105.61 &  0.44\z &    1.1\z & \cite{crl08} \\ 
    J1906+0746   &   144.072 &  5.1 & 12.2 &   4.53 &   ? &     0.17 &     1.42 & 0.085   &    1.37  & \cite{lsf06,kas08} \\ 
    B1913+16     &    59.030 &  8.0 & 10.4 &   7.13 & 100 &     0.32 &     2.34 &  0.62\z &    1.4\z & \cite{tw82, tw89} \\
    B2303+46     &  1066.371 &  7.5 & 11.9 &   4.35 &   ? &    12.34 &    32.69 &  0.66\z &    1.2\z & \cite{lb90, vk99} \\
    \bottomrule
  \end{tabular}
\end{table}

\end{landscape}

\begin{landscape}

\ifpdf
\begin{longtable}{l|rrrrrrrrrr}
\else
\begin{table}[htbp]
\fi
  \caption[Parameters for 64 low-eccentricity binary pulsars currently
  known in the Galactic disk.]{Parameters for 64 low-eccentricity
  binary pulsars currently known in the Galactic disk. The parameters
  listed are defined in Tables~\ref{tab:imsps} and~\ref{tab:ebpsrs}.}
  \label{tab:bmsps}
\ifpdf\\\else
\begin{tabular}{l|rrrrrrrrrr}
\fi
\toprule
  \vsp PSR &
  \multicolumn{1}{c}{$P$} &
  \multicolumn{1}{c}{$\log \tau_\mathrm{c}$} &
  \multicolumn{1}{c}{$\log B$} &
  \multicolumn{1}{c}{$d$} &
  \multicolumn{1}{c}{$v_\mathrm{t}$} &
  \multicolumn{1}{c}{$P_\mathrm{b}$} &
  \multicolumn{1}{c}{$x$} &
  \multicolumn{1}{c}{$e$} &
  \multicolumn{1}{c}{$m_2$} &
  \multicolumn{1}{c}{Refs.} \\
  &
  \multicolumn{1}{c}{[ms]} &
  \multicolumn{1}{c}{[$\log(\mathrm{yr})$]} &
  \multicolumn{1}{c}{[$\log(\mathrm{G})$]} &
  \multicolumn{1}{c}{[kpc]} &
  \multicolumn{1}{c}{[km/s]} &
  \multicolumn{1}{c}{[days]} &
  \multicolumn{1}{c}{[s]} & &
  \multicolumn{1}{c}{[$M_\odot$]} & \\ [0.4 em]
\midrule
\ifpdf
\endfirsthead
\multicolumn{11}{c}{\small{\tablename} \thetable{} -- {\it Continued}}
\\[4mm]
\toprule
  \vsp PSR &
  \multicolumn{1}{c}{$P$} &
  \multicolumn{1}{c}{$\log \tau_\mathrm{c}$} &
  \multicolumn{1}{c}{$\log B$} &
  \multicolumn{1}{c}{$d$} &
  \multicolumn{1}{c}{$v_\mathrm{t}$} &
  \multicolumn{1}{c}{$P_\mathrm{b}$} &
  \multicolumn{1}{c}{$x$} &
  \multicolumn{1}{c}{$e$} &
  \multicolumn{1}{c}{$m_2$} &
  \multicolumn{1}{c}{Refs.} \\
  &
  \multicolumn{1}{c}{[ms]} &
  \multicolumn{1}{c}{[$\log(\mathrm{yr})$]} &
  \multicolumn{1}{c}{[$\log(\mathrm{G})$]} &
  \multicolumn{1}{c}{[kpc]} &
  \multicolumn{1}{c}{[km/s]} &
  \multicolumn{1}{c}{[days]} &
  \multicolumn{1}{c}{[s]} & &
  \multicolumn{1}{c}{[$M_\odot$]} & \\ [0.4 em]
\midrule
\endhead
\fi
  J0034$-$0534 &    1.877 &  9.9 &  7.9\z &  0.98 &    71\dz &    1.59 &    1.44 & $<$~0.00002\zzz & 0.1\z & \cite{bhl94} \\
  J0218+4232  &    2.323 &  8.7 &  8.6\z &  5.85 &     ?\dz &    2.03 &    1.98 & $<$~0.00002\zzz & 0.2\z & \cite{nbf95} \\
  J0407+1607  & 25.702\z &  9.7 &  9.2\z &  1.30 &    35\dz & 669.1\z & 106.5\z &    0.000937\zz & 0.2\z & \cite{lxf05} \\
  J0437$-$4715 &    5.757 &  9.7 &  8.5\z &  0.18 &   121\dz &    5.74 &    3.37 &    0.000019\zz & 0.1\z & \cite{jlh93} \\
  J0613$-$0200 &    3.062 &  9.7 &  8.2\z &  2.19 &    77\dz &    1.20 &    1.09 &    0.000007\zz & 0.1\z & \cite{lnl95, tsb99} \\ [1.0 em]
  J0621+1002  &   28.854 & 10.1 &  9.0\z &  1.88 &    31\dz &    8.31 &   12.03 &    0.0025\zzzz & 0.5\z & \cite{cnst96,sna02} \\
  B0655+64    &  195.671 &  9.7 & 10.1\z &  0.48 &    32\dz &    1.03 &    4.13 &    0.000008\zz & 0.7\z & \cite{jl88, vk95} \\
  J0751+1807  &    3.479 &  9.8 &  8.2\z &  2.02 &     ?\dz &    0.26 &    0.40 & $>$~0.0\zzzzzzz & 0.1\z & \cite{lzc95} \\
  B0820+02    &  864.873 &  8.1 & 11.5\z &  1.43 &    35\dz & 1232.47 &  162.15 &    0.011868\zz & 0.2\z & \cite{mncl80, vk95} \\
  J0900$-$3144 &   11.110 &  9.4 &  8.9\z &  0.82 &     ?\dz &   18.74 &   17.25 &    0.0000103\z & 0.4\z & \cite{bjd06} \\ [1.0 em]
  J1012+5307  &    5.256 &  9.8 &  8.4\z &  0.52 &   102\dz &    0.60 &    0.58 & $<$~0.0000008\z & 0.1\z & \cite{nll95, lcw01} \\ 
  J1022+1001  &   16.453 &  9.8 &  8.9\z &  0.60 & $>$~50\dz &    7.81 &   16.77 &    0.000098\zz & 0.7\z & \cite{cnst96, kxc99} \\
  J1045$-$4509 &    7.474 & 10.0 &  8.5\z &  3.25 &    52\dz &    4.08 &    3.02 &    0.000024\zz & 0.2\z & \cite{lnl95, tsb99} \\
  J1125$-$6014 &     2.63 & 10.0 &  8.0\z &  1.94 &     ?\dz &    8.75 &    8.34 &    0.0000007913\z & 0.3\z & \cite{lfl06} \\
  J1157$-$5114 &   43.589 &  9.7 &  9.4\z &  1.88 &     ?\dz &    3.51 &   14.29 &    0.00040\zzz & 1.2\z & \cite{eb01} \\ [1.0 em]
  J1216$-$6410 &    3.539 & 10.5 &  7.9\z &  1.72 &     ?\dz &    4.04 &    2.94 &    0.000006809 & 0.2\z & \cite{lfl06} \\
  J1232$-$6501 &   88.282 &  9.2 &  9.9\z & 10.00 &     ?\dz &    1.86 &    1.61 &    0.00011\zzz & 0.1\z & \cite{clm01} \\
  B1257+12    &    6.219 &  9.5 &  8.6\z &  0.62 &   281\dz & \multicolumn{4}{c}{planetary system} & \cite{wf92, wol94} \\ 
  J1420$-$5625 &   34.117 &  9.9 &  9.2\z &  1.74 &     ?\dz &  40.3\z &  29.5\z &    0.0035\zzzz & 0.4\z & \cite{hfs04} \\
  J1435$-$6100 &    9.348 &  9.8 &  8.7\z &  3.25 &     ?\dz &    1.35 &    6.18 &    0.00001\zzz & 0.9\z & \cite{clm01} \\[1.0 em]
  J1439$-$5501 &   28.635 &  9.5 &  9.3\z &  0.77 &     ?\dz &    2.11 &    9.83 &    0.00004985 & 1.4\z & \cite{lfl06} \\
  J1454$-$5846 &   45.249 &  9.0 &  9.8\z &  3.32 &     ?\dz &   12.42 &   26.52 &    0.0019\zzzz & 0.9\z & \cite{clm01} \\
  J1455$-$3330 &    7.987 &  9.9 &  8.5\z &  0.74 &   100\dz &   76.17 &   32.36 &    0.000170\zz & 0.3\z & \cite{lnl95, tsb99} \\
  J1528$-$3146 &   60.822 &  9.6 &  9.6\z &  0.99 &     ?\dz &    3.18 &   11.45 &    0.0002130\z & 1.2\z & \cite{jbo07} \\
  J1600$-$3053 &    3.598 &  9.8 &  8.3\z &  2.67 &    86\dz &   14.35 &    8.80 &    0.0001737\z & 0.2\z & \cite{jbo07} \\ 
  J1603$-$7202 &   14.842 & 10.2 &  8.7\z &  1.64 &    27\dz &    6.31 &    6.88 & $>$~0.0\zzzzzzz & 0.3\z & \cite{llb96, tsb99} \\ 
  J1614$-$2230 &    3.151 &  9.7 &  8.2\z &  1.28 &     ?\dz &    8.69 &   11.29 & 0.00000122  &  0.5\z & \cite{hrr05,ran08a} \\
  J1614$-$2318 &   33.504 & 10.0 &  9.1\z &  1.80 &     ?\dz &    3.15 &    1.33 &     0           & 0.09\z & \cite{hrr05,ran08a} \\
  J1618$-$3921 &   11.987 &  9.5 &  8.9\z &  4.76 &     ?\dz &   22.8\z &  10.2\z &    0.027\zzzzz & 0.2\z & \cite{eb01b,bai08} \\  
  J1640+2224  &    3.163 & 10.2 &  8.0\z &  1.18 &    76\dz &  175.46 &   55.33 &     0.0008\zzzz & 0.3\z & \cite{wdk00} \\  [1.0 em]
  J1643$-$1224 &    4.622 &  9.7 &  8.4\z &  4.86 &   159\dz &  147.02 &   25.07 &     0.000506\zz & 0.1\z & \cite{lnl95, tsb99} \\
  J1709+2313  &    4.631 & 10.3 &  8.1\z &  1.83 &    89\dz &   22.7\z &  15.3\z &    0.0000187\z & 0.3\z & \cite{lwf04} \\ 
  J1711$-$4322 &  102.618 &  7.8 & 10.7   &  4.18 &     ?\dz &  922.47 & 139.62  &     0.002376\zz & 0.2\z & \cite{lfl06} \\
  J1713+0747  &    4.570 & 10.0 &  8.3\z &  0.89 &    27\dz &   67.83 &   32.34 &     0.000075\zz & 0.3\z & \cite{fwc93, cfw94} \\ 
  J1732$-$5049 &    5.313 &  9.8 &  8.4\z &  1.81 &     ?\dz &    5.26 &    3.98 &     0.0000098\z & 0.2\z & \cite{eb01b} \\  [1.0 em]
  J1738+0333  &    5.850 &  9.6 &  8.6\z &  1.97 &    64\dz &    0.35 &    0.34 &     0.000004\zz & 0.1\z & \cite{jac04} \\
  J1744$-$3922 &  172.444 &  9.2 &  10.2  &  4.60 &     ?\dz &    0.19 &    0.21 &     0.001281\zz & 0.1\z & \cite{fsk04,brr07} \\ 
  J1745$-$0952 &   19.376 &  9.5 &  9.1\z &  2.38 &     ?\dz &    4.94 &    2.38 &     0.0000180\z & 0.1\z & \cite{eb01b} \\
  J1751$-$2857 &    3.915 &  9.7 &  8.3\z &  1.40 &     ?\dz &  110.75 &   32.53 &     0.000128\zz & 0.2\z & \cite{sfl05} \\
  J1757$-$5322 &    8.870 &  9.7 &  8.7\z &  1.36 &     ?\dz &    0.45 &    2.09 &     0.000004\zz & 0.6\z & \cite{eb01} \\  [1.0 em]
  B1800$-$27   &  334.415 &  8.5 & 10.9\z &  3.62 &     ?\dz &  406.78 &   58.94 &     0.000507\zz & 0.1\z & \cite{jlm92} \\ 
  J1802$-$2124 &   12.648 &  9.4 &  9.0\z &  3.33 &     ?\dz &    0.70 &    3.72 &     0.000003162 & 1.0\z & \cite{fsk04} \\ 
  J1804$-$2717 &    9.343 &  9.5 &  8.8\z &  1.17 &     ?\dz &   11.13 &    7.28 &     0.000035\zz & 0.2\z & \cite{llb96} \\
  J1810$-$2005 &   32.822 &  9.6 &  9.3\z &  4.04 &     ?\dz &   15.01 &   11.98 &     0.000025\zz & 0.3\z & \cite{clm01} \\
  B1831$-$00   &  520.954 &  8.8 & 10.9\z &  2.63 &     ?\dz &    1.81 &    0.72 & $>$~0.0\zzzzzzz & 0.1\z & \cite{dmr86} \\  [1.0 em]
  J1841+0130  &   29.773 &  7.8 & 10.2\z &  3.19 &     ?\dz &   10.47 &    3.50 &     0.00008194\z & 0.1\z & \cite{lfl06} \\
  J1853+1303  &    4.092 &  9.9 &  8.3\z &  1.60 &     ?\dz &  115.65 &  40.8\z &     0.0000237\z & 0.3\z & \cite{sfl05} \\ 
  B1855+09    &    5.362 &  9.7 &  8.5\z &  1.00 &     29.2 &   12.33 &    9.23 &     0.000022\zz & 0.2\z & \cite{srs86, ktr94} \\ 
  J1904+0412  &   71.095 & 10.1 &  9.4\z &  4.01 &     ?\dz &   14.93 &    9.63 &     0.00022\zzz & 0.2\z & \cite{clm01} \\
  J1909$-$3744 &    2.947 &  9.5 &  8.3\z &  0.82 &   143\dz &    1.53 &    1.89 &     0.00000026  & 0.2\z & \cite{jbv03} \\  
  J1910+1256  &    4.984 &  9.9 &  8.3\z &  1.90 &     ?\dz &   58.47 &  21.1\z &     0.0002302\z & 0.2\z & \cite{sfl05} \\
  J1911$-$1114 &    3.626 &  0.0 &  0.0\z &  1.59 &   183\dz &    2.72 &    1.76 & $>$~0.0\zzzzzzz & 0.1\z & \cite{llb96, tsb99} \\ 
  J1918$-$0642 &    7.646 &  9.7 &  8.6\z &  1.40 &     ?\dz &   10.91 &    8.35 &     0.0000222\z & 0.3\z & \cite{eb01b} \\ 
  J1933$-$6211 &    3.543 & 10.2 &  8.1\z &  0.63 &     ?\dz &   12.82 &   12.28 &     0.000001230\z & 0.4\z & \cite{jbo07} \\
  B1953+29    &    6.133 &  9.5 &  8.6\z &  5.39 &    98\dz &  117.35 &   31.41 &     0.00033\zzz & 0.2\z & \cite{bbf83, wdk00} \\ [1.0 em]
  B1957+20    &    1.607 &  9.4 &  8.1\z &  1.53 &   190\dz &    0.38 &    0.09 & $>$~0.0\zzzzzzz &  0.02 & \cite{fst88, aft94} \\
  J2016+1948  &   64.940 &  9.4 &  9.7\z &  1.84 &     ?\dz &  635.0\z & 150.7\z &    0.00128\zzz &       & \cite{naf03, lf05} \\ 
  J2019+2425  &    3.935 & 10.4 &  8.0\z &  0.91 &    83\dz &   76.51 &   38.77 &     0.000111\zz & 0.3\z & \cite{nt95, nss01} \\
  J2033+1734  &    5.949 &  9.9 &  8.4\z &  1.38 &     ?\dz &   56.31 &   20.16 &     0.00013\zzz & 0.2\z & \cite{rtj96} \\ 
  J2051$-$0827 &    4.509 &  9.7 &  8.4\z &  1.28 &    14\dz &    0.10 &    0.05 & $>$~0.0\zzzzzzz &  0.03 & \cite{sbl96, tsb99} \\ [1.0 em]
  J2129$-$5721 &    3.726 &  9.5 &  8.4\z &  2.55 &    56\dz &    6.63 &    3.50 & $>$~0.0\zzzzzzz & 0.1\z & \cite{llb96, tsb99} \\
  J2145$-$0750 &   16.052 & 10.3 &  8.6\z &  0.50 &    38\dz &    6.84 &   10.16 &     0.000019\zz & 0.4\z & \cite{bhl94, tsb99} \\ 
  J2229+2643  &    2.978 & 10.4 &  7.9\z &  1.43 &   113\dz &   93.02 &   18.91 &     0.00026\zzz & 0.1\z & \cite{wdk00} \\
  J2317+1439  &    3.445 & 10.6 &  7.9\z &  1.89 &    68\dz &    2.46 &    2.31 & $>$~0.0\zzzzzzz & 0.2\z & \cite{cnt93, cam95a} \\
\bottomrule
\ifpdf
\end{longtable}
\else
\end{tabular}
\end{table}
\fi

\end{landscape}
\newpage


\bibliography{refs}

\end{document}